\documentclass[12pt]{article}
\pdfoutput=1
\usepackage[a4paper]{geometry}
\usepackage{jheppub, amsmath,amssymb,amsfonts,amsxtra, mathrsfs, makeidx,graphics,graphicx,amsthm,epsfig, ytableau,bm,longtable,float, color,tikz,mathtools,xfrac,footnote,rotating, lscape, makecell, environ,mathtools, empheq}

\usetikzlibrary{positioning}
\usetikzlibrary{chains}
\usetikzlibrary{arrows,fit,decorations.pathreplacing}
\tikzstyle{every picture}+=[remember picture]
\tikzstyle{na} = [baseline]

\usetikzlibrary{arrows, decorations.markings, calc, fadings, decorations.pathreplacing, patterns, decorations.pathmorphing, positioning}

\def\node#1#2{\overset{#1}{\underset{#2}{{\color{gray} \bullet}}}}

\def\sqwnode#1#2{\overset{#1}{\underset{#2}{{\square}}}}

\def\node#1#2{\overset{#1}{\underset{#2}{\circ}}}

\def\sqwnode#1#2{\overset{#1}{\underset{#2}{{ \square}}}}

\def\wver#1#2{\overset{{\llap{$\scriptstyle#1$}\displaystyle{\square}{\rlap{$\scriptstyle#2$}}}}{\scriptstyle\vert}}

\tikzstyle{every picture}+=[remember picture]
\tikzstyle{na} = [baseline=-.5ex]

\usepackage{bbm}
\usepackage{subfigure}

\providecommand{\abs}[1]{\lvert#1\rvert}

\newcommand{\eg}{\textit{e.g.}}

\newcommand{\ie}{\textit{i.e.}}

\numberwithin{equation}{section}

\newcommand{\nn}{\nonumber}

\newcommand{\be}{\begin{equation}} \newcommand{\ee}{\end{equation}}
\newcommand{\bea}{\begin{equation} \begin{aligned}} \newcommand{\eea}{\end{aligned} \end{equation}}

\def\tilde{\widetilde}

\def\hat{\widehat}

\def\bar{\overline}

\def\cm{\mathsf{m}}

\def\rt2{\sqrt{2}}

\def\det{\mathop{\rm det}}

\def\tr{\mathop{\rm tr}}

\def\abs#1{\left|#1\right|}

\def\CC{{\cal C}}

\def\CH{{\cal H}}

\def\CK{{\cal K}}

\def\CM{{\cal M}}
\def\CN{{\cal N}}

\def\1{{\ds 1}}

\newcommand{\cA}{\mathcal{A}}
\newcommand{\cB}{\mathcal{B}}

\newcommand{\cN}{\mathcal{N}}

\def\repa{\raise4pt\hbox{$\square$}\mkern-14mu\raise-4pt\hbox{$\square$}}
\def\repab{\overline{\raise4pt\hbox{$\square$}\mkern-14mu\raise-4pt\hbox{$\square$}\mkern-1mu}}

\def\smileface{\ensuremath{\hbox{\large$\bigcirc$}\mkern-15mu\raise-1pt\hbox{\scriptsize$\smallsmile$}%
\mkern-10mu\raise4pt\hbox{..}\mkern4mu}}
\def\frownface{\ensuremath{\hbox{\large$\bigcirc$}\mkern-15mu\raise-1pt\hbox{\scriptsize$\smallfrown$}%
\mkern-10mu\raise4pt\hbox{..}\mkern4mu}}

\newcommand{\ba}{\begin{array}}
\newcommand{\ea}{\end{array}}
\newcommand{\bi}{\begin{itemize}}
\newcommand{\ei}{\end{itemize}}
\def\vec#1{\bm{#1}}
\def\bea#1\eea{\allowdisplaybreaks \begin{align}#1\end{align}}
 \newcommand{\ben}{\begin{enumerate}}
\newcommand{\een}{\end{enumerate}}
\newcommand{\bean}{\begin{eqnarray*}}
\newcommand{\eean}{\end{eqnarray*}}
\newcommand{\eref}[1]{(\ref{#1})}

\newcommand{\PE}{\mathop{\rm PE}}

\newcommand{\tQ}{\widetilde{Q}}

\newcommand{\tA}{\widetilde{A}}

\newcommand{\tB}{\widetilde{B}}

\newcommand{\BC}{\mathbb{C}}

\newcommand{\BZ}{\mathbb{Z}}

\newcommand{\BH}{\mathbb{H}}

\newcommand{\comment}[1]{}
\newcommand{\diag}{\mathrm{diag}}

\newcommand{\Sym}{\mathrm{Sym}}

\definecolor{light-gray}{gray}{0.7}

\newcommand{\purple}{\color{purple}}
\newcommand{\blue}{\color{blue}}
\newcommand{\gray}{\color{light-gray}}
\newcommand{\red}{\color{red}}

\def\aup#1 {\overset{#1}{\uparrow} \, \overset{\tilde{#1}}{\downarrow}}

\title{The moduli spaces of $S$-fold CFTs}
\author[a,b]{Ivan Garozzo,} 
\author[a,b]{Gabriele Lo Monaco,} 
\author[b,c]{and Noppadol Mekareeya}
\affiliation[a]{Dipartimento di Fisica, Universit\`a di Milano-Bicocca, \\ Piazza della Scienza 3, I-20126 Milano, Italy}
\affiliation[b]{INFN, sezione di Milano-Bicocca, \\Piazza della Scienza 3, I-20126 Milano, Italy}
\affiliation[c]{Department of Physics, Faculty of Science, \\
Chulalongkorn University, Phayathai Road, \\
Pathumwan, Bangkok 10330, Thailand}
\emailAdd{ivangarozzo@gmail.com}
\emailAdd{gabriele.lomonaco92@gmail.com}
\emailAdd{n.mekareeya@gmail.com}
\abstract{An $S$-fold has played an important role in constructing supersymmetric field theories with interesting features.  It can be viewed as a type of $\mathrm{AdS}_4$ solutions of Type IIB string theory where the fields in overlapping patches are glued by elements of $SL(2,\BZ)$. This paper examines three dimensional quiver theories that arise from brane configurations with an inclusion of the $S$-fold. An important feature of such a quiver is that it contains a link, which is the $T(U(N))$ theory, between two $U(N)$ groups, along with bifundamental and fundamental hypermultiplets. We systematically study the moduli spaces of those quiver theories, including the cases in which the non-zero Chern--Simons levels are turned on.  A number of such moduli spaces turns out to have a very rich structure and tells us about the brane dynamics in the presence of an $S$-fold.}

\begin{document}
\maketitle

\section{Introduction}
Mirror symmetry \cite{Intriligator:1996ex} in three dimensional $\CN=4$ gauge theories is one of the most important dualities that relates theories with non-trivial infrared fixed point. For a pair of theories that are related by mirror symmetry, the duality exchanges the Higgs and Coulomb branches of such theories. Quantum effects on the Coulomb branch arise classically on the Higgs branch of the dual theory.  This symmetry admits realisations in string theory \cite{deBoer:1996mp, Porrati:1996xi, Hanany:1996ie}; one of which involves $S$-duality on Type IIB brane systems, consisting of D3, NS5 and D5 branes, preserving eight supercharges \cite{Hanany:1996ie}.   This type of brane systems (which we shall refer to as the Hanany--Witten brane configuration) gives rise to three dimensional quiver theories, and the mirror theory can be easily derived by considering the $S$-dual of the aforementioned brane system.  This provides a very powerful method in obtaining a large class of mirror theories in three dimensions.   An interesting generalisation to this is to consider, not just $S$-duality, but the action of full $SL(2,\BZ)$ duality group inherited from Type IIB string theory on the quiver theories \cite{Witten:2003ya, Gaiotto:2008ak}. The more general dualities relate, for example, 3d $\CN=4$ gauge theories with zero Chern--Simons levels to Chern--Simons--matter theories \cite{Jafferis:2008em, Gaiotto:2008sd, Hosomichi:2008jd, Imamura:2008nn, Assel:2014awa}.  In general, the latter theories admit $\CN=3$ supersymmetric Lagrangian descriptions; however, the amount of supersymmetry at the fixed point can get enhanced and range from $\CN=4$ to $\CN=8$ \cite{Aharony:2008ug, Hosomichi:2008jb, Aharony:2008gk}.

A certain class of 3d $\CN=4$ superconformal field theories can be realised on the half-BPS domain wall, also known as the ``Janus domain wall'' or ``Janus interface'', of the four dimensional $\CN=4$ super-Yang-Mills theory \cite{DHoker:2006qeo, Gaiotto:2008ak}.  One that plays an important role in this paper is known as $T(U(N))$.  We summarise a necessary detail of this theory in section \ref{sec:review}.   The theory $T(U(N))$ is invariant under mirror symmetry and has a global symmetry $U(N) \times U(N)$, where one of the two $U(N)$ is manifest in the Lagrangian description, whereas the other is not but gets enhanced in the infrared.  We may gauge such $U(N)$ symmetries and couple them to matter.  In this way, we can form a quiver theory such that $T(U(N))$ is a link connecting two $U(N)$ gauge groups; an example of this is depicted in the right diagram of \eref{T1NN1}.  We may also turn on a non-zero Chern--Simons level for either or both $U(N)$ gauge groups; an example is depicted in \eref{NkN011}, where the Chern--Simons level $k$ is turned on for one of the $U(N)$ gauge groups.   The main aim of this paper is to study the moduli space of such theories.  It should be noted that for $N=1$, the quiver that contains only $T(U(1))$ links between $U(1)$ gauge groups (possibly with non-zero Chern--Simons levels), but without bi-fundamental and fundamental matters, gives rise to a abelian pure Chern--Simons theory with mixed Chern--Simons terms between gauge groups.  Such abelian theories were studied in detailed in \cite{Ganor:2014pha}. 

One important motivation to study quiver theories with $T(U(N))$ links (with or without non-trivial Chern--Simons levels for the $U(N)$ gauge groups) is because they have interesting holographic duals \cite{Assel:2018vtq}.  The construction involves $\mathrm{AdS}_4 \times K_6$ Type IIB string solutions with monodromies\footnote{It should be mentioned that a similar solution in $\mathrm{AdS}_5$ was considered in \cite{Garcia-Etxebarria:2015wns, Aharony:2016kai}, and those in $\mathrm{AdS}_3$ were considered in \cite{Couzens:2017way, Couzens:2017nnr}.} in $K_6$ in the $S$-duality group $SL(2,\BZ)$. These solutions were obtained by quotienting the solutions corresponding to the holographic dual of Janus interfaces in 4d $\CN=4$ super--Yang--Mills \cite{DHoker:2007zhm, DHoker:2007hhe}.  The former type of solutions is referred to as the {\bf $S$-fold} in \cite{Inverso:2016eet, Assel:2018vtq}.\footnote{Its supersymmetry and relation to a singular limit of previously known Janus solutions  \cite{DHoker:2007zhm, DHoker:2007hhe} were also found in \cite{Inverso:2016eet}.} The $S$-fold solutions can be divided into two classes, known as the {\bf $J$-fold} and the {\bf $S$-flip}.  

The $J$-fold solutions are those associated with a monodromy given by an element $J \in SL(2,\BZ)$ with $\tr J >2$.  The corresponding geometry can be constructed by using $\mathrm{AdS}_4 \times S^2 \times S^2 \times \Sigma_2$, where $\Sigma_2$ is a non-compact Riemann surface with the topology of a strip. The ends of the strip are then identified with a $J$-twisted boundary condition.  It was shown in \cite{Assel:2018vtq} that this type of solutions preserve $OSp(4|4)$ symmetry and thus are dual to 3d $\CN=4$ superconformal field theories. The $J$-fold solutions can, in fact, be obtained as a quotient of a Janus interface solution.  As a result, the quiver field theory dual of such a solution contains a component corresponding to such an interface, namely the $T(U(N))$ theory.  From the brane perspective, one can introduce a five-dimensional surface implementing the monodromy under the action of $J$ into the brane system.   Among the possible choices of the $SL(2,\BZ)$ elements, we may take the monodromy to be associated with $J_k = -S T^k$ in this case, the corresponding $J$-fold gives rise to a Chern--Simons level $k$ to one of the $U(N)$ gauge groups.  An example of such a configuration and the corresponding quiver theory is given by \eref{introJfold} and \eref{NkN011}.  

The $S$-flip solutions can be discussed in a similar way as for the $J$-folds.  In this case, the $SL(2,\BZ)$ element implementing the monodromy is taken to be $S$.  Geometrically, we need to perform an exchange of coordinates corresponding to the two $S^2$ in $\mathrm{AdS}_4 \times S^2 \times S^2 \times \Sigma_2$, together with a flip at the S-interface such that $\Sigma_2$ becomes a M\"obius strip topologically.  Similarly to the $J$-fold, the insertion of the $S$-flip into a brane system gives rise to a $T(U(N))$ link between two $U(N)$ gauge groups, where the Chern--Simons levels of those are zero.  It was shown in \cite{Assel:2018vtq} that the $S$-fold solutions preserve $OSp(3|4)$ and the dual superconformal field theory is expected to have $\CN = 3$ supersymmetry.

In this paper, we consider the Hanany--Witten brane systems with an insertion of $S$-flips or $J$-folds, as well as the three dimensional quiver theories that arise on the worldvolume of the D3 branes.  Let us summarise the main points.  For the system with an $S$-flip, the quiver consists of a $T(U(N))$ link between two $U(N)$ gauge groups with zero Chern--Simons level.  We find that such a theory has two branches of the moduli space, namely the Higgs and the Coulomb branches.  The Higgs branch of such theories is given by a hyperK\"ahler quotient described at the beginning of section \ref{sec:zeroCS}.  The Coulomb branch, on the other hand, can be computed in a very similar way to the usual 3d $\CN=4$ gauge theories \cite{Cremonesi:2013lqa}, with the remark that the Coulomb branch dynamics does not receive a contribution from the vector multiplet from the gauge groups that are linked by $T(U(N))$.  In other words, the segment of the D3 branes passing through the $S$-flip does not move along the Coulomb branch directions. We also check that these results are consistent with mirror symmetry, namely the Higgs ({\it resp.} Coulomb) branch of a given theory agrees with the Coulomb ({\it resp.} Higgs) branch of the mirror theory, obtained by applying $S$-duality to the original brane system.  Subsequently, we turn on non-zero Chern--Simons levels for the $U(N)$ gauge groups in the quiver.  We focus on the abelian theories in section \ref{sec:abel}.  The models analysed in this section are, in fact, a generalisation of those studied in \cite{Ganor:2014pha, Assel:2017eun, Assel:2018vtq} in the sense that we also include bifundamental and fundamental matter, along with the $J$-fold, in the quivers.  This makes the moduli space become highly non-trivial; for example, it may contains many non-trivial branches.  We, however, do not have a general prescription to compute the moduli space for non-abelian theories with $T(U(N))$ links and non-zero CS levels.  Nevertheless, in section \ref{sec:nonabel}, we show that, for theories that arise from $N$ M2-branes probing Calabi-Yau 4-fold singularities, it is possible to compute the Hilbert series for each configuration of magnetic fluxes. 

The paper is organised as follows. In section \ref{sec:review}, we give a brief summary of the brane configurations for linear quivers and compact models, as well as a brief review on the $S$-fold solutions and $(p,q)$ fivebranes.  In section \ref{sec:zeroCS}, quiver theories corresponding to the brane systems with $S$-flips are examined.  The Higgs and the Coulomb branches of the moduli space are studied using the Hilbert series.  We also provide a consistency check of our results against mirror symmetry.  In section \ref{sec:abel}, we then consider abelian theories arise from the brane systems with $J$-folds, along with NS5 and D5 branes.  We systematically analyse various branches of the moduli space.  In section \ref{sec:nonabel}, we examine an example of non-abelian theory with $T(U(N))$ links that can be realised on M2-branes on a Calabi--Yau four fold singularity.  In this example, we compute the Hilbert series of the moduli space and analyse the contribution from each configuration of magnetic fluxes.  We conclude the paper in section \ref{sec:conclusion} and discuss about some open problems for future work.  The technical analysis for theories with many $J$-folds is collected in Appendix \ref{app:multiJ}.

\section{$S$-fold solutions and their SCFT duals} \label{sec:review}
A large class of $\cN=4$ quiver gauge theories in three dimensions can be engineered using brane systems involving D3, D5, NS5 branes \cite{Hanany:1996ie}.  Each type of branes spans the following directions:
\be
\begin{tabular}{c||cccccc|c|cccc}
\hline
~     & 0 & 1 &2 &3 & 4 & 5 & 6 & 7 &8 &9 &\\
\hline
D3   & X & X & X&  &  &   & X&  & & & \\ 
NS5 & X & X & X& X & X & X &    &   & & &\\
D5   & X & X & X&  &  &  &  &X &X &X & \\
\end{tabular}
\ee
The $x^6$ direction can be taken to be compact or non-compact.  

\subsection{Linear quivers: $T^{\vec \sigma}_{\vec \rho} (SU(N))$ and its variants}
If $x^6$ direction is non-compact, we obtain a linear quiver of the form
\be \label{quivTsigrhoSUN}
\begin{tikzpicture}[font=\footnotesize]
\begin{scope}[auto,%
  every node/.style={draw, minimum size=1.1cm}, node distance=0.6cm];
\node[circle] (UN1) at (0, 0) {$N_1$};
\node[circle, right=of UN1] (UN2) {$N_2$};
\node[draw=none, right=of UN2] (dots) {$\cdots$};
\node[circle, right=of dots] (UNlm1) {$N_{\ell'-2}$};
\node[circle, right=of UNlm1] (UNl) {$N_{\ell'-1}$};
\node[rectangle, below=of UN1] (UM1) {$M_1$};
\node[rectangle, below=of UN2] (UM2) {$M_2$};
\node[rectangle, below=of UNlm1] (UMlm1) {$M_{\ell'-2}$};
\node[rectangle, below=of UNl] (UMl) {$M_{\ell'-1}$};
\end{scope}
\draw (UN1) -- (UN2)
(UN2)--(dots)
(dots)--(UNlm1)
(UNlm1)--(UNl)
(UN1)--(UM1)
(UN2)--(UM2)
(UNlm1)--(UMlm1)
(UNl)--(UMl);
\end{tikzpicture}
\ee
where a circular node with a label $N$ denotes a $U(N)$ gauge group and a square node with a label $M$ denotes a $U(M)$ flavour symmetry.  This class of linear quivers was studied in \cite{Gaiotto:2008ak} and each of the theories in this class is represented by $T^{\vec \sigma}_{\vec \rho} (SU(N))$ for some $N$, with $\vec \sigma$ and $\vec \rho$ partitions of $N$.   

From the brane perspective, if we move the D5-branes to one side and the NS5-branes to the other side, $N$ is the total number of D3-branes in the middle, $\vec \sigma$ contains the differences between the number of D3-branes on the left and on the right of each D5-brane, and $\vec \rho$ contains the differences between the number of D3-branes on the left and on the right of each NS5-brane.  Let us provide an example for $N=6$, $\vec \sigma = (3,2,1)$ and $\vec \rho=(2^2,1^2)$:
\be
\begin{tikzpicture}[baseline]
\draw (0,0)--(0,2.5) node[black,midway, xshift =0cm, yshift=1.5cm]{\footnotesize };
\draw (1,0)--(1,2.5) node[black,midway, xshift =0cm, yshift=1.5cm]{\footnotesize }; 
\draw (2,0)--(2,2.5) node[black,midway, xshift =-0.5cm, yshift=-1.7cm] {\footnotesize NS5} node[black,midway, xshift =0cm, yshift=1.5cm]{\footnotesize};
\draw (3,0)--(3,2.5) node[black,midway, xshift =0cm, yshift=1.5cm]{\footnotesize }; 
\draw [dashed,red] (-3,0)--(-3,2.5) node[black,midway, xshift =0cm, yshift=1.5cm] {\footnotesize};
\draw [dashed,blue] (-4,0)--(-4,2.5) node[black,midway, xshift =0cm, yshift=-1.7cm] {\footnotesize D5} node[black,midway, xshift =0cm, yshift=1.5cm] {\footnotesize};
\draw [dashed,purple] (-5,0)--(-5,2.5)  node[black,midway, xshift =0cm, yshift=1.5cm] {\footnotesize};
\draw [red] (0,2)--(-3,2) node[black,midway, yshift=0.4cm] {\tiny D3}; \draw [red] (1,1.8)--(-3,1.8); \draw [red](2,1.6)--(-3,1.6);
\draw [blue] (0,1)--(-4,1); \draw [blue](1,0.8)--(-4,0.8);
\draw [purple](0,0.4)--(-5,0.4);
\draw (0,0.6)--(1,0.6); \draw (1,1.1)--(2,1.1); \draw (2,1.4)--(3,1.4); 
\end{tikzpicture}  
\ee
To read off the quiver gauge theory, it is convenient to move the D5-branes inside the NS5-brane intervals as follows:
\be
\begin{tikzpicture}[baseline]
\draw (0,0)--(0,2.5) node[black,midway, xshift =0cm, yshift=1.5cm] {\footnotesize};
\draw (1,0)--(1,2.5) node[black,midway, xshift =0cm, yshift=1.5cm] {\footnotesize}; 
\draw (2,0)--(2,2.5) node[black,midway, xshift =0cm, yshift=1.5cm] {\footnotesize };
\draw (3,0)--(3,2.5) node[black,midway, xshift =0cm, yshift=1.5cm] {\footnotesize};
\draw [dashed,purple] (0.5,0)--(0.5,2.5) node[black,midway, xshift =0cm, yshift=-1.5cm] {\footnotesize};
\draw [dashed,blue] (1.5,0)--(1.5,2.5) node[black,midway, xshift =0cm, yshift=-1.5cm] {\footnotesize};
\draw [dashed,red] (2.5,0)--(2.5,2.5) node[black,midway, xshift =0cm, yshift=-1.5cm] {\footnotesize};
\draw (0,0.6)--(1,0.6); \draw (1,1.1)--(2,1.1); \draw (2,1.4)--(3,1.4); 
\end{tikzpicture}  
\hspace{3cm}
\begin{tikzpicture}[scale=0.7, transform shape]
\begin{scope}[auto,%
  every node/.style={draw, minimum size=1.1cm}, node distance=0.6cm];
\node[circle] (UN1) at (0, 0) {$1$};
\node[circle, right=of UN1] (UN2) {$1$};
\node[circle, right=of UN2] (UN3) {$1$};
\node[rectangle, below=of UN1] (UM1) {$1$};
\node[rectangle, below=of UN2] (UM2) {$1$};
\node[rectangle, below=of UN3] (UM3) {$1$};
\end{scope}
\draw (UN1) -- (UN2)
(UN2)--(UN3)
(UN1)--(UM1)
(UN2)--(UM2)
(UN3)--(UM3);
\end{tikzpicture}
\ee
Since three dimensional mirror symmetry \cite{Intriligator:2013lca} exchanges D5-brane and NS5-branes \cite{Hanany:1996ie}, it also exchanges $\vec \sigma$ and $\vec \rho$.
A quiver description of $T^{\vec \sigma}_{\vec \rho} (SU(N))$ for a general $\vec \sigma$ and $\vec \rho$ can be found in, for example, \cite[sec. 2]{Assel:2011xz} or \cite[sec 2.1]{Cremonesi:2014uva}.

%
\paragraph{The $T(SU(N))$ theory.} A theory that plays an important role in this paper is that with $\vec \sigma = \vec \rho = [1^N]$. Such a theory is denoted by $T(SU(N))$ and its quiver description is
\be \label{12N} 
\node{}{1}-\node{}{2}- \cdots - \node{}{N-1}-\sqwnode{}{N}~.
\ee
As an explicit example, the brane configurations for $T(SU(3))$ are as follows:
\be\label{TSU3}
\begin{tikzpicture}[baseline]
\draw (0,0)--(0,2.5) node[black,midway, xshift =0cm, yshift=1.5cm]{\footnotesize };
\draw (1,0)--(1,2.5) node[black,midway, xshift =0cm, yshift=1.5cm]{\footnotesize }; 
\draw (2,0)--(2,2.5) node[black,midway, xshift =-0.5cm, yshift=-1.7cm] {\footnotesize NS5} node[black,midway, xshift =0cm, yshift=1.5cm]{\footnotesize};
\draw [dashed,red] (-3,0)--(-3,2.5) node[black,midway, xshift =0cm, yshift=1.5cm] {\footnotesize};
\draw [dashed,blue] (-4,0)--(-4,2.5) node[black,midway, xshift =0cm, yshift=-1.7cm] {\footnotesize D5} node[black,midway, xshift =0cm, yshift=1.5cm] {\footnotesize};
\draw [dashed,purple] (-5,0)--(-5,2.5)  node[black,midway, xshift =0cm, yshift=1.5cm] {\footnotesize};
\draw [red] (0,1.5)--(-3,1.5) node[black,midway, yshift=0.4cm] {\tiny D3}; 
\draw [blue] (0,1)--(-4,1); 
\draw [purple](0,0.4)--(-5,0.4);
\draw (0,0.6)--(1,0.6); \draw (1,1.1)--(2,1.1); \draw (0,1.3)--(1,1.3); 
\end{tikzpicture}  
\ee 
In general $T(SU(N))$ is invariant under mirror symmetry.  The Higgs and the Coulomb branches of this theory are both isomorphic to the closure of the maximal nilpotent orbit of $SU(N)$ \cite{Gaiotto:2008ak}, which is denoted by $\CN_{SU(N)}$. We can conveniently define $\CN_{SU(N)}$ as a set of $N \times N$ complex matrices $M$ such that $\tr(M^p) =0$, for $p=1, \ldots, N$; the quaternionic dimension of this space is therefore $\frac{1}{2}N(N-1)$.  For quiver \eref{12N}, the symmetries of the Higgs and Coulomb branch are thus both $SU(N)$; the former is manifest in the Lagrangian (or quiver) description as a flavour symmetry, whereas the latter is not manifest but gets enhanced from the topological symmetry $U(1)^{N-1}$ in the infrared.

\paragraph{The $T(U(N))$ theory.}  An important variant of the $T(SU(N))$ theory is the $T(U(N))$ theory \cite[sec 4.4]{Gaiotto:2008ak}.   The latter is defined as a product between the $T(SU(N))$ theory and an ``almost trivial'' $T(U(1))$ theory, where the latter can be characterised as follows.   The Coulomb and Higgs branches of $T(U(1))$ are trivial; each of them consists of only one point.  Nevertheless, $T(U(1))$ comes with a $U(1) \times U(1)$ background vector multiplet, along with an $\cN = 4$ background mixed Chern--Simons term with level $1$ between such $U(1)$ vector multiplets.   Explicitly, the action for the following quiver
\be
\begin{tikzpicture}[baseline]
\tikzstyle{every node}=[font=\footnotesize]
\node[draw, circle] (node1) at (-1.5,0) {$1_{k_1}$};
\node[draw, circle] (node2) at (1.5,0) {$1_{k_2}$};
\draw[draw=red,solid,thick,-]  (node1) to  node[midway,above] {{\red $T(U(1))$}}  (node2) ; 
\end{tikzpicture}
\ee
in the $\CN=2$ notation is given by (see \eg~ \cite[(4.4)]{Kapustin:1999ha})
\be \label{actionTU1}
\begin{split}
& \int \mathrm{d}^3 x \mathrm{d}^4 \theta \, \left(\frac{k_1}{4\pi}  \Sigma_1\,V_1 + \frac{k_2}{4\pi} \Sigma_2\,V_2 {\blue - \frac{1}{4\pi}\Sigma_1 V_2 - \frac{1}{4\pi}\Sigma_2 V_1} \right)  \\
 & \quad - \int \mathrm{d}^3 x \mathrm{d}^2 \theta \, \left(\frac{k_1}{4\pi}  \Phi_1^2 + \frac{k_2}{4\pi}  \Phi_2^2  {\blue - \frac{1}{2 \pi} \Phi_1 \Phi_2 }+ {\text{c.c.}} \right)~.
\end{split}
\ee
where $\Sigma_i, V_i$ (with $i=1,2$) are, respectively, the $\cN=2$ linear multiplet and vector multiplet of the $i$-th gauge node, and $\Phi_i$ are the $\cN=2$ chiral multiplets of the $\CN=4$ vector multiplets of the $i$-th gauge group.  In the above equation, we highlight the contribution from the mixed Chern--Simons terms due to $T(U(1))$ in blue.  We emphasise that the mixed Chern--Simons terms come with the level $-1$ in our convention for $T(U(1))$.  Thus, one may view the $T(U(N))$ theory as having a global symmetry $U(N) \times U(N)$, such that the two $U(1)$ subgroups of each $U(N)$ acts trivially on the theory, and that an $\cN = 4$ background mixed Chern--Simons term with level $-N$ is added for the two corresponding $U(1)$ background vector multiplets.

It should be mentioned that there is a close cousin of the $T(U(1))$ theory.  This theory is called $\bar{T(U(1))}$ in \cite{Assel:2014awa}. This theory can be defined almost in the same way as above, except that the minus signs in the blue terms of \eref{actionTU1} are changed to plus signs.  In other words, the level of the mixed Chern--Simons terms is $+1$.  One can then define $\bar{T(U(N))}$ theory as a product between $T(SU(N))$ and $\bar{T(U(1))}$.  As a consequence, $\bar{T(U(N))}$ has a global symmetry $U(N) \times U(N)$, such that the two $U(1)$ subgroups of each $U(N)$ acts trivially on the theory, and that an $\cN = 4$ background mixed Chern--Simons term with level $N$ is added for the two corresponding $U(1)$ background vector multiplets.

\subsection{Compact models}
Let us now take $x^6$ to be a circular direction.  We refer to this type of configurations as compact models.  An example of this is as follows: 
\be \label{introBranes}
\begin{tikzpicture}[baseline]
\tikzstyle{every node}=[font=\footnotesize, node distance=0.45cm]
\tikzset{decoration={snake,amplitude=.4mm,segment length=2mm,
                       post length=0mm,pre length=0mm}}
\draw[blue,thick] (0,0) circle (1.5cm) node[midway, right] {$N$ D3};
\draw[black,thick] (0,1) -- (0,2) node[right] {NS5};
\def \n {6}
\def \radius {1.2cm}
\def \margin {0} 
\foreach \s in {1,...,10}
{
	\node[draw=none] (\s) at ({360/\n * (\s - 2)+30}:{\radius-10}) {};
}
\node[draw=none, circle] (last) at ({360/3 * (3 - 1)+30}:{\radius-10}) {};
\node[draw=none,  below right= of 1] (f1) {$\bullet$};
\node[draw=none, above right= of 2] (f2) {};
\node[draw=none, above = of 3] (f3) {};
\node[draw=none, above left= of 4] (f4) {};
\node[draw=none,  below left= of 5] (f5) {$\bullet$};
\node[draw=none,  below = of last] (f6) {$\bullet$};
\node[draw=none] at (-0.9,-1.3) {{\Large $\mathbf{\ddots}$}};
\node[draw=none] at (0,-2.4) {{$n$ D5s}};
\end{tikzpicture}
\qquad \qquad \qquad 
\begin{minipage}{0,2\textwidth}
\begin{tikzpicture}[baseline]
\tikzstyle{every node}=[font=\footnotesize]
\node[draw, circle] (node1) at (0,2) {$N$};
\draw[thick] (node1) edge [out=45,in=135,loop,looseness=5]  (node1);
\node[draw=none] at (1.3,2.5) {{}};
\node[draw, rectangle] (sqnode) at (0,0) {$n$};
\draw (node1)--(sqnode);
\end{tikzpicture}
\end{minipage}
\ee
where the loop around the node denotes a hypermultiplet in the adjoint representation of the $U(N)$ gauge group.  The mirror theory can be obtained simply by applying $S$-duality to the above brane system in the usual way:
\be \label{TUNloopmirr}
\begin{tikzpicture}[baseline]
\tikzstyle{every node}=[font=\footnotesize]
\tikzset{decoration={snake,amplitude=.4mm,segment length=2mm,
                       post length=0mm,pre length=0mm}}
\draw[blue,thick] (0,0) circle (1.5cm) node[midway, right] {$N$ D3};
\draw node at (0,1.8) {\large{$\bullet$}} node at (0.4,1.8) {D5};
\def \n {6}
\def \radius {1.2cm}
\def \margin {0} 
\foreach \s in {1,...,10}
{
	\node[draw=none] (\s) at ({360/\n * (\s - 2)+30}:{\radius-10}) {};
}
\node[draw=none, circle] (last) at ({360/3 * (3 - 1)+30}:{\radius-10}) {};
\node[draw=none,  below right= of 1] (f1) {};
\node[draw=none, above right= of 2] (f2) {};
\node[draw=none, above = of 3] (f3) {};
\node[draw=none, above left= of 4] (f4) {};
\node[draw=none,  below left= of 5] (f5) {};
\node[draw=none,  below = of last] (f6) {};
\node[draw=none] at (-1,-1.5) {{\Large $\mathbf{\ddots}$}};
\node[draw=none] at (0,-2.4) {{$n$ NS5s}};
\draw[-, >=latex,black, thick] (1) to (f1);
\draw[-, >=latex,black, thick] (5) to (f5);
\draw[-, >=latex,black, thick] (last) to (f6);
\end{tikzpicture}
\qquad \qquad
\begin{tikzpicture}[baseline, scale=0.6,font=\scriptsize]
\def \n {6}
\def \radius {2.4cm}
\def \margin {15} 
\node[draw, circle] at ({360/\n * (1 - 2)}:\radius) {$N$};
\draw[-, >=latex] ({360/\n * (1 - 3)+\margin}:\radius);
arc ({360/\n * (1 - 3)+\margin}:{360/\n * (1-2)-\margin}:\radius);
\node[draw, circle] at ({360/\n * (2 - 2)}:\radius) {$N$};
\draw[-, >=latex] ({360/\n * (2 - 3)+\margin}:\radius);
arc ({360/\n * (2 - 3)+\margin}:{360/\n * (2-2)-\margin}:\radius);
\foreach \s in {3,...,5}
\node[draw, circle] at ({360/\n * (\s - 2)}:\radius) {$N$};
\foreach \s in {1,...,5}
{	
	\draw[-, >=latex] ({360/\n * (\s - 3)+\margin}:\radius) 
	arc ({360/\n * (\s - 3)+\margin}:{360/\n * (\s-2)-\margin}:\radius);
}
\node[draw, circle] at ({360/3 * (3 - 1)}:\radius) {$N$};
\node[draw, rectangle] (sq) at (5,0) {$1$};
\draw (3,0)--(4.65,0);
\draw[-, >=latex] ({360/6 * (5 -2)+\margin}:\radius) 
arc ({360/6 * (5 -2)+\margin}:{360/6 * (5-1)-\margin}:\radius);
\node[draw=none] at (4.8,-1.8) {$n$ circular nodes};
\end{tikzpicture}
\ee

%

\subsection{The holographic duals of linear quivers and compact models}
Both linear quivers and compact models have known holographic duals in sting theory.  Type IIB supergravity solutions have been found in \cite{Assel:2011xz, Assel:2012cj}. Historically, these solutions descend from the seminal work \cite{DHoker:2007zhm, DHoker:2007hhe}, where $\text{AdS}_4\times S^2\times S^2\times \Sigma_2$ backgrounds have been found, with $\Sigma_2$ a non-compact Riemann surface with the topology of infinite strip $\mathbb{R}\times\,I$ with coordinates $(y, x)$,  where $I$ is an interval. The dual field theory is supposed to be four-dimensional SYM with space-dependent coupling constant, since the ten-dimensional metric is actually asymptotically $\mathrm{AdS}_5 \times S^5$ in the limit $y\,\to\,\infty$. The metric, the dilaton and the fluxes are completely determined in terms of two harmonic functions $\cA_i$ on $\Sigma_2$.  These functions can admit suitable singularities on the boundary of the strip.  Those are interpreted as the singularities coming from D5 and NS5 branes, like those presented in example \eref{TSU3}.  We illustrate this in figure \eref{AdS5S5}.
\be \label{AdS5S5}
  \includegraphics[scale=0.6]{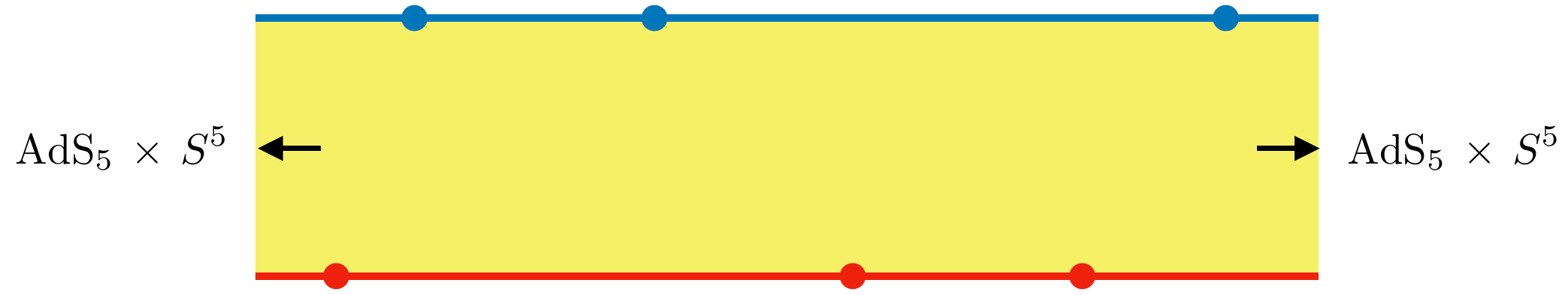}
\ee

Backgrounds dual to 3d $\cN=4$ linear quiver theories can be obtained by picking suitable harmonic functions on $\Sigma_2$: specifically, we can make a choice of harmonic functions such that $I$ shrinks to zero as $y\,\to\,\pm\infty$. The resulting topology is $\text{AdS}_4\,\times\,\cB_6$ where $\cB_6\,\approx\,S^5\,\times\, I$ is the six-dimensional ball.  This is illustrated in \eref{AdS4B6}.
\be \label{AdS4B6}
  \includegraphics[scale=0.6]{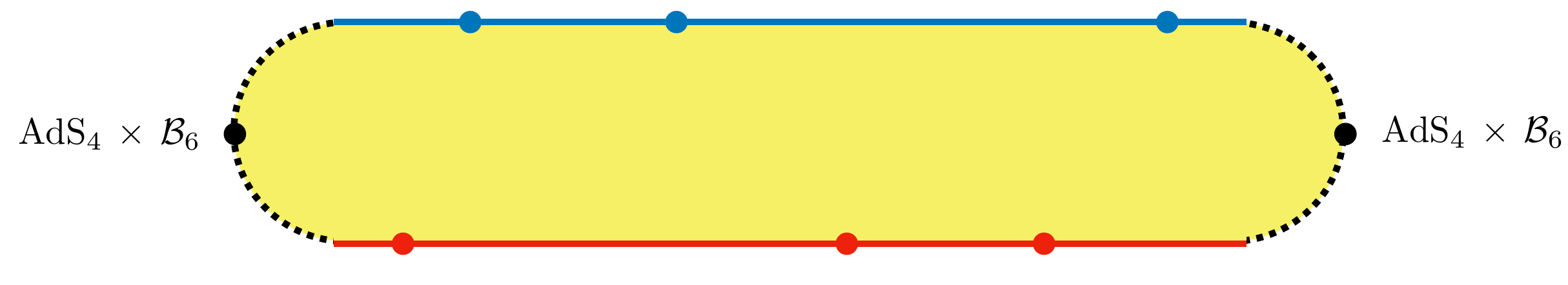}
\ee

Getting holographic duals of 3d $\cN=4$ compact models is more subtle and a quotient procedure is involved. Harmonic functions on $\Sigma_2$ can be chosen to have an infinite number of singularities, but in such a way to be periodic along the infinite direction with period $T$:
\be
\nonumber
\cA_i(y+T)\,=\,\cA_i(y)\,.
\ee
The whole solution is invariant under this translation, being completely determined by $\cA_i$. At this stage, we can perform a quotient with respect to ``$T$-symmetry'' ending with a configuration where points $(x, y)$ and $(x, y+T)$ of the Riemann surface are identified; we end up with a surface with the topology of the annulus; see figure \eref{annulus}.
\be \label{annulus}
  \includegraphics[scale=0.4]{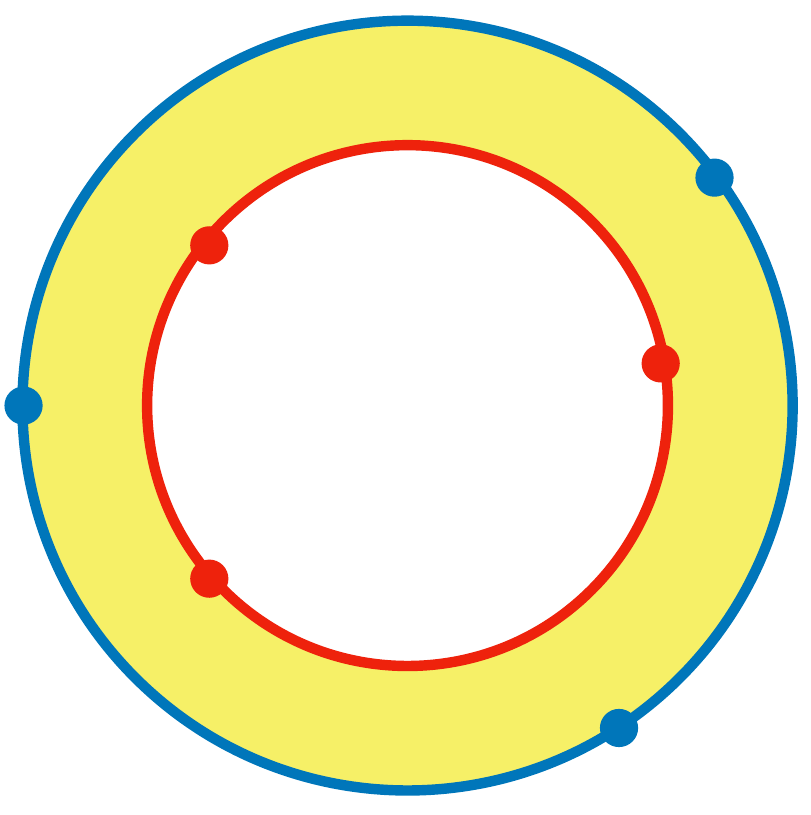}
\ee

\subsection{$J$-folds}
A more general quotient procedure can, in fact, be implemented.  In particular, one may introduce an $SL(2,\BZ)$ duality-twisted boundary condition \cite{Ganor:2014pha, Assel:2018vtq} upon identifying the two ends of the aforementioned Riemann surface.   This can be done as follows.  As before, the starting point is a choice of harmonic functions $\cA_i$, that completely fixes the physical fields of the solution. For instance, let us focus on the axio-dilaton $\tau\,=\,C_0+i\,e^{-2\phi}$ where $C_0$ is the potential of the one-form flux $F_1$ and $\phi$ is the dilaton. As it is well-known, Type IIB supergravity admits a non-trivial action of $SL(2, \mathbb{Z})$, generating orbits of equivalent solutions; the axio-dilaton is not invariant under this $SL(2, \mathbb{Z})$ action. We can imagine to pick harmonic functions $\cA_i$ such:
\be
\tau(y+T)\,=\,M\,\tau(y)
\ee
where $M$ represents the action of $SL(2, \mathbb{Z})$ on the axio-dilaton and we require that similar relations hold for all other fluxes, with an appropriate element of $SL(2,\BZ)$ acting on them. If such a choice can be performed, we can imagine to quotient with respect to the joint action of $SL(2,\mathbb{Z})$ and translation by $T$ along the non-compact direction $y$. Points $(x, y+T)$ and $(x, y)$ are again identified; the Riemann surface has a cut along $(x, T)$, passing through the fields undergo an $SL(2, \mathbb{Z})$ transformation. We end up with a Riemann surface with the topology of the annulus and a non-trivial monodromy under $SL(2, \mathbb{Z})$.  This is illustrated in \eref{jfold}.
\be \label{jfold}
  \includegraphics[width=0.8\textwidth]{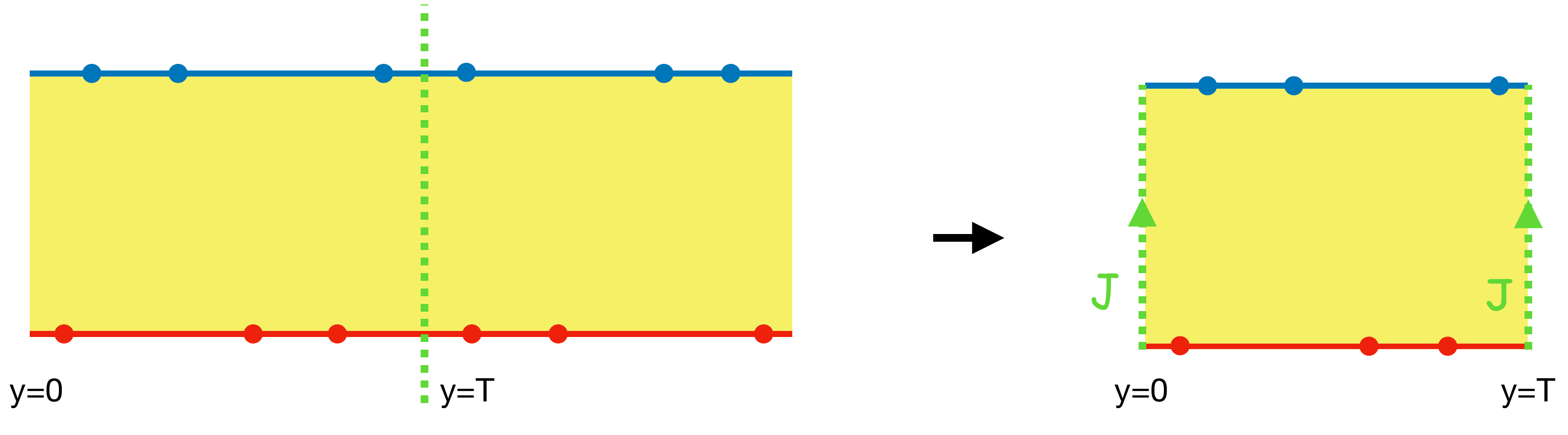}
\ee
It turns out that such a quotient is related to a particular choice of $SL(2, \mathbb{Z})$ element. Let
\be
S=\left(\begin{matrix} 0 &-1 \\ 1 &0  \end{matrix}\right) 
\qquad
T=\left(\begin{matrix} 1 &0 \\ 1 &1  \end{matrix}\right)\,~,
\ee
satisfying $S^2=-1$ and $(ST)^3=1$, be the generators of $SL(2,\BZ)$. Then the aforementioned quotient can be performed for every element of $SL(2,\BZ)$ of the form:
\be
J_k\,=\,-S\,T^k =\left(\begin{matrix} k &1 \\ -1 &0  \end{matrix}\right)  \,,\qquad \bar{J}_k\,=\, -J_{-k}\,.
\ee
This kind of solutions was studied in the context of abelian theories in \cite{Ganor:2014pha} and is referred to as the {\bf $J$-fold} in \cite{Assel:2018vtq}. These are often regarded as non-geometrical, in the sense that we performed a quotient with respect to some symmetry of the theory not descending from isometries of the metric. 

The quotient also admits a realisation at the level of brane configurations: it corresponds to a five-dimensional surface implementing the aforementioned monodromy under $SL(2, \mathbb{Z})$ action. As we have seen, $\Sigma_2$ has the topology of the annulus, thus corresponding to circular brane configuration with an insertion of $J$-folds.  An example of a brane configuration with a $J$-fold is as follows:
\be \label{introJfold}
\begin{tikzpicture}[baseline]
\tikzstyle{every node}=[font=\footnotesize]
\draw[blue,thick] (0,0) circle (1.5cm) node[midway, right] {$N$ D3};
\draw[thick] (0,-1)--(0,-2) node[right] {NS5};
\node[draw=none] at (-1.7,0) {\large{$\bullet$}};
\node[draw=none] at (-2.3,0) {D5};
\node[draw=none] at (1.7,0) {\large{$\bullet$}};
\tikzset{decoration={snake,amplitude=.4mm,segment length=2mm,
                       post length=0mm,pre length=0mm}}
\draw[decorate,red,thick] (0,1) -- (0,2) node[right] {$J_k$};
\end{tikzpicture}
\ee
The insertion of the $J_k$-fold in such a brane system can be viewed as introducing a 3d interface, with a non-trivial $SL(2,\BZ)$ action $J_k$, to the 4d $\CN=4$ super-Yang-Mills theory living on the D3-branes on the circle.  The theory on such a 3d interface was studied in \cite[sec. 8]{Gaiotto:2008ak}.  This is, in fact, the $T(U(N))$ theory with a Chern--Simons level $k$ for one of the flavour $U(N)$ symmetry, whereas the other $U(N)$ flavour symmetry has Chern--Simons level zero.  One can then couple this 3d theory to the theory on the D3-brane on a circle.  The $U(N)_k$ and the $U(N)_0$ flavour symmetries\footnote{Unless specified otherwise, we denote the Chern--Simons level as the subscript.} are then coupled to the $U(N)_L$ and $U(N)_R$ gauge fields on the left and on the right of the interface, respectively\footnote{As pointed out in \cite{Assel:2014awa, Assel:2018vtq}, there are two possibilities for coupling the $U(N)$ flavour symmetry to the $U(N)$ gauge field on each side, namely $U(N)_+ = \diag(U(N) \times U(N))$ or $U(N)_- = \diag(U(N) \times U(N)^\dagger)$. For $T(U(N))$, the gauging is chosen to be $U(N)_+$ on both sides, whereas for $\bar{T(U(N))}$, the gauging is chosen to be $U(N)_+$ on one side and $U(N)_-$ on the other side.}. For instance, the three dimensional quiver theory associated to the brane system \eref{introJfold} is
\be \label{NkN011}
\begin{tikzpicture}[baseline]
\tikzstyle{every node}=[font=\footnotesize]
\node[draw, circle] (node1) at (-1.5,0) {$N_k$};
\node[draw, circle] (node2) at (1.5,0) {$N_0$};
\node[draw, rectangle] (sqnode1) at (-2.8,0) {$1$};
\node[draw, rectangle] (sqnode2) at (2.8,0) {$1$};
\draw[draw=black,solid,thick,-]  (node1) to[bend right=20]   (node2) ; 
\draw[draw=black,solid,thick,-]  (node1)--(sqnode1) ; 
\draw[draw=black,solid,thick,-]  (node2)--(sqnode2) ; 
\draw[draw=red,solid,thick,-]  (node1) to[bend left=20]  node[midway,above] {{\red $T(U(N))$}}  (node2) ; 
\end{tikzpicture}
\ee
where $N_k$ and $N_0$ denotes gauge groups $U(N)$ with Chern--Simons levels $k$ and $0$ respectively.  We emphasise that there is a mixed CS term with level $-N$ between the two gauge groups.  Due to the presence of the $T(U(N))$ theory as a link, this is not a conventional Lagrangian theory, because only one $U(N)$ symmetry is manifest in the Lagrangian description of the $T(U(N))$ theory, whereas the other $U(N)$ symmetry emerges in the infrared\footnote{It should be mentioned that similar quiver theories, with special unitary gauge groups and $T(SU(N))$ links, were studied in \cite[sec. 4.1]{Terashima:2011qi} and \cite[sec. 5.2]{Gang:2015wya} in the context of 3d-3d correspondence and the twisted compactification of the 6d $\CN=(2,0)$ theory on a torus bundle over $S^1$.}.

\subsection{$S$-flips}
Another type of quotients that is similar to the $J$-fold is possible. In this case we select the $SL(2,\mathbb{Z})$ element implementing the monodromy to be $S$. However, in order to have a desired symmetry of the supergravity solution, we have to perform an exchange of coordinates corresponding to the two $S^2$ in $\text{AdS}_4\,\times\,S^2\times\,S^2\,\times\,\Sigma_2$ and a reflection of $x$ coordinate, being identified at the $S$-interface in an antipodal way, as depicted in \eref{fig:sflip}.
\be \label{fig:sflip}
  \includegraphics[width=0.3\textwidth]{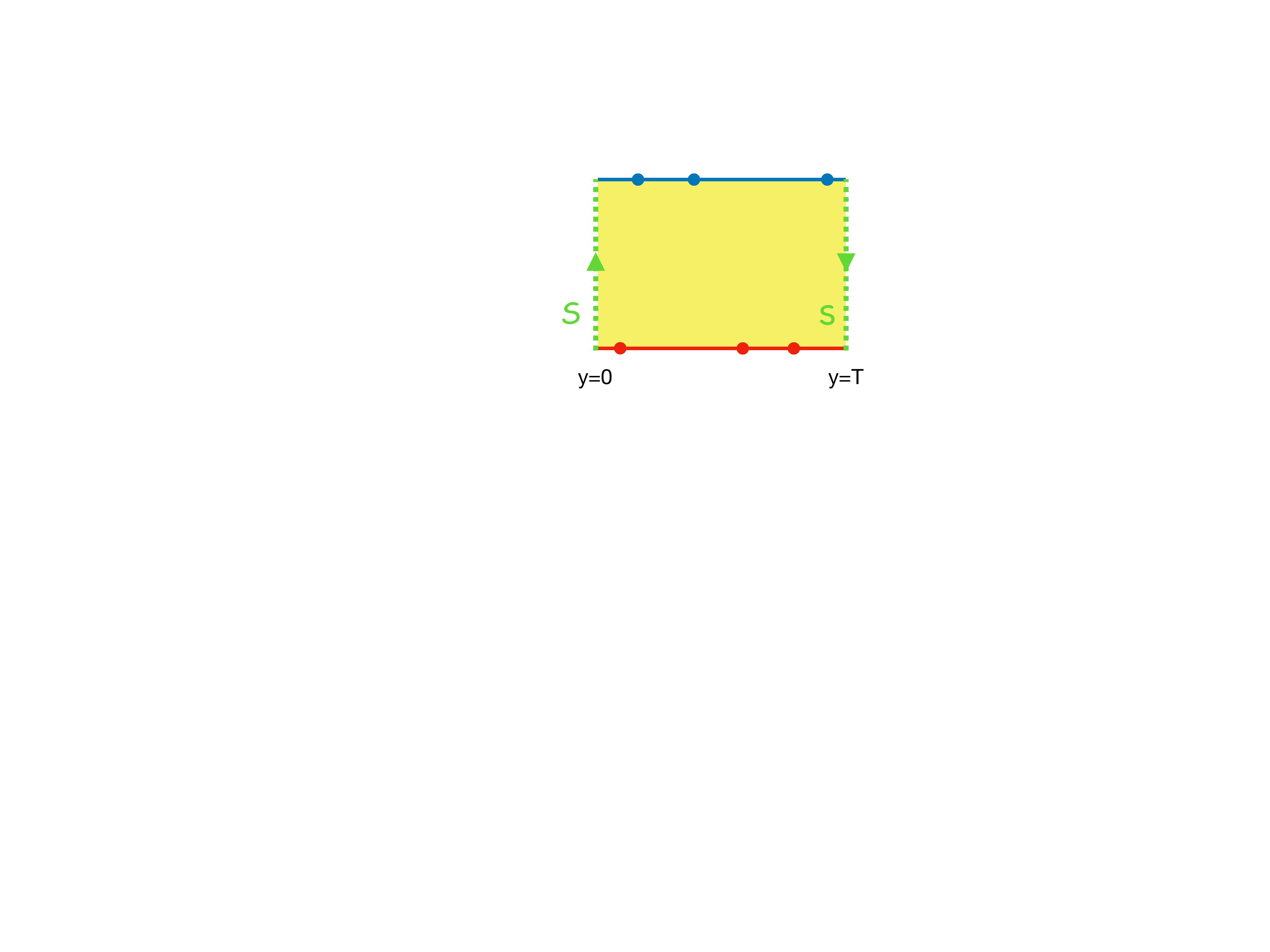}
\ee
The Riemann surface now has the topology of the M\"obius strip.  This type of solutions is referred to as an {\bf $S$-flip} in \cite{Assel:2018vtq}.  Similarly to the $J$-fold, the S-flip has an avatar at the level of circular brane configuration, as five-dimensional surface passing through the configuration undergoes an $SL(2, \mathbb{Z})$ transformation and a rotation of coordinates such that $(x^{3,4,5}\,,\,x^{7,8,9})\,\to\,(x^{7,8,9}\,,-\,x^{3,4,5})$.    When an $S$-flip is inserted into a brane system, the corresponding quiver diagram can be obtained in the same way as that with the $J$-fold, except that the Chern--Simons level is set to zero.  An example for this type of configurations is depicted in \eref{T2NN}.

\subsection{$(p,q)$ fivebranes}
Let us now consider $(p,q)$ fivebranes \cite{Aharony:1997ju, Aharony:1997bh}, where $(1,0)$ denotes an NS5 brane and $(0,1)$ denotes a D5 brane.  For a given ordered pair $(p,q)$, we can write this as
\be
(p,q)\,=\,\bar{J}_{k_1}\,\bar{J}_{k_2}\,\dots\, \bar{J}_{k_r}\,(1,0)
\ee
for some $k_1\,,k_2\,,\dots\,k_r$.   Thus, any $(p,q)$ brane is related to an NS5 brane by an $SL(2, \mathbb{Z})$ transformation. Using this realisation, we can convert a $(p,q)$ brane to an equivalent configuration involving $J$-folds as follows:
\be \label{JJJ}
\begin{tikzpicture}[baseline, scale=0.8]
\tikzstyle{every node}=[font=\footnotesize, node distance=0.45cm]
\draw[blue,thick] (0,0)--(10,0) node[right] {$N$ D3};
\draw[red,dotted, thick] (1.5,0.5)--(2.5,0.5);
\draw[purple,dotted, thick] (7.5,0.5)--(8.5,0.5);
\draw[thick] (5,1)--(5,-1) node[right] {NS5};
\tikzset{decoration={snake,amplitude=.4mm,segment length=2mm,
                       post length=0mm,pre length=0mm}}
\draw[decorate,red,thick] (1,1)--(1,-1)node[left] {$\bar{J}_{k_1}^{-1}$};
\draw[decorate,red,thick] (3,1)--(3,-1)node[left] {$\bar{J}_{k_r}^{-1}$};
\draw[decorate,purple,thick] (7,1)--(7,-1)node[right] {$\bar{J}_{k_r}$};
\draw[decorate,purple,thick] (9,1)--(9,-1)node[right] {$\bar{J}_{k_1}$};
\end{tikzpicture}
\ee
From the perspective of the quiver diagram, each $\bar{J}_k$ gives rise to a $T(U(N))$ link with a Chern--Simons level $k$ for the $U(N)$ group on the left, whereas each $\bar{J}^{-1}_{-k}$ gives rise to a $\bar{T(U(N))}$ link with a Chern--Simons level $k$ for the $U(N)$ group on the right.   In particular, the corresponding quiver theory for the following $SL(2,\BZ)$-equivalent brane systems 
\be \label{JJJ1}
\scalebox{0.8}{
\begin{tikzpicture}[baseline, scale=0.8]
\tikzstyle{every node}=[font=\footnotesize, node distance=0.45cm]
\draw[blue,thick] (0,0)--(2,0) node[right] {$N$ D3};
\draw[thick] (1,1)--(1,-1) node[below] {$(p,q)$};
\draw[thick] (2,1)--(2,-1) node[right] {NS5};
\draw[thick] (0,1)--(0,-1) node[left] {NS5};
\end{tikzpicture}}
\qquad  \qquad
\scalebox{0.8}{
\begin{tikzpicture}[baseline, scale=0.8]
\tikzstyle{every node}=[font=\footnotesize, node distance=0.45cm]
\draw[blue,thick] (-1,0)--(11,0) node[right] {$N$ D3};
\draw[red,dotted, thick] (1.5,0.5)--(2.5,0.5);
\draw[purple,dotted, thick] (7.5,0.5)--(8.5,0.5);
\draw[thick] (5,1)--(5,-1) node[right] {NS5};
\draw[thick] (11,1)--(11,-1) node[right] {NS5};
\draw[thick] (-1,1)--(-1,-1) node[left] {NS5};
\tikzset{decoration={snake,amplitude=.4mm,segment length=2mm,
                       post length=0mm,pre length=0mm}}
\draw[decorate,red,thick] (1,1)--(1,-1)node[left] {$\bar{J}_{k_1}^{-1}$};
\draw[decorate,red,thick] (3,1)--(3,-1)node[left] {$\bar{J}_{k_r}^{-1}$};
\draw[decorate,purple,thick] (7,1)--(7,-1)node[right] {$\bar{J}_{k_r}$};
\draw[decorate,purple,thick] (9,1)--(9,-1)node[right] {$\bar{J}_{k_1}$};
\end{tikzpicture}}
\ee
is as follows:
\be
\scalebox{0.7}{
\begin{tikzpicture}[baseline, scale=0.6]
\tikzstyle{every node}=[font=\footnotesize, minimum size=1.2cm]
\node[draw, circle] (node1p) at (2,2) {$N_{k_r}$};
\node[draw=none] (node2p) at (6,2) {};
\node[draw=none] (node3p) at (6,2) {{\Large $\cdots$}};
\node[draw,circle] (node4p) at (10,2) {$N_{k_2}$};
\node[draw,circle] (node5p) at (14,2) {$N_{k_1}$};
\node[draw,circle] (node6p) at (18,2) {$N_{0}$};
\node[draw, circle] (node1n) at (-2,2) {$N_{-k_r}$};
\node[draw=none] (node2n) at (-6,2) {};
\node[draw=none] (node3n) at (-6,2) {{\Large $\cdots$}};
\node[draw,circle] (node4n) at (-10,2) {$N_{-k_2}$};
\node[draw,circle] (node5n) at (-14,2) {$N_{-k_1}$};
\node[draw,circle] (node6n) at (-18,2) {$N_{0}$};
\draw[draw=red,dashed,thick,-]  (node1n) to node[midway, above]{\red $\bar{T(U(N))}$}(node2n) ; 
\draw[draw=red,dashed,thick,-]  (node3n) to node[midway, above]{\red $\bar{T(U(N))}$}(node4n) ; 
\draw[draw=red,dashed,thick,-]  (node4n) to node[midway, above]{\red $\bar{T(U(N))}$}(node5n) ; 
\draw[draw=red,dashed,thick,-]  (node5n) to node[midway, above]{\red $\bar{T(U(N))}$}(node6n) ; 
\draw[draw=black,solid,thick,-]  (node1p) to  (node1n) ; 
\draw[draw=purple,dashed,thick,-]  (node1p) to node[midway, above]{\purple $T(U(N))$}  (node2p) ;  
\draw[draw=purple,dashed,thick,-]  (node3p) to node[midway, above]{\purple $T(U(N))$}  (node4p) ;  
\draw[draw=purple,dashed,thick,-]  (node4p) to node[midway, above]{\purple $T(U(N))$}  (node5p) ;  
\draw[draw=purple,dashed,thick,-]  (node5p) to node[midway, above]{\purple $T(U(N))$}  (node6p) ;  
\end{tikzpicture}}
\ee
This agrees with the description provided in \cite[fig. 75]{Gaiotto:2008ak} and \cite[fig. 6]{Assel:2014awa}.

\section{Models with zero Chern--Simons levels} \label{sec:zeroCS}
In this section, we consider theories with zero Chern--Simons (CS) levels and with certain links between gauge nodes in the quiver being $T(U(N))$.  From the brane perspective, such a theory arises from the Hanany--Witten brane configuration \cite{Hanany:1996ie}, namely a system of D3, NS5 and D5 branes that preserves eight supercharges, with an insertion of $S$-flips \cite{Assel:2018vtq}.   The presence of an $S$-flip gives rise to the aforementioned $T(U(N))$ link in the quiver.   The moduli space of such quiver theories is studied below.  The main result can be summarised as follows.  

We find that these theories have two branches of the moduli space, namely the Higgs and the Coulomb branches.  Let us first discuss about the Higgs branch. We propose that this is given by the hyperK\"ahler quotient of a product of each component in the quiver by the gauge symmetry.  By each component, we mean a bi-fundamental hypermultiplet, a fundamental hypermultiplet and a $T(U(N))$ link that connects two $U(N)$ groups together.  The former two can be treated in the usual way as in a Lagrangian theory.  whereas each $T(U(N))$ link contributes two copies of the closure of the {\it maximal nilpotent orbit} of $SU(N)$, denoted by $\CN_{SU(N)}$.  The reason for latter is two-fold: (1) the Higgs and the Coulomb branches of $T(U(N))$ are both isomorphic to $\CN_{SU(N)}$, and (2) in order to realise the two $U(N)$ groups connected by $T(U(N))$, we need two copies of $SU(N)$ subgroups, one arises from the Higgs branch and the other arises from the Coulomb branch of $T(U(N))$.

The Coulomb branch is similar to the usual 3d $\CN=4$ gauge theories, but with the following important remark.  We propose that the scalars in the vector multiplets of any two gauge nodes that are connected by a $T(U(N))$ link are frozen and do {\it not} contribute to the Coulomb branch.  The other gauge nodes in the quivers still give rise to vector multiplets that contribute to the Coulomb branch.  From the brane perspective, this proposal implies that the D3-brane segment between two NS5-branes that is stretched through the $S$-flip cannot move along the NS5-brane directions (\ie~ the Coulomb branch directions).

We check that the descriptions of the Higgs and the Coulomb branches mentioned above are consistent with $S$-duality and mirror symmetry.  Given a brane system, say of theory $A$, we can obtain a brane system of the mirror theory, say theory $B$, using $S$-duality.  We find that the moduli space of theories $A$ and $B$ are related by mirror symmetry \cite{Intriligator:1996ex, Hanany:1996ie}. in the following sense. The Higgs branch ({\it resp.} Coulomb branch) of theory $A$ computed by using the above proposal is in an agreement with the Coulomb branch ({\it resp.} Higgs branch) of theory $B$.  

Below we provide examples to demonstrate the above discussion.
\subsection{Example 1: A flavoured affine $A_1$ quiver}
Let us consider the following brane set-up and the following theory.
\be \label{T2NN}
\scalebox{0.8}{
\begin{tikzpicture}[baseline]
\tikzstyle{every node}=[font=\footnotesize]
\draw[blue,thick] (0,0) circle (1.5cm) node[midway, right] {$N$ D3};
\draw[thick] (0,-1)--(0,-2) node[right] {NS5};
\node[draw=none] at (-1.5,1) {\large{$\bullet$}};
\node[draw=none] at (-2,1) {D5};
\node[draw=none] at (-1.5,-1) {\large{$\bullet$}};
\tikzset{decoration={snake,amplitude=.4mm,segment length=2mm,
                       post length=0mm,pre length=0mm}}
\draw[decorate,red,thick] (0,1) -- (0,2) node[right] {$S$};
\end{tikzpicture}}
\qquad\qquad\qquad
\begin{tikzpicture}[baseline]
\tikzstyle{every node}=[font=\footnotesize]
\node[draw, circle] (node1) at (-1.5,0) {$N$};
\node[draw, circle] (node2) at (1.5,0) {$N$};
\node[draw, rectangle] (sqnode) at (-3,0) {$2$};
\draw[draw=black,solid,thick,-]  (node1) to[bend right=20]   (node2) ; 
\draw[draw=black,solid,thick,-]  (node1)--(sqnode) ; 
\draw[draw=red,solid,thick,-]  (node1) to[bend left=20]  node[midway,above] {{\red $T(U(N))$}}  (node2) ; 
\end{tikzpicture}
\ee
where, throughout this section, we denote a gauge group $U(N)$ with zero CS level by a circular node with the label $N$.  The flavour symmetry $U(N_f)$ is denoted by a square node with the label $N_f$.

The mirror theory can be derived by applying the $S$-duality to the brane system \eref{T2NN} which yields
\be \label{mirrT2NN}
\scalebox{0.9}{
\begin{tikzpicture}[baseline]
\tikzstyle{every node}=[font=\footnotesize]
\draw[blue,thick] (0,0) circle (1.5cm) node[midway, right] {$N$ D3};
\draw[thick] (-0.75,-0.5)--(-1.5,-1);
\draw[thick] (-0.75,0.5)--(-1.5,1);
\node[draw=none] at (-2,1) {NS5};
\node[draw=none] at (0,-2) {D5};
\node[draw=none] at (0,-1.7) {\large{$\bullet$}};
\tikzset{decoration={snake,amplitude=.4mm,segment length=2mm,
                       post length=0mm,pre length=0mm}}
\draw[decorate,red,thick] (0,1) -- (0,2) node[right] {$S$};
\end{tikzpicture}}
\qquad\qquad\qquad \scalebox{1}{
\begin{tikzpicture}[baseline]
\tikzstyle{every node}=[font=\footnotesize]
\node[draw, circle] (node1) at (-1.5,0) {$N$};
\node[draw, circle] (node2) at (1.5,0) {$N$};
\node[draw, circle] (node3) at (0,1) {$N$};
\node[draw, rectangle] (sqnode) at (0,2) {$1$};
\draw[draw=black,solid,thick,-]  (node1)-- (node2) ; 
\draw[draw=black,solid,thick,-]  (node2)--(node3) ; 
\draw[draw=red,solid,thick,-]  (node1)--(node3) node at (-1.5,0.7) {{\red $T(U(N))$}} ; 
\draw[draw=black,solid,thick,-]  (node3)--(sqnode) ; 
\end{tikzpicture}}
\ee

\subsubsection*{The Higgs branches}
We claim that the Higgs branch of \eref{T2NN} is given by
\be \label{HiggsT2NN}
\CH_{\eref{T2NN}} = \frac{\CH\left( [U(2)]-[U(N)_1] \right) \times \CN_{SU(N)_1} \times \CN_{SU(N)_2} \times \CH([U(N)_1]-[U(N)_2])}{U(N)_1 \times U(N)_2}~,
\ee
where $\CN_{SU(N)}$ denotes the closure of the maximal nilpotent orbit of $SU(N)$.  Throughout this paper, we shall use shorthand notations $\CH$ and $\CC$ to stand for the Higgs branch and the Coulomb branch respectively.
The quaternionic dimension of \eref{HiggsT2NN} is
\be
\dim_\BH \, \CH_{\eref{T2NN}} = 2N + 2 \left[ \frac{1}{2}(N-1)N\right] + N^2 - N^2 - N^2 = N~.
\ee

Similarly, we claim that the Higgs branch of \eref{mirrT2NN}  is 
{\small
\be
\begin{split}
\CH_{\eref{mirrT2NN}} &= \Big[ \CH([U(N)_1]-[U(N)_3]) \times \CH([U(N)_2]-[U(N)_3]) \times \CH\left( [U(1)]-[U(N)_2] \right)  \\
& \quad \times \CN_{SU(N)_1} \times \CN_{SU(N)_2} \Big] / \left(U(N)_1 \times U(N)_2 \times U(N)_3 \right)~.
\end{split}
\ee}
The dimension of this space is
\be
\dim_\BH \, \CH_{\eref{mirrT2NN}} = N^2 +N^2 + N +2 \left[ \frac{1}{2}(N-1)N\right]  - 3 N^2 = 0~.
\ee

\subsubsection*{The Coulomb branches}
Since mirror symmetry identifies the Coulomb branch $\CC_{\eref{T2NN}}$ of \eref{T2NN} with the Higgs branch $\CH_{\eref{mirrT2NN}}$ of \eref{mirrT2NN}, it follows that 
\be
\dim_\BH \, \CC_{\eref{T2NN}} = \dim_\BH \, \CH_{\eref{mirrT2NN}} = 0~,
\ee
and hence $\CC_{\eref{T2NN}}$ is trivial.
We see that even though the theory \eref{T2NN} has gauge group $U(N) \times U(N)$, its Coulomb branch is trivial.  This is consistent with our proposal: the scalars in the vector multiplets of $U(N) \times U(N)$ gauge group in \eref{T2NN} are frozen to a particular value, because they are linked by $T(U(N))$.  From the brane perspective, this means that the D3-branes do {\it not} move along the direction of the S-flip, but {\it get stuck} at a particular position in the $x^{3,4,5}$ directions.  On the other hand, since the Higgs branch of \eref{T2NN} is non-trivial, this means that the D3-branes that align along the direction of the S-fold and NS5-branes can move along the $x^{7,8,9}$ directions.

By the same token,
\be
\dim_\BH \, \CC_{\eref{mirrT2NN}} = \dim_\BH \, \CH_{\eref{T2NN}} = N~.
\ee
We see that even though \eref{mirrT2NN} has gauge group $U(N) \times U(N) \times U(N)$, its Coulomb branch has dimension $N$, rather than $3N$ (which is the sum of the ranks of the gauge groups).  This is indeed again consistent with our proposal: the scalars of the two $U(N)$ gauge groups connected by $T(U(N))$ are frozen, but those of the remaining $U(N)$ gauge group can acquire VEVs.  The latter gauge group has rank $N$ and contributes $N$ to $\dim_\BH \, \CC_{\eref{mirrT2NN}}$.   From the brane perspective, the D3-brane segment between two NS5 branes that stretch across the S-flip get stuck at a particular position along the $x^{3,4,5}$ directions.  On the other hand, the segment that does not stretch across the S-flip can move along the latter. 

\subsubsection*{The Hilbert series}
To confirm these statements, we compute the Hilbert series of the Higgs branch of \eref{T2NN} using the description \eref{HiggsT2NN}:\footnote{The plethystic exponential (PE) of a multivariate function $f(x_1,x_2, \ldots, x_n)$ such that $f(0,0,\ldots, 0)=0$ is defined as $\PE[f(x_1,x_2, \ldots, x_n)] = \exp \left( \sum_{k=1}^\infty \frac{1}{k} f(x_1^k, x_2^k,\ldots, x_n^k) \right).$
} 
\small{\be \label{HSEx1}
\begin{split}
& H[\CH_{\eref{T2NN}} ](t, x) = \\
& \int \mathrm{d}\mu_{U(N)} ( \vec u)  \int \mathrm{d}\mu_{U(N)} ( \vec w) \\ 
& \times \PE \left[ - t^2 (u_1+u_2)(u_1^{-1} +u_2^{-1}) -t^2 (w_1+w_2)(w_1^{-1} +w_2^{-1}) \right] \\
& \times\PE\left[t (x+x^{-1}) \left \{ \sum_{i=1}^N u^{-1}_i +\sum_{i=1}^N u_i \right \}\right]  \\
& \times H[\CN_{SU(N)}](t, \vec u)  H[\CN_{SU(N)}](t, \vec w) \\
& \times \PE \left[ \left(\sum_{i=1}^N u_i \right)  \left(\sum_{i=1}^N w^{-1}_i \right) t +  \left(\sum_{i=1}^N u^{-1}_i \right)  \left(\sum_{i=1}^N w_i \right) t  \right]~,
\end{split}
\ee}
where the $U(N)$ Haar measure is given by
\be
 \int \mathrm{d}\mu_{U(N)} ( \vec z)  =\left( \prod_{i=1}^N \oint_{|z_i| =1} \frac{\mathrm{d} z_i}{2 \pi i z_i} \right) \prod_{1\leq i < j \leq N} \left( 1- \frac{z_i}{z_j} \right)~,
\ee
and the Hilbert series of the closure of the maximal orbit of $SU(N)$ is (see \cite[(3.4)]{Hanany:2011db} and \cite{Hanany:2016gbz}):
\be
H[\CN_{SU(N)}](t, \vec z) =\left[ \prod_{j=2}^N (1-t^{2j}) \right] \times \PE \left[ t^2 \chi^{SU(N)}_{\mathbf{adj}} (\vec z) \right]~,
\ee
with $ \chi^{SU(N)}_{\mathbf{adj}} (\vec z)$ the character of the adjoint representation of $SU(N)$:
\be
\chi^{SU(N)}_{\mathbf{adj}} (\vec z)  = (z_1 +z_2) (z^{-1}_1 +z^{-1}_2)-1~.
\ee
Let us now explain the contribution of each line in \eref{HSEx1}.  The first two lines describe the gauging of the symmetry $U(N) \times U(N)$.  The second line is the contribution of the fundamental hypermultiplets.  The third line is contribution of two copies of $\CN_{SU(N)}$; one is the Higgs branch and the other is the Coulomb branch of $T(U(N))$.  The last line is the contribution of the bi-fundamental hypermultiplets.  Here $x$ is a fugacity for the $SU(2)$ global symmetry.

The integrals in \eref{HSEx1} can be evaluated in an exact manner and yield
\be
H[\CH_{\eref{T2NN}} ](t, x) =  \PE \left[ \chi^{SU(2)}_{\mathbf{adj}}(x) \sum_{j=1}^N t^{2j} -\sum_{j=1}^N t^{2N+2j}\right]~.
\ee
where
\be
 \chi^{SU(2)}_{\mathbf{adj}} (x) = x^2+1+x^{-2}~.
\ee
The Higgs branch of \eref{T2NN} thus has an $SU(2)$ isometry; this is manifest as a flavour symmetry in the quiver.  In fact, this Hilbert series is equal to that of the Coulomb branch of $U(N)$ gauge theory with $2N$ flavours (also known as the $T^{[1^{2N}]}_{[N^2]}(SU(2N))$ theory \cite{Gaiotto:2008ak}) \cite[(5.6)]{Cremonesi:2013lqa}, where the $U(1)$ topological symmetry gets enhanced to $SU(2)$ at strong coupling :
\be
\begin{split}
\CH_{\eref{T2NN}} &= \CC\left( \text{$U(N)$ gauge theory with $2N$ flavours} \right) \\
&= \CC\left( T^{[1^{2N}]}_{[N^2]}(SU(N)) \right) \\
&=  \text{the intersection between the Slodowy slice }\\
& \quad \,\, \text{transverse to the nilpotent orbit associated with $[N,N]$} \\
& \quad \,\, \text{and the nilpotent cone of $SL(2N,\BC)$ \cite{Gaiotto:2008ak}},
\end{split}
\ee 

Indeed, we can see an effective $U(N)$ gauge theory with $2N$ flavours from \eref{mirrT2NN} as follows.  Since the two $U(N)$ gauge groups connected by the red line do not contribute to the Coulomb branch, we can effectively think of them as flavour symmetries, and so the $U(N)$ gauge group on the lower right hand corner has effectively $2N$ flavours transformed under it.


\subsection{Example 2: Another flavoured affine $A_1$ quiver}
Let us now consider the following theory:
\be \label{T1NN1}
\begin{tikzpicture}[baseline]
\tikzstyle{every node}=[font=\footnotesize]
\draw[blue,thick] (0,0) circle (1.5cm) node[midway, right] {$N$ D3};
\draw[thick] (0,-1)--(0,-2) node[right] {NS5};
\node[draw=none] at (-1.7,0) {\large{$\bullet$}};
\node[draw=none] at (-2.3,0) {D5};
\node[draw=none] at (1.7,0) {\large{$\bullet$}};
\tikzset{decoration={snake,amplitude=.4mm,segment length=2mm,
                       post length=0mm,pre length=0mm}}
\draw[decorate,red,thick] (0,1) -- (0,2) node[right] {$S$};
\end{tikzpicture}
\qquad\qquad
\begin{tikzpicture}[baseline]
\tikzstyle{every node}=[font=\footnotesize]
\node[draw, circle] (node1) at (-1.5,0) {$N$};
\node[draw, circle] (node2) at (1.5,0) {$N$};
\node[draw, rectangle] (sqnode1) at (-2.8,0) {$1$};
\node[draw, rectangle] (sqnode2) at (2.8,0) {$1$};
\draw[draw=black,solid,thick,-]  (node1) to[bend right=20]   (node2) ; 
\draw[draw=black,solid,thick,-]  (node1)--(sqnode1) ; 
\draw[draw=black,solid,thick,-]  (node2)--(sqnode2) ; 
\draw[draw=red,solid,thick,-]  (node1) to[bend left=20]  node[midway,above] {{\red $T(U(N))$}}  (node2) ; 
\end{tikzpicture}
\ee

The mirror theory can be obtained by applying the S-duality to the brane system \eref{mirrT1NN1}:
\be \label{mirrT1NN1}
\begin{tikzpicture}[baseline]
\tikzstyle{every node}=[font=\footnotesize]
\draw[blue,thick] (0,0) circle (1.5cm) node[midway, right] {$N$ D3};
\draw[thick] (-0.75,-0.5)--(-1.5,-1);
\draw[thick] (-0.75,0.5)--(-1.5,1);
\node[draw=none] at (-2,1) {NS5};
\node[draw=none] at (0,-2) {D5};
\node[draw=none] at (0,-1.7) {\large{$\bullet$}};
\tikzset{decoration={snake,amplitude=.4mm,segment length=2mm,
                       post length=0mm,pre length=0mm}}
\draw[decorate,red,thick] (0,1) -- (0,2) node[right] {$S$};
\end{tikzpicture}
\qquad\qquad\qquad
\begin{tikzpicture}[baseline]
\tikzstyle{every node}=[font=\footnotesize]
\node[draw, circle] (node1) at (-1.5,0) {$N$};
\node[draw, circle] (node2) at (1.5,0) {$N$};
\node[draw, circle] (node3) at (0,1) {$N$};
\node[draw, rectangle] (sqnode) at (1.5,-1) {$1$};
\draw[draw=black,solid,thick,-]  (node1)-- (node2) ; 
\draw[draw=black,solid,thick,-]  (node2)--(node3) ; 
\draw[draw=red,solid,thick,-]  (node1)--(node3) node at (-1.5,0.7) {{\red $T(U(N))$}} ; 
\draw[draw=black,solid,thick,-]  (node2)--(sqnode) ; 
\end{tikzpicture}
\ee
 
We claim that the Higgs branch of \eref{T1NN1} is given by the following quotient:
{\small
\be \label{HiggsT1NN1}
\begin{split}
\CH_{\eref{T1NN1}} &= \Big[ \CH\left( [U(1)]-[U(N)_1] \right) \times \CN_{SU(N)_1} \times \CN_{SU(N)_2} \times \CH([U(N)_1]-[U(N)_2]) \\
& \qquad  \times \CH\left( [U(1)]-[U(N)_2] \right) \Big]/ \left(U(N)_1 \times U(N)_2 \right)~,
\end{split}
\ee}
The quaternionic dimension of $\CH_{\eref{T1NN1}}$ is $N$.  

Similarly, the Higgs branch of \eref{mirrT1NN1} is given by
{\small
\be
\begin{split}
\CH_{\eref{mirrT1NN1} } &= \Big[ \CH([U(N)_1]-[U(N)_3]) \times \CH([U(N)_2]-[U(N)_3]) \times \CH\left( [U(1)]-[U(N)_3] \right)  \\
& \quad \times \CN_{SU(N)_1} \times \CN_{SU(N)_2} \Big] / \left(U(N)_1 \times U(N)_2 \times U(N)_3 \right)~.
\end{split}
\ee}
The dimension of this space is $0$.  

Since mirror symmetry identifies the Higgs branch of \eref{mirrT1NN1} with the Coulomb branch of \eref{T1NN1}, this means that the Coulomb branch of theory \eref{T1NN1} is trivial.  This supports our proposal that the scalars in the vector multiplets of the gauge groups connected by $T(U(N))$ are frozen and do not contribute to the Coulomb branch.

Similarly to the previous example, the Higgs branch Hilbert series of \eref{T1NN1} is equal to 
\be \label{HSEx2}
\begin{split}
& H[\CH_{\eref{T1NN1}} ](t, x,y) = \\
& \int \mathrm{d}\mu_{U(N)} ( \vec u)  \int \mathrm{d}\mu_{U(N)} ( \vec w) \\ 
& \times \PE \left[ - t^2 (u_1+u_2)(u_1^{-1} +u_2^{-1}) -t^2 (w_1+w_2)(w_1^{-1} +w_2^{-1}) \right] \\
& \times\PE\left[t \left \{ x \sum_{i=1}^N u^{-1}_i + x^{-1} \sum_{i=1}^N u_i \right \}+ t \left \{ y \sum_{i=1}^N v^{-1}_i + y^{-1} \sum_{i=1}^N v_i \right \}\right]  \\
& \times H[\CN_{SU(N)}](t, \vec u)  H[\CN_{SU(N)}](t, \vec w) \\
& \times \PE \left[ \left(\sum_{i=1}^N u_i \right)  \left(\sum_{i=1}^N w^{-1}_i \right) t +  \left(\sum_{i=1}^N u^{-1}_i \right)  \left(\sum_{i=1}^N w_i \right) t  \right]~,
\end{split}
\ee
where $x$ and $y$ are the two $U(1)$ flavour fugacities.
This turns out to be equal to 
\be
H[\CH_{\eref{T1NN1}} ](t, x,y) = \PE \left[ \sum_{j=1}^{N} \left \{ t^{2j} + \left( \frac{x}{y} + \frac{y}{x} \right) t^{N+3-2j} - t^{2N+4-2j} \right \} \right]~.
\ee
The Higgs branch of \eref{T1NN1} thus has a $U(1)$ isometry. This Hilbert series, in fact, is equal to that of the Coulomb branch Hilbert series of the $U(N)$ gauge theory with $2N+1$ flavours (\ie~ the $T^{[1^{2N+1}]}_{[N+1,N]}(SU(2N+1))$ theory in the notation of \cite{Gaiotto:2008ak}).  This suggests that
\be \label{HiggsT1NN1}
\begin{split}
\CH_{\eref{T1NN1}} 
&= \CC\left(\text{$U(N)$ with $2N+1$ flavours} \right) \\
&= \CC\left( T^{[1^{2N+1}]}_{[N+1,N]}(SU(2N+1)) \right) \\
&= \text{the intersection between the Slodowy slice }\\
& \quad \,\, \text{transverse to the nilpotent orbit associated with $[N+1,N]$} \\
& \quad \,\, \text{and the nilpotent cone of $SL(2N+1,\BC)$ \cite{Gaiotto:2008ak}}~.
\end{split}
\ee
This, again, confirms the statement that the scalars in the vector multiplet of the gauge groups connected by the red line $T(U(N))$ are frozen and do not contribute to the Coulomb branch dimension.  This statement can be clearly seen in quiver \eref{mirrT1NN1}: since the two $U(N)$ gauge groups connected by the red line do not contribute to the Coulomb branch, we can effectively think of them as flavour symmetries, and so the $U(N)$ gauge group on the lower right hand corner has effectively $2N+1$ flavours transforming under it. In terms of branes, the segment of the D3-branes between two NS5 branes that is cut by the $S$-flip does not have any motion along the $x^{3,4,5}$ directions, whereas the other D3-brane segment still has a motion along those directions.

\subsection{Example 3: Quivers with a $T(U(N))$ loop}
We consider the following brane set-up and the following corresponding theory.
\be \label{TUNloop}
\begin{tikzpicture}[baseline]
\tikzstyle{every node}=[font=\footnotesize, node distance=0.45cm]
\tikzset{decoration={snake,amplitude=.4mm,segment length=2mm,
                       post length=0mm,pre length=0mm}}
\draw[blue,thick] (0,0) circle (1.5cm) node[midway, right] {$N$ D3};
\draw[decorate,red,thick] (0,1) -- (0,2) node[right] {$S$};
\def \n {6}
\def \radius {1.2cm}
\def \margin {0} 
\foreach \s in {1,...,10}
{
	\node[draw=none] (\s) at ({360/\n * (\s - 2)+30}:{\radius-10}) {};
}
\node[draw=none, circle] (last) at ({360/3 * (3 - 1)+30}:{\radius-10}) {};
\node[draw=none,  below right= of 1] (f1) {$\bullet$};
\node[draw=none, above right= of 2] (f2) {};
\node[draw=none, above = of 3] (f3) {};
\node[draw=none, above left= of 4] (f4) {};
\node[draw=none,  below left= of 5] (f5) {$\bullet$};
\node[draw=none,  below = of last] (f6) {$\bullet$};
\node[draw=none] at (-0.9,-1.3) {{\Large $\mathbf{\ddots}$}};
\node[draw=none] at (0,-2.4) {{$n$ D5s}};
\end{tikzpicture}
\qquad \qquad \qquad
\begin{tikzpicture}[baseline]
\tikzstyle{every node}=[font=\footnotesize]
\node[draw, circle] (node1) at (0,1) {$N$};
\draw[red,thick] (node1) edge [out=45,in=135,loop,looseness=5]  (node1);
\node[draw=none] at (1.3,1.5) {{\red $T(U(N))$}};
\node[draw, rectangle] (sqnode) at (0,-1) {$n$};
\draw (node1)--(sqnode);
\end{tikzpicture}
\ee
The mirror theory can be obtained by applying $S$-duality to the above system:
\be \label{TUNloopmirr}
\begin{tikzpicture}[baseline]
\tikzstyle{every node}=[font=\footnotesize]
\tikzset{decoration={snake,amplitude=.4mm,segment length=2mm,
                       post length=0mm,pre length=0mm}}
\draw[blue,thick] (0,0) circle (1.5cm) node[midway, right] {$N$ D3};
\draw[decorate,red,thick] (0,1) -- (0,2) node[right] {$S$};
\def \n {6}
\def \radius {1.2cm}
\def \margin {0} 
\foreach \s in {1,...,10}
{
	\node[draw=none] (\s) at ({360/\n * (\s - 2)+30}:{\radius-10}) {};
}
\node[draw=none, circle] (last) at ({360/3 * (3 - 1)+30}:{\radius-10}) {};
\node[draw=none,  below right= of 1] (f1) {};
\node[draw=none, above right= of 2] (f2) {};
\node[draw=none, above = of 3] (f3) {};
\node[draw=none, above left= of 4] (f4) {};
\node[draw=none,  below left= of 5] (f5) {};
\node[draw=none,  below = of last] (f6) {};
\node[draw=none] at (-1,-1.5) {{\Large $\mathbf{\ddots}$}};
\node[draw=none] at (0,-2.4) {{$n$ NS5s}};
\draw[-, >=latex,black, thick] (1) to (f1);
\draw[-, >=latex,black, thick] (5) to (f5);
\draw[-, >=latex,black, thick] (last) to (f6);
\end{tikzpicture}
\qquad \qquad
\begin{tikzpicture}[baseline, scale=0.6,font=\scriptsize]
\def \n {6}
\def \radius {2.4cm}
\def \margin {15} 
\node[draw, circle] at ({360/\n * (1 - 2)}:\radius) {$N$};
\draw[-, >=latex] ({360/\n * (1 - 3)+\margin}:\radius);
arc ({360/\n * (1 - 3)+\margin}:{360/\n * (1-2)-\margin}:\radius);
\node[draw, circle] at ({360/\n * (2 - 2)}:\radius) {$N$};
\draw[-, >=latex] ({360/\n * (2 - 3)+\margin}:\radius);
arc ({360/\n * (2 - 3)+\margin}:{360/\n * (2-2)-\margin}:\radius);
\foreach \s in {3,...,5}
\node[draw, circle] at ({360/\n * (\s - 2)}:\radius) {$N$};
\foreach \s in {1,...,5}
{	
	\draw[-, >=latex] ({360/\n * (\s - 3)+\margin}:\radius) 
	arc ({360/\n * (\s - 3)+\margin}:{360/\n * (\s-2)-\margin}:\radius);
}
\node[draw, circle] at ({360/3 * (3 - 1)}:\radius) {$N$};
\draw[thick, red,-, >=latex] ({360/6 * (5 -2)+\margin}:\radius) 
arc ({360/6 * (5 -2)+\margin}:{360/6 * (5-1)-\margin}:\radius);
\node[draw=none] at (4.8,-1.8) {$(n+1)$ nodes};
\node[draw=none] at (-3.2,-1.5) {{\red $T(U(N))$}};
\end{tikzpicture}
\ee
The Higgs branch of \eref{TUNloop} is given by the following description
\be
\CH_{\eref{TUNloop}} = \frac{\CN_{SU(N)} \times \CN_{SU(N)} \times \CH\left( [U(N)] - [U(n)]\right)}{U(N)}~.
\ee
The quaternionic dimension of which is equal to
\be \label{dimHTUNloop}
\dim_\BH \, \CH_{\eref{TUNloop}} =  \left[ 2 \times \frac{1}{2}(N-1)(N) \right] + n N - N^2  = (n-1)N~.
\ee
Observe that for $n=1$, the Higgs branch is trivial for any $N$.  On the other hand, the Higgs branch of \eref{TUNloopmirr} is given by the following description
\be
\CH_{\eref{TUNloopmirr}} = \frac{\CN_{SU(N)} \times \CN_{SU(N)} \times \CH[U(N)-U(N)]^n}{U(N)^{n+1}/U(1)^{N}}~,
\ee
where we quotiented by $U(N)^{n+1}/U(1)^{N}$ because at a generic point on the Higgs branch, the gauge symmetry $U(N)^{n+1}$ is not completely broken but it is broken to $U(1)^N$ (see \eg~ \cite{Hanany:2010qu}).  The dimension of this space is actually zero:
\be
\dim_\BH \, \CH_{\eref{TUNloopmirr}} =   \left[ 2 \times \frac{1}{2}(N-1)(N) \right]  + n N^2 - \left[ (n+1)N^2 - N\right] = 0~.
\ee

From mirror symmetry, $\CC_{\eref{TUNloop}}$ is identified with  $\CH_{\eref{TUNloopmirr}}$, and so
\be
\dim_\BH \, \CC_{\eref{TUNloop}} = \dim_\BH \, \CH_{\eref{TUNloopmirr}} =0~.
\ee
This is consistent with our proposal because \eref{TUNloop} has a single circular node that is connected by the $T(U(N))$ link and so it does not contribute to the Coulomb branch dynamics.

On the other hand, it can be checked using the Hilbert series that the Higgs branch $\CH_{\eref{TUNloop}}$ is in fact isomorphic to the Coulomb branch of the following quiver\footnote{For example, the Hilbert series of the Higgs branch $\CH_{\eref{TUNloop}}$ for $N=n=2$ is precisely the Coulomb branch Hilbert series of 3d $\CN=4$ $U(2)$ gauge theory with $4$ flavours. These can be computed similarly as in the preceding subsections.}
\be \label{NNN}
\sqwnode{}{N}- \underbrace{\node{}{N} - \cdots -\node{}{N}}_{(n-1)\,\,\text{nodes}} - \sqwnode{}{N}~.
\ee
This quiver can be derived from \eref{TUNloopmirr} using our proposal:  since the vector multiplets two gauge nodes linked by $T(U(N))$ in \eref{TUNloopmirr} are frozen, we can take them to be flavour nodes, and quiver \eref{NNN} thus follows. 

Amusingly, using brane and mirror symmetry (see \cite[(2.5)]{Hanany:2018vph}), we also know that 
\be \label{moremirr}
\CH_{\eref{TUNloop}}  = \CC_{\eref{NNN}} = \CH \left[ \node{}{1}-\node{}{2}- \cdots- \node{}{N-1}-\node{\wver{}{n}}{N}- \node{}{N-1}- \cdots-\node{}{2}- \node{}{1}  \right]~.
\ee
In a special case of or $n=1$, the quiver on right of the above equation is the star-shaped quiver that is mirror \cite{Benini:2010uu} to the $S^1$ compactification of a clsss $\mathsf{S}$ theory of type $A_{N-1}$ associated with a sphere with two maximal and one minimal puncture.  The latter is actually a theory of free hypermultiplets.  Thus, the spaces in \eref{moremirr} are zero dimensional; this is in agreement with \eref{dimHTUNloop}. 


\section{Abelian theories with non-zero Chern--Simons levels} \label{sec:abel}
In this section, we focus on field theories that arise from Hanany--Witten brane configurations, with a single D3-brane on $S^1$ and with an inclusion of $J$-folds.  These can be represented as abelian quiver theories  with non-zero Chern--Simons (CS) levels\footnote{We denote the CS level by a subscript, for example $U(N)_k$ denotes a group $U(N)$ with CS level $k$.  In a quiver node, we abbreviate this as $N_k$.}, and $T(U(1))$ connected between quiver nodes.  The presence of a $T(U(1))$ link between two quiver nodes gives rise to a mixed CS level between them.  In fact, the systems consisting only a D3-brane on the circle and $J$-folds (but with no D5 and no NS5 brane) were studied in \cite{Ganor:2014pha}. Such systems give rise to pure CS theories.  In order to make the moduli space more interesting, we may also include NS5 and D5 branes in the system.  These introduce bi-fundamental and fundamental hypermultiplets into the quiver theory.  The moduli space of theories in this section is more sophisticated to analyse than those in section \ref{sec:zeroCS}.  This is because the vacuum equations may admit many sets of non-trivial solutions, in which case the moduli space has many branches. Below we systematically analyse such branches, and provide necessary conditions on the CS levels in order to have a non-trivial moduli space.

As a warm-up, we first analyse linear quivers without a $T(U(1))$ link in section \ref{sec:linquivwoJ}. This also serves as a generalise of the analysis in \cite{Cremonesi:2016nbo} and a complement to the analysis of \cite{Assel:2017eun}, where in this paper we provide direct analyses of the moduli space from the vacuum equations and compute the Hilbert series.  Subsequently in section \ref{sec:oneJfoldnoflv}, we introduce a $J$-fold in to the brane system. Finally, in section \eref{sec:addflavours}, we add flavours in to the quiver.  In the latter, under some conditions, the fundamental hypermultiplets may contribute non-trivially to the moduli space.  The analysis for theories with more than one $J$-fold is more technical and we postpone the discussion to Appendix \ref{app:multiJ}.  

\subsection{Warm-up: Theories without a $J$-fold} \label{sec:linquivwoJ}
Before adding a $J$-fold to the brane systems, it is instructive studying in a systematic way the moduli space of linear quivers without fundamental matter. 
\be \label{genlinquiv}
\begin{tikzpicture}[baseline, scale=0.7]
\tikzstyle{every node}=[font=\footnotesize, minimum size=1.2cm]
\node[draw, circle] (node1) at (-2,2) {$1_{k_1}$};
\node[draw, circle] (node2) at (2,2) {$1_{k_2}$};
\node[draw, circle] (node3) at (6,2) {$1_{k_{n-1}}$};
\node[draw, circle] (node4) at (10,2) {$1_{k_n}$};
\draw[draw=black,solid,thick,-]  (node1) to  (node2) ; 
\draw[draw=black,dashed,thick,-]  (node2) to  (node3) ; 
\draw[draw=black,solid,thick,-]  (node3) to  (node4) ; 
\end{tikzpicture}
\ee
This is made up of $n$ $U(1)$ gauge nodes with Chern-Simons levels $k_i\,,\,\,i\,=\,1,\dots\,n$. The $i$-th node is connected to the $(i-1)$-th one by an hyper-multiplet $(A_i, \tA_i)$. In $\CN=2$ language, the quiver appears as:
\be 
\begin{tikzpicture}[baseline, scale=0.7]
\tikzstyle{every node}=[font=\footnotesize, minimum size=1.2cm]
\node[draw, circle] (node1) at (-2,2) {$1_{k_1}$};
\node[draw, circle] (node2) at (2,2) {$1_{k_2}$};
\node[draw, circle] (node3) at (6,2) {$1_{k_{n-1}}$};
\node[draw, circle] (node4) at (10,2) {$1_{k_n}$};
\draw[draw=black,solid,thick,<->]  (node1) to node[pos=0.2, above=-0.2cm] {$\tA_1$} node[pos=0.8, above=-0.25cm] {$A_1$}  (node2) ; 
\draw[draw=black,solid,thick,<->]  (node3) to node[pos=0.2, above=-0.2cm] {$\tA_{n-1}$} node[pos=0.8, above=-0.25cm] {$A_{n-1}$}  (node4) ; 
\draw[draw=black,dashed,thick,-]  (node2) to  (node3) ;
\draw[black] (node1) edge [out=-45,in=-135,loop,looseness=4] node[midway,below=-0.2cm] {$\varphi_1$}  (node1);
\draw[black] (node2) edge [out=-45,in=-135,loop,looseness=4] node[midway,below=-0.2cm] {$\varphi_2$}  (node2);
\draw[black] (node3) edge [out=-45,in=-135,loop,looseness=4] node[midway,below=-0.3cm] {$\varphi_{n-1}$}  (node3);
\draw[black] (node4) edge [out=-45,in=-135,loop,looseness=4] node[midway,below=-0.3cm] {$\varphi_n$}  (node4);
\end{tikzpicture}
\ee
with the superpotential
\be
W = \sum_{i=1}^{n-1} (\tilde{A}_i \varphi_i A_i - A_i \varphi_{i+1} \tilde{A}_i) +\frac{1}{2}\sum_{i=1}^n k_i \varphi_i^2~.
\ee

Due to $\CN=3$ supersymmetry of the theory, we are allowed to collect at the same time both F-terms and D-terms, in such a way we really need to solve a unique set of  equations. Let us call $\Phi_i\,=\, (\varphi_i\,,\, \sigma_i)$, $\mu_i\,=\, (A_i\,\tA_i\,,\, |A_i|^2 - |\tA_i|^2)$: the whole set of $F$-terms and $D$-terms now read
\begin{equation}
\label{linearA}
A_i(\Phi_{i+1}-\Phi_i)\,=\,0\,,\quad \tA_i(\Phi_{i+1}-\Phi_i)\,=\,0\quad i=1\,,\dots\,,n-1
\end{equation}
\begin{equation}
\label{quadraticA}
\begin{split}
&k_1\,\Phi_1\,=\, \mu_1\\
& k_i\,\Phi_i\,=\, \mu_i-\mu_{i-1}\quad i=2\,,\dots\,,n-1\\
&k_n\,\Phi_n\,=\, - \mu_{n-1}
\end{split}
\end{equation}
Moreover, the R-charge and gauge charges of the monopole operators with flux $(m_1, \ldots, m_n)$ read, respectively:
\begin{equation}
R[V_{(m_1, \ldots, m_n)} ]\,=\, \dfrac{1}{2}\sum_{i=1}^{n-1}|m_{i+1}-m_i| \,,\quad q_i[V_{(m_1, \ldots, m_n)} ]\,=\, -k_i\,m_i
\end{equation}
where $m_i$ is the magnetic flux of the $i$-th gauge group. 

\subsubsection*{Cutting the quiver}

It is convenient to study the solutions to the vacuum equations according to the vanishing of the VEVs of the bi-fundamental hypermultiplets.  In particular, the vacuum equations may admit the solutions in which 
\be
\begin{split}
&A_{l_1}=\tilde{A}_{l_1}=A_{l_2}=\tilde{A}_{l_2}= \cdots =A_{l_m}=\tilde{A}_{l_m}=0~, \,\, \text{for some $l_1 < l_2 < \cdots <l_m$} \\ 
&\text{and}\,\,\, A_{p}, \, \tilde{A}_{p} \neq 0 \,\, \text{for $p\notin \{l_1,l_2, \ldots, l_m \}$}~,
\end{split}
\ee
In which case, the quiver diagram in question is naturally divided into sub-quivers, and we shall henceforth say that the quiver is ``cut'' at the positions $l_1, l_2, \cdots, l_m$.  If the vacuum equations do not admit such a solution, we say that the quiver cannot be cut.  As we shall see in explicit examples below, the vacuum equations of certain quivers may admit more than one option of cuts, in which case, each option gives rise to a branch of the moduli space.

In order to determine whether we need to cut the quiver, we can proceed as follows.  Suppose that the quiver cannot be cut, \ie~ all $A_i$ and $\tilde{A}_i$ are non-zero. This implies that $\Phi_i = \Phi \neq 0$ for all $i$.  If the system of equations \eref{quadraticA} admits a solution in which $\mu_j =0$ for some $j$, then our initial assumption that the quiver cannot be cut is contradicted, and we need to cut a quiver somewhere.  However, it should be emphasised that if the aforementioned system of equations have a solution in which $\mu_j \neq 0$ for all $j$, what we can infer is that there is a branch of the moduli space corresponding to no cut; however, there may exist another branch of the moduli space corresponding to a cut in the quiver.

Let us now cut the quiver in question at two positions, namely $l$ and $m$ with $m>l$.  This divides the the orginal quiver into three sub-quivers that we will denote as: ``left", collecting the nodes first $l$ nodes, ``central", collecting the node $l+1\,,\dots,\,l+m$, and finally ``right" encoding the last $n-l-m$ nodes, as depicted below.
\be \label{cutillustrated}
\scalebox{0.8}{
\begin{tikzpicture}[baseline, scale=0.7]
\tikzstyle{every node}=[font=\scriptsize, minimum size=1.4cm]
\node[draw, circle] (node1) at (-3,2) {$1_{k_1}$};
\node[draw, circle] (node2) at (0,2) {$1_{k_l}$};
\node[draw, circle] (node3) at (5,2) {$1_{k_{l+1}}$};
\node[draw, circle] (node4) at (8,2) {$1_{k_{l+m}}$};
\node[draw, circle] (node5) at (13,2) {$1_{k_{l+m+1}}$};
\node[draw, circle] (node6) at (17,2) {$1_{k_n}$};
\draw[draw=black,dashed,thick,-]  (node1) to  (node2) ; 
\draw[draw=blue,solid,thick,-]  (node2) to node[midway, above] {\blue{$A_l=\tilde{A}_l=0$}}  (node3); 
\draw[draw=black,dashed,thick,-]  (node3) to  (node4); 
\draw[draw=blue,solid,thick,-]  (node4) to node[midway, above] {\blue{$A_{l+m}=\tilde{A}_{l+m}=0$}}  (node5);
\draw[draw=black,dashed,thick,-]  (node5) to  (node6); 
\end{tikzpicture}}
\ee
Below we derive necessary conditions for each sub-quivers to contribute non-trivially to the moduli space.

Let us consider the left sub-quiver.  We fix $A_l\,=\,\tA_l\,=\,0$ and assume that $A_i$ and $\tilde{A}_i$ are non-vanishing for all $i=1, 2, \ldots, l$.  Then \eqref{linearA} implies that $\Phi_i\,=\,\Phi\,=\,(\varphi\,,\,\sigma)\,\,\forall\.i=1,2, \ldots, l$. The sum of the first $l$ equations in $\eqref{quadraticA}$ provides the following constraint
\be
\left(\sum_{i=1}^l k_i\right)\,\varphi\,=\, A_l\tA_l\,=\,0~.
\ee
Since $\varphi \neq 0$ (otherwise $A_{l-1}\tilde{A}_{l-1}$ would be zero, contradicting our assumption), we see that a {\it necessary condition} for the left sub-quiver to contribute non-trivially to the moduli space of vacua is
\be
\sum_{i=1}^l k_i =0~.
\ee

A similar argument also applies for the right sub-quiver.  We fix  $A_{l+m}=\tilde{A}_{l+m}=0$ and assume that $A_i$ and $\tilde{A}_i$ are non-vanishing for all $i=l+m+1,\ldots, n$.  A necessary condition for this sub-quiver to contribute non-trivially to the moduli space is 
\be \label{lastnodes}
\sum_{i=l+m+1}^{n} k_i=0~,
\ee

If the central sub-quiver contains a sub-quiver whose CS levels sum to zero, we may cut the former further into smaller sub-quivers.  Otherwise, a necessary condition for the central sub-quiver to contribute non-trivially to the moduli space is
\be
\sum_{i=l+1}^{l+m} k_i =0~.
\ee
This again follows from the sum of the $(l+1)$-th to the $(l+m)$-th equations in \eref{quadraticA}, with $\mu_{l}=\mu_{l+m}=0$. 

Note that there can be many ways in cutting a given quiver into sub-quivers.  Consider the following gauge theory as an example
\be \label{modelIV}
\begin{tikzpicture}[baseline, scale=0.7]
\tikzstyle{every node}=[font=\scriptsize, minimum size=1cm]
\node[draw, circle] (node1) at (-6,0) {$1_{-1}$};
\node[draw, circle] (node2) at (-3,0) {$1_{+1}$};
\node[draw, circle] (node3) at (0,0) {$1_{-1}$};
\node[draw, circle] (node4) at (3,0) {$1_{+1}$};
\draw[draw=black,solid,thick,-]  (node1) to  (node2) ; 
\draw[draw=black,solid,thick,-]  (node2) to  (node3) ; 
\draw[draw=black,solid,thick,-]  (node3) to  (node4) ; 
\end{tikzpicture}
\ee
There are two ways in cutting such a quiver in order to obtain a non-trivial moduli space, namely
\be \label{cutsmodelIV}
\begin{split}
\mathrm{I}: &\qquad \begin{tikzpicture}[baseline, scale=0.7]
\tikzstyle{every node}=[font=\scriptsize, minimum size=1cm]
\node[draw, circle] (node1) at (-6,0) {$1_{-1}$};
\node[draw, circle] (node2) at (-3,0) {$1_{+1}$};
\node[draw, circle] (node3) at (0,0) {$1_{-1}$};
\node[draw, circle] (node4) at (3,0) {$1_{+1}$};
\draw[draw=black,solid,thick,-]  (node1) to  (node2) ; 
\draw[draw=blue,solid,thick,-]  (node2) to node[midway, above=0.2] {\blue{$A_2=\tilde{A}_2=0$}}  (node3) ; 
\draw[draw=black,solid,thick,-]  (node3) to  (node4) ; 
\end{tikzpicture}\\
\mathrm{II}: &\qquad \begin{tikzpicture}[baseline, scale=0.7]
\tikzstyle{every node}=[font=\scriptsize, minimum size=1cm]
\node[draw, circle] (node1) at (-6,0) {$1_{-1}$};
\node[draw, circle] (node2) at (-3,0) {$1_{+1}$};
\node[draw, circle] (node3) at (0,0) {$1_{-1}$};
\node[draw, circle] (node4) at (3,0) {$1_{+1}$};
\draw[draw=blue,solid,thick,-]  (node1) to node[midway, above=0.2] {\blue{$A_1=\tilde{A}_1=0$}}  (node2) ; 
\draw[draw=black,solid,thick,-]  (node2) to  (node3) ; 
\draw[draw=blue,solid,thick,-]  (node3) to node[midway, above=0.2] {\blue{$A_3=\tilde{A}_3=0$}}  (node4) ; 
\end{tikzpicture}
\end{split}
\ee
In case I, both left and right sub-quivers contribute non-trivially to the moduli space, whereas in case II, only the central sub-quiver contributes non-trivially.  We shall refer to the vacuum spaces corresponding to these two options as {\it branches} of the moduli space for \eref{modelIV}.  We shall go over the detailed computation of the moduli space later.

\subsubsection*{The Hilbert series}
Let us consider quiver \eref{cutillustrated} and assume that the left, central and right sub-quivers cannot be cut further.  Using \eref{linearA}, we see that $\sigma_1=\sigma_2=\ldots=\sigma_l$, \,\, $\sigma_{l+1} = \sigma_{l+2} = \ldots = \sigma_{l+m}$, and $\sigma_{l+m+1} = \ldots = \sigma_{n}$.  In other words, the magnetic fluxes for the monopole operators for all nodes in each sub-quiver are equal:
\be
\begin{split}
m_1=m_2=\ldots=m_l &\equiv m_L~, \\
m_{l+1} = m_{l+2} = \ldots = m_{l+m} &\equiv m_{C}, \\
m_{l+m+1} = m_{l+m+2} = \ldots = m_{n} &\equiv m_{R}~.
\end{split}
\ee
The $R$-charge of the monopole operator with the flux $(m_1, \ldots, m_n)$ is therefore
\be
R[V_{(m_1, \ldots, m_n)}] = \frac{1}{2} \sum_{i=1}^{n-1} |m_i - m_{i+1}| = \dfrac{1}{2}\left(|m_L-m_C|+|m_C-m_R|\right)~.
\ee

The Hilbert series can be computed using the same procedure as presented in \cite[sec. 4--sec. 6]{Cremonesi:2016nbo}.  The idea is to count the monopole operators dressed by appropriate chiral fields in the theory such that the combination is gauge invariant.  The appropriate combination of chiral fields that are used to dress the monopole operators are counted by the baryonic generating function  \cite{Forcella:2007wk}.

Let $g_{L}(t, \vec B)$, $g_{C}(t, \vec B)$ and $g_{R}(t, \vec B)$ be baryonic generating functions for the left, central and right sub-quivers, respectively.  Then, the Hilbert series for the moduli space for quiver \eref{cutillustrated} is given by
\be
\begin{split}
H(t; z_L, z_C, z_R) &=\, \sum_{m_L \in \mathbb Z}\sum_{m_C\in \mathbb Z}\sum_{m_R\in \mathbb Z}  t^{|m_L-m_C|+|m_C-m_R|} z_L^{m_L} z_C^{m_C} z_R^{m_R} \times \\
& \qquad g_{L}(t, \{ k_1 m_{L},\ldots, k_l m_L \} )\,g_{C}(t, \{ k_{l+1} m_{C},\ldots, k_{m-1} m_C \}) \times \\
& \qquad g_{R}(t, (k_m m_{R}, \ldots, k_n m_R \})~,
\end{split}
\ee
where $z_{L, C, R}$ are fugacities for the topological symmetries.  The first line is the contribution from the monopole operators and the second and third lines are the contribution from an appropriate combination of chiral fields in the quiver that will be used to dress the monopole operators.

\subsubsection*{Example 1: Quiver \eref{modelIV}}  
The two non-trivial cuts depicted in \eref{cutsmodelIV} corresponds to two non-trivial branches of the moduli space.

\paragraph{{\it Branch I.}} This corresponds to the top diagram in \eref{cutsmodelIV}, where the VEVs of $A_2$ and $\tilde{A}_2$ are zero, and the VEVs of other bifundamentals are non-zero.  The cut splits the quiver \eref{modelIV} into two sub-quivers, each of which can be identified as the half-ABJM theory\footnote{We define the {\bf half-ABJM theory} by a theory with $U(1)_k \times U(1)_{-k}$ gauge symmetry with a single bi-fundamental hypermultiplet.} \cite[sec. 4.1.3]{Cremonesi:2016nbo}.  Let us denote the magnetic fluxes associated with the four nodes of the quiver from left to right by $(m_L, m_L, m_R, m_R)$.  The Hilbert series for this branch of the moduli space is then given by
\be
\begin{split}
H^{\mathrm{(I)}}_{\eref{modelIV}} (t;z_1, z_2) &= \sum_{m_L \in \BZ} \,\, \sum_{m_R \in \BZ} t^{|m_L-m_R|} g_{\text{ABJM/2}}(t; m_L) g_{\text{ABJM/2}}(t; m_R) \\ 
&= \sum_{m_L \in \BZ} \,\, \sum_{m_R \in \BZ} t^{|m_L-m_R|}  \frac{t^{|m_L|}}{1-t^2} \frac{t^{|m_R|}}{1-t^2} z_1^{m_L} z_2^{m_R}  \\
&= \sum_{m=0}^\infty \chi^{SU(3)}_{[m,m]}(z_1, z_2) t^{2m}~.
\end{split}
\ee
where $g_{\text{ABJM/2}}(t; B)$ is the baryonic generating function of the half-ABJM theory
\be
g_{\text{ABJM/2}}(t; B) = \oint_{|u_1=1} \frac{\mathrm{d} u_1}{ 2 \pi i u_1^{B+1}} \oint_{|u_2=1} \frac{\mathrm{d} u_2}{ 2 \pi i u_2^{-B+1}} \PE \left[(u_1 u_2^{-1} + u_1^{-1} u_2)t  \right] = \frac{t^{|B|}}{1-t^2}~,
\ee
and the character of the adjoint representation $[1,1]$ of $SU(3)$ is
\be
\chi^{SU(3)}_{[1,1]}(z_1, z_2) = 2+z_1 z_2+\frac{1}{z_1 z_2}+z_1+\frac{1}{z_1}+z_2+\frac{1}{z_2}~.
\ee
The last line indicates that this branch is isomorphic to the reduced moduli space of one $SU(3)$ instanton on $\BC^2$ \cite{Benvenuti:2010pq}, or equivalently the closure of the minimal nilpotent orbit of $SU(3)$.  The eight generators can be written in terms of a traceless $3 \times 3$ matrix as
\be
M = \begin{pmatrix}
\varphi_L         & V_{(1,1,0,0)} & V_{(1,1,1,1)} \\
V_{(-1,-1,0,0)}   & \varphi_R     &  V_{(0,0,1,1)} \\
V_{(-1,-1,-1,-1)}   & V_{(0,0,-1,-1)}      &  - \varphi_L - \varphi_R \\
\end{pmatrix}
\ee
where $\varphi_L  = \varphi_1 = \varphi_2$ and $\varphi_R = \varphi_3 = \varphi_4$.  The Hilbert series indicates that the matrix $M$ satisfies the following conditions \cite{Gaiotto:2008nz}:
\be
\mathrm{rank}\, M \leq 1~, \qquad M^2=0~.
\ee

\paragraph{{\it Branch II.}} This corresponds to the bottom diagram in \eref{cutsmodelIV}, where the VEVs of $A_1$, $\tilde{A}_1$, $A_3$ and $\tilde{A}_3$ are zero, and the VEVs of other bifundamentals are non-zero.  In this case, only the central sub-quiver contributes to the computation of the Hilbert series.  The magnetic fluxes associated with the four nodes of the quiver from left to right can be written as $(0, m, m, 0)$, with $m\in \BZ$, where the zeros follow from the $D$-term equations.  The Hilbert series for this branch of the moduli space is then given by
\be \label{C2Z3HS}
\begin{split}
H^{\mathrm{(II)}}_{\eref{modelIV}} (t;z) &= \sum_{m \in \BZ}  t^{|0-m|+|m-m|+|m-0|} g_{\text{ABJM/2}}(t; m) z^m \\ 
&= \sum_{m \in \BZ}  t^{2|m|}  \frac{t^{m}}{1-t^2} = \PE \left[ t^2 +(z+z^{-1})t^3 - t^6\right]~.
\end{split}
\ee
This indicates that this branch is isomorphic to $\BC^2/\BZ_3$.  The generators of this moduli space are $V_{(0,1,1,0)}$, $V_{(0,-1,-1,0)}$ and $\varphi\equiv \varphi_2= \varphi_3$, satisfying the relation
\be \label{C2Z3ring}
V_{(0,1,1,0)} V_{(0,-1,-1,0)} = \varphi^3~.
\ee  

Branches I and II of \eref{modelIV} are indeed the Higgs and Coulomb branches of 3d $\CN=4$ $U(1)$ gauge theory with 3 flavours, as pointed out in \cite[sec. 4.2]{Jafferis:2008em}.  The brane system of the former can be obtained by applying the $SL(2,\BZ)$ action $T^T$ to the brane system of the latter.

\subsubsection*{Example 2: No cut in the quiver \eref{genlinquiv}}
We assume that $A_i$ and $\tilde{A}_i$ are non-vanishing for all $i=1,\ldots, n$, \ie~ there is no cut in the quiver.
In this case, \eqref{linearA} implies that
\begin{equation}
\Phi_i\,=\, \Phi\,=\,(\varphi\,,\, \sigma)\quad \forall\,i=1,\ldots,n
\end{equation}
As a consequence, the magnetic fluxes are constrained to be all equal $m_1\,=\,m_2\,=\,\dots\,=\,m$. The equations \eqref{quadraticA}, instead, simply constrain the bilinears $\mu_i$ in terms of $\varphi$. Summing over the $n$ equations, we obtain the following condition
\begin{equation} \label{sumD}
(k_1\,+\,k_2\,+\dots+\, k_n)\,\Phi=0
\end{equation}
Note that $\Phi=0$ would imply $\mu_i=1\,\,\forall\,\,i$ contradicting the initial assumption that all $A_i,\tA_i\,\neq\,0$. Thus, as we discuss before, the moduli space is non-trivial if
\begin{equation}
\label{linearConstr}
\sum_{i=1}^n\,k_i\,=\,0
\end{equation}
Let us assume \eref{linearConstr} in the subsequent discussion. 

The bare monopoles $V_{(m,\ldots,m)}$, with flux ${(m,\ldots,m)}$, have $R$-charge $R[V_{(m,\ldots,m)}]\,=\,0$.  They need to be dressed in order to make them gauge invariant, because of their gauge charge under the $i$-th gauge group is $q_i[V_{(m,\ldots,m)}]\,=\, -k_i\,m$.   Let us define for convenience
\be
K_i\,=\, \sum_{j=1}^i\,k_j
\ee
If $K_i \geq 0$ for all $i=1, \ldots, n-1$, we can form the following gauge invariant dressed monopole operator:
\begin{equation}
\begin{split}
&\bar V_+\,\equiv\, V_{(1,\ldots, 1)}\,A_1^{K_1}\,A_2^{K_2}\dots A_{n-1}^{K_{n-1}}~,\\
&\bar V_-\,\equiv\, V_{(-1,\ldots, -1)}\,\tA_1^{K_1}\,\tA_2^{K_2}\dots \tA_{n-1}^{K_{n-1}}~.
\end{split}
\end{equation}
Note that if $K_j< 0$ for some $j$, we replace $A_j^{K_j}$ in the first equation by $\tilde{A}_j^{-K_j}$, and $\tilde{A}_j^{K_j}$ in the second equation by $A_j^{-K_j}$.  In any case, the $R$-charges of the above dressed monopole operators are
\be
R[\bar V_{\pm}]\,=\, \dfrac{1}{2}\sum_{i=1}^{n-1}\,|K_i| = \frac{1}{2} K
\ee
with 
\be
K \equiv \sum_{i=1}^{n-1}\,|K_i|~.
\ee
The chiral ring is generated by the three operators $\{\varphi\,,\,\bar V_+\,,\,\bar V_-\}$, statisfying the following relation:
\be
\bar V_+\,\bar V_-\,=\, \varphi^K~.
\ee
Thus, the variety associated to this branch is:
\be
\mathbb{C}^2/\mathbb{Z}_{K}~.
\ee

We can obtain the same result using the Hilbert series. Let us call $\{q_1\,,q_2\,,\dots,\,q_n\}$ the fugacities associated to the $n$ gauge nodes and $t$ the fugacity associated to the $R$-symmetry. The ingredients entering the Hilbert series are:
\bi
\item The $n-1$ bifundamental hypermultiplets contribute as:
\be
\PE[t(q_1q_2^{-1}+q_1^{-1}q_2)] \PE[t(q_2q_3^{-1}+q_2^{-1}q_3)]\dots\,\PE[t(q_{n-1}q_n^{-1}+q_{n-1}^{-1}q_n)]
\ee
\item There is also a contribution from $\varphi$ which gives $\PE[t^2]$.
\item The $F$-terms \eref{quadraticA} impose further $(n-1)$ constraints on the former, after taking into account the condition \eref{sumD}, which is the overall sum of \eref{quadraticA}.  These contribute $\PE[-(n-1)t^2]$ to the Hilbert series.
\ei
The baryonic generating function is thus:
\be
g(t; \vec B)\,=\PE[-(n-1)t^2] \PE[t^2] \oint \dfrac{dq_1}{2\pi iq_1^{1+B_1}}\dots\oint \dfrac{dq_n}{2\pi iq_n^{1+B_n}}\prod_{i=1}^{n-1} \PE[t(q_iq_{i+1}^{-1}+q_i^{-1}q_{i+1})]
\ee
and can perform a change of variable:
\be
\{y_1\,,y_2\,,\dots,\,y_n\}\,=\,\{ q_1q_2^{-1}\,,\,q_2q_3^{-1}\,\dots\,,\,q_{n-1}q_n^{-1}\,,\,q_n \}
\ee
Thus, the baryonic function becomes:
\be
\PE[-(n-2)t^2]\prod_{i=1}^{n-1}\oint \dfrac{dy_i}{2\pi iy_i^{1+\tB_i}} \PE[t(y_i+y_i^{-1})]\oint \dfrac{dy_n}{2\pi iy_i^{1+\tB_n}}
\ee
where we defined $\tB_i\,=\,\sum_{j=1}^i\,B_j$. The previous integrals are known:
\be
\oint \dfrac{dy_i}{2\pi iy_i^{1+\tB_i}} \PE[t(y_i+y_i^{-1})]\,=\, \dfrac{t^{\abs{\tB_i}}}{1-t^2}\,\,,\quad \oint \dfrac{dy_n}{2\pi iy_i^{1+\tB_n}}\,=\, \delta_{\tB_n\,,\,0}
\ee
and then the baryonic generating function simplifies to
\be \label{barylinquiv}
g(t; \vec B)\,=\, \dfrac{t^{\sum_{i=1}^{n-1}|\tB_i |}}{1-t^2}\delta_{\tB_n\,,\,0}~, ~ \text{with} \,\, \tB_i\,=\,\sum_{j=1}^i\,B_j~.
\ee
Recall that the charge of the monopole operator under the $U(1)_i$ gauge symmetry is  $q_i[V_{(m,\ldots,m)}]\,=\,-k_i m$. As a consequence, the Hilbert series reads:
\be \label{HSnocutlin}
\begin{split}
H(t; z)\,&= \sum_{m \in \BZ} g(t; \{k_1 m, \ldots, k_n m\} )  z^m  \\
&= \frac{1}{1-t^2}\sum_{m\in\mathbb Z} t^{|m| \sum_{i=1}^n |\sum_{j=1}^i k_j |} z^m \\
&= \frac{1}{1-t^2}\sum_{m\in\mathbb Z} t^{K\abs{m}} z^m\\
&= \PE\left[ t^2 + (z+z^{-1}) t^K - t^{2K}  \right] \, ,
\end{split}
\ee
where $\tB_n$ in \eref{barylinquiv} is $m \sum_{i=1}^nk_i\,=\,0$ and hence the Kronecker delta gives 1.  Here $z$ is the fugacity for the topological symmetry. We obtained exactly the Hilbert series of $\mathbb{C}^2/\mathbb{Z}_{K}$.

\paragraph{Example.}  Let us consider the following quiver.
\be \label{modelIVswap}
\begin{tikzpicture}[baseline, scale=0.7]
\tikzstyle{every node}=[font=\scriptsize, minimum size=1cm]
\node[draw, circle] (node1) at (-6,0) {$1_{-1}$};
\node[draw, circle] (node2) at (-3,0) {$1_{-1}$};
\node[draw, circle] (node3) at (0,0) {$1_{+1}$};
\node[draw, circle] (node4) at (3,0) {$1_{+1}$};
\draw[draw=black,solid,thick,-]  (node1) to  (node2) ; 
\draw[draw=black,solid,thick,-]  (node2) to  (node3) ; 
\draw[draw=black,solid,thick,-]  (node3) to  (node4) ; 
\end{tikzpicture}
\ee
This quiver has two non-trivial branches.  One corresponds to no cut at all and the other corresponds to the cuts in the first and the third position.  As we discussed above, the former branch is isomorphic to $\BC^2/\BZ_4$.  The second branch is the same as that discussed around \eref{C2Z3HS} and \eref{C2Z3ring}; it is isomorphic to $\BC^2/\BZ_3$.

\subsection{Theories with one $J$-fold} \label{sec:oneJfoldnoflv}
In this section we want to present the analysis of moduli space of a class of theories dual to a brane configurations with one $J$-fold and a collection of $(1,k)$ branes. The associated quiver is
\be 
\begin{tikzpicture}[baseline, scale=0.7]
\tikzstyle{every node}=[font=\footnotesize, minimum size=1.2cm]
\node[draw, circle] (node1) at (-2,2) {$1_{k_1}$};
\node[draw, circle] (node2) at (2,2) {$1_{k_2}$};
\node[draw, circle] (node3) at (6,2) {$1_{k_3}$};
\node[draw, circle] (node4) at (10,2) {$1_{k_n}$};
\draw[draw=black,solid,thick,-]  (node1) to  (node2) ; 
\draw[draw=black,solid,thick,-]  (node2) to  (node3) ; 
\draw[draw=black,dashed,thick,-]  (node3) to  (node4) ;
\draw[purple] (node1) edge [out=90,in=90,looseness=0.3] node[midway,above=-0.3cm] {$T(U(1))$}  (node4);
\end{tikzpicture}
\ee
In the 3d $\cN=2$ notation, this can be rewritten as
\be 
\begin{tikzpicture}[baseline, scale=0.7]
\tikzstyle{every node}=[font=\footnotesize, minimum size=1.2cm]
\node[draw, circle] (node1) at (-2,2) {$1_{k_1}$};
\node[draw, circle] (node2) at (2,2) {$1_{k_2}$};
\node[draw, circle] (node3) at (6,2) {$1_{k_3}$};
\node[draw, circle] (node4) at (10,2) {$1_{k_n}$};
\draw[draw=black,solid,thick,<->]  (node1) to node[pos=0.2, above=-0.2cm] {$\tA_1$} node[pos=0.8, above=-0.2cm] {$A_1$}  (node2) ; 
\draw[draw=black,solid,thick,<->]  (node2) to node[pos=0.2, above=-0.2cm] {$\tA_2$} node[pos=0.8, above=-0.2cm] {$A_2$}  (node3) ; 
\draw[draw=black,dashed,thick,-]  (node3) to  (node4) ;
\draw[black] (node1) edge [out=-45,in=-135,loop,looseness=4] node[midway,below=-0.2cm] {$\varphi_1$}  (node1);
\draw[black] (node2) edge [out=-45,in=-135,loop,looseness=4] node[midway,below=-0.2cm] {$\varphi_2$}  (node2);
\draw[black] (node3) edge [out=-45,in=-135,loop,looseness=4] node[midway,below=-0.3cm] {$\varphi_3$}  (node3);
\draw[black] (node4) edge [out=-45,in=-135,loop,looseness=4] node[midway,below=-0.3cm] {$\varphi_n$}  (node4);
\draw[purple] (node1) edge [out=90,in=90,looseness=0.3] node[midway,above=-0.3cm] {$T(U(1))$}  (node4);
\end{tikzpicture}
\ee
with the superpotential
\be
W= \sum_{i=1}^{n-1} (-\tA_i \varphi_i A_i + A_i \varphi_{i+1} \tA_i) +\left( \sum_{j=1}^n \frac{1}{2} k_j \varphi_j^2 \right) {\blue - \varphi_1 \varphi_n}~.
\ee
where we emphasise the contribution from the mixed CS term due to the $T(U(1))$ theory in blue.
Let us write $\Phi_i\,=\, (\varphi_i\,,\, \sigma_i)$, $\mu_i\,=\, (A_i\,\tA_i\,,\, |A_i|^2 - |\tA_i|^2)$. The vacuum equations are
\be
\label{linearA1}
A_i(\Phi_{i+1}-\Phi_i)\,=\,0\,,\quad \tA_i(\Phi_{i+1}-\Phi_i)\,=\,0\quad i=1\,,\dots\,,n-1
\ee
\begin{equation}
\label{quadraticA1}
\begin{split}
&k_1\,\Phi_1 {\blue -\Phi_n} \,=\, \mu_1\\
& k_i\,\Phi_i\,=\, \mu_i-\mu_{i-1}\quad i=2\,,\dots\,,n-1\\
& k_n\,\Phi_n {\blue -\Phi_1}\,=\, - \mu_{n-1}
\end{split}
\end{equation}
The charges of the monopole operators $V_{(m_1, \ldots, m_n)}$ under the $i$-th $U(1)$ gauge group are
\begin{equation}
\begin{split}
&q_1[V_{(m_1, \ldots, m_n)}]\,=\, -(k_1\,m_1{\blue -m_n})\\
&q_i[V_{(m_1, \ldots, m_n)}]\,=\, -k_i\,m_i~, \quad \quad i\,=\,2\,,\dots,\, n-1\\
&q_n[V_{(m_1, \ldots, m_n)}]\,=\, -(k_n\,m_n {\blue -m_1})~.
\end{split}
\end{equation}
The $R$-charges of $V_{(m_1, \ldots, m_n)}$ is given by
\be
R[V_{(m_1, \ldots, m_n)}]\,=\, \dfrac{1}{2}\sum_{i=1}^{n-1}|m_{i}-m_{i+1}| ~.
\ee

\subsubsection{Cutting the quiver}
The process of cutting the quiver works similarly as in the previous subsection.  However, since there are non-trivial contributions from the $T(U(1))$ theory, some conditions must be modified.

\subsubsection*{Cutting at one point}  
Let us consider a case in which $A_l=\tilde{A}_l=0$ and other bifundamental hypermultiplets are non-zero.  In other words, we cut the quiver precisely at one point where $A_l$ and $\tilde{A}_l$ are located.  In this case equations \eqref{linearA1} implies
\be
\Phi_1\,=\dots=\,\Phi_l\,=\,\Phi\,=\,(\varphi\,,\,\sigma)\,,\quad \Phi_{l+1}\,=\dots=\,\Phi_n\,=\,\tilde \Phi\,=\,(\tilde \varphi\,,\,\tilde \sigma)
\ee
The system \eqref{quadraticA1} then becomes:
\be
\begin{array}{llll}
k_1 \Phi-\tilde \Phi\,=\, \mu_1\,,\,\, &k _2\Phi\,=\,\mu_2-\mu_1\,,\,\, &\ldots, \,\, & k_l \Phi \,=\,-\mu_{l-1} \\
k_{l+1}\tilde \Phi\,=\, \mu_{l+1}\,,\,\,&k_{l+2}\tilde\Phi\,=\,\mu_{l+2}-\mu_{l+1}\,,\,\, &\ldots,\,\, &k_n \tilde\Phi-\Phi\,=\,-\mu_{n}
\end{array}
\ee
The sum of the first $l$ equations and the sum of the remaining $n-l$ ones provide two constraints:
\be
\label{cond1}
\left( \sum_{i=1}^l\,k_i \right) \Phi-\tilde\Phi\,=\,0\,,\quad  \left( \sum_{i=l+1}^n\,k_i \right) \tilde\Phi-\Phi\,=\,0
\ee
Since $\Phi$ and $\tilde{\Phi}$ are non-zero (otherwise, this would violate the assumption that $A_j$ and $\tilde{A}_j$ are non-zero for $j \neq l$), we arrive at the following necessary condition for the existence of a non-trivial solution of the vacuum equation:
\be 
\left(\sum_{i=1}^l\,k_i\right)\left(\sum_{i=l+1}^n\,k_i\right)\,=\,1
\ee
Since all Chern-Simons levels are integers, the above equation is equivalent to
\be \label{twochoiceskk}
\sum_{i=1}^l\,k_i\,=\sum_{i=l+1}^n\,k_i\,=\,\pm1
\ee
The system of equations \eqref{cond1} is now simply solved by $\tilde\Phi\,=\,\pm\Phi$. Let us analyse separately the two cases:
\bi
\item $\Phi\,=\,\tilde\Phi\,$: In this case we choose
\be
\sum_{i=1}^l k_i =\sum_{i=l+1}^n k_i = 1~.
\ee
This moduli space is parametrised by $\varphi$ and the two basic dressed monopole operators.  Let us define for convenience
\be
\begin{split}
\tilde k_j\, &=\,(k_1-1\,,\,k_2\,,\dots,\,k_{n-1}\,,\,k_n-1)~, \\
\tilde{K}_i\,&=\,\sum_{j=1}^i\,\tilde k_j~. 
\end{split}
\ee
If $\tilde{K}_i \geq 0$ for all $i=1,\ldots, l-1, l+1,\ldots, n-1$, the basic dressed monopole operators are
\be
\begin{split}
&\bar V_+\,=\, V_{(1,1,\ldots,1)}\,A_1^{\tilde K_1}\,\dots A_{l-1}^{\tilde K_{l-1}}\, A_{l+1}^{\tilde K_{l+1}}\dots A_{n-1}^{\tilde K_{n-1}}\\
&\bar V_-\,=\, V_{-(1,1,\ldots,1)} \,\tA_1^{\tilde K_1}\,\dots \tA_{l-1}^{\tilde K_{l-1}}\, \tA_{l+1}^{\tilde K_{l+1}}\dots \tA_{n-1}^{\tilde K_{n-1}}~,
\end{split}
\ee
If $\tilde{K}_j< 0$ for some $j$, we replace $A_j^{\tilde{K}_j}$ in the first equation by $\tilde{A}_j^{-\tilde{K}_j}$, and $\tilde{A}_j^{\tilde{K}_j}$ in the second equation by $A_j^{-\tilde{K}_j}$.  In any case, the $R$-charge of the above dressed monopole operators are
\be
R[\bar V_{\pm}]\,=\frac{1}{2} \sum_{\substack{1\leq i \leq n-1 \\ i \neq l}} |\tilde K_i| \equiv \frac{1}{2} \tilde{K}
\ee
where the bare monopole operators have R-charge $R[V_{\pm( 1, 1,\ldots,1)}]\,=\,0$, and we define
\be
\tilde{K} =\sum_{\substack{1\leq i \leq n-1 \\ i \neq l}} |\tilde K_i|~.
\ee
 Thus, $\bar{V}_\pm$ satisfy
\be
\bar{V}_+ \bar{V}_- = \varphi^{\tilde{K}}~.
\ee
This branch of the moduli space is therefore
\be
\mathbb{C}^2/\mathbb Z_{\tilde{K}}~.
\ee

Let $g_{L}(t, \vec B)$ and $g_{R}(t, \vec B)$ be baryonic generating functions for the left sub-quiver (containing nodes $1, \ldots, l$) and the right sub-quivers (containing nodes $l+1, \ldots, n$), respectively.  Then, the Hilbert series for this case is given by
\be
\begin{split}
H(t; z) &=\, \sum_{m \in \mathbb Z}  z^m g_{L}(t, \{( k_1-1) m, k_2 m,\ldots, k_l m \} )  \times \\
& \hspace{2cm} g_{R}(t, \{ k_{l+1} m, \ldots, k_{n-1} m, (k-1) m\}) (1-t^2)~,
\end{split}
\ee
where $z$ is a fugacity for the topological symmetry.  Using the expressions for $g_L$ and $g_R$ given by \eref{barylinquiv}. we obtain
\be
\begin{split}
H(t; z) &= \sum_{m \in \mathbb Z}  z^m  \dfrac{t^{|m|  \sum_{i=1}^{l-1}|\tilde{K}_i| }}{1-t^2}\delta_{\sum_{i=1}^l k_i, 1} \times  \dfrac{t^{|m|  \sum_{i=l+1}^{n-1}|\tilde{K}_i| }}{1-t^2}\delta_{\sum_{i=l+1}^n k_i, 1} (1-t^2)\\
&= \begin{cases} 
\PE \left[ t^2 + (z+z^{-1}) t^{\tilde{K}} - t^{2 \tilde{K}} \right] & \text{if}\, \sum_{i=1}^l k_i = \sum_{i=l+1}^n k_i= 1 \\
0 & \text{otherwise}~.
\end{cases}
\end{split}
\ee
The Hilbert series in the first line in the second equality is indeed that of $\BC^2/\BZ_{\tilde{K}}$.
\item  $\Phi\,=\,-\,\tilde\Phi\,$: In this case, we choose
\be
\sum_{i=1}^l k_i =\sum_{i=l+1}^n k_i = -1~.
\ee
The basic monopole operators are $V_{+-} \equiv V_{(1^l,(-1)^{n-l})}$ and $V_{-+} \equiv V_{((-1)^l,1^{n-l})}$, whose $R$-symmetry are 
\be
R[V_{+-}] = R[V_{-+}]  = 1~.
\ee
Let us define for convenience
\be
\begin{split}
\tilde k'_j\, &=\,(k_1+1\,,\,k_2\,,\dots,\,k_{n-1}\,,\,k_n+1)~, \\
\tilde{K}'_i\,&=\,\sum_{j=1}^i\,\tilde k'_j~. 
\end{split}
\ee
For $\tilde K'_i>0$ for $i=1, \ldots, l-1$ and $\tilde K'_{j} <0$ for $j=l+1, \ldots, n-1$, the basic dressed monopole operators can be written as
\be
\begin{split}
&\bar V_{+-}\,=\, V_{+-}\,A_1^{\tilde K'_1}\,\dots A_{l-1}^{\tilde K'_{l-1}}\, A_{l+1}^{-\tilde K'_{l+1}}\dots A_{n-1}^{-\tilde K'_{n-1}}\\
&\bar V_{-+}\,=\, V_{-+} \,\tA_1^{\tilde K'_1}\,\dots \tA_{l-1}^{\tilde K'_{l-1}}\, \tA_{l+1}^{-\tilde K'_{l+1}}\dots \tA_{n-1}^{-\tilde K'_{n-1}}~,
\end{split}
\ee
where it should be noted that in this case $\sum_{i=1}^l k_i =\sum_{i=l+1}^n k_i = -1$.   Similarly as before, $\bar{V}_\pm$ satisfy
\be
\bar{V}_{+-} \bar{V}_{-+} = \varphi^{\tilde{K}'+2}~,
\ee
where we define
\be
\tilde{K}' =\sum_{\substack{1\leq i \leq n-1 \\ i \neq l}} |\tilde K'_i|~.
\ee
This branch of the moduli space is therefore
\be
\mathbb{C}^2/\mathbb Z_{\tilde{K}'+2}~.
\ee
The Hilbert series for this case is given by
\be
\begin{split}
H(t; z) &=\, \sum_{m \in \mathbb Z}  t^{|m-(-m)|} z^m g_{L}(t, \{( k_1+1) m, k_2 m,\ldots, k_l m \} )  \times \\
& \hspace{2cm} g_{R}(t, \{ -k_{l+1} m, \ldots, -k_{n-1} m, -(k+1) m\}) (1-t^2),
\end{split}
\ee
where $z$ is a fugacity for the topological symmetry.  Using the expressions for $g_L$ and $g_R$ given by \eref{barylinquiv}. we obtain
\be
\begin{split}
H(t; z) &= \sum_{m \in \mathbb Z}  t^{2|m|} z^m  \dfrac{t^{|m|  \sum_{i=1}^{l-1}|\tilde{K}'_i| }}{1-t^2}\delta_{\sum_{i=1}^l k_i, -1} \times  \dfrac{t^{|m|  \sum_{i=l+1}^{n-1}|\tilde{K}'_i| }}{1-t^2}\delta_{\sum_{i=l+1}^n k_i, -1}  (1-t^2)\\
&=\sum_{m \in \mathbb Z}  z^m  \dfrac{t^{|m|  (2+\sum_{i=1}^{l-1}|\tilde{K}'_i|+\sum_{i=l+1}^{n-1}|\tilde{K}'_i| )}}{1-t^2}  \delta_{\sum_{i=1}^l k_i, -1} \delta_{\sum_{i=l+1}^n k_i, -1}  \\
&= \begin{cases} 
\PE \left[ t^2 + (z+z^{-1}) t^{\tilde{K}+2} - t^{2 (\tilde{K}+2)} \right] & \text{if}\, \sum_{i=1}^l k_i = \sum_{i=l+1}^n k_i= -1 \\
0 & \text{otherwise}~.
\end{cases}
\end{split}
\ee
The Hilbert series in the first line in the third equality is indeed that of $\BC^2/\BZ_{\tilde{K}+2}$.
\ei

\subsubsection*{Cutting at two points} 
Let us consider a case in which $A_l=\tilde{A}_l=A_m = \tilde{A}_m=0$ (with $m>l$) and other bifundamental hypermultiplets are non-zero.  In other words, we cut the quiver precisely at one point where $A_l, \tilde{A}_l$ and $A_m, \tilde{A}_m$  are located.  This naturally divides the quiver in question into 3 sub-quivers, which we shall refer to as left (L), central (C) and right (R).  The central sub-quiver is the same as that is considered in section \ref{sec:linquivwoJ}.  In this case equations \eqref{linearA1} implies
\be
\begin{split}
\Phi_1\,=\dots=\,\Phi_l\,=\,\Phi_L \,=\,(\varphi_L\,,\,\sigma_L)\,~, \\
\Phi_{l+1}\,=\dots=\,\Phi_{m-1}\,=\, \Phi_C \,=\,(\varphi_C\,,\, \sigma_C)~, \\
\Phi_{m+1}\,=\dots=\,\Phi_{n}\,=\, \Phi_R \,=\,(\varphi_R\,,\, \sigma_R)~.
\end{split}
\ee
The system \eqref{quadraticA1} then becomes:
\be
\begin{array}{llll}
k_1 \Phi_L- \Phi_R\,=\, \mu_1\,,\,\, &k _2\Phi_L\,=\,\mu_2-\mu_1\,,\,\, &\ldots, \,\, & k_l \Phi_L \,=\,-\mu_{l-1} \\
k_{l+1} \Phi_C\,=\, \mu_{l+1}\,,\,\,&k_{l+2}\Phi_C\,=\,\mu_{l+2}-\mu_{l+1}\,,\,\, &\ldots,\,\, &k_{m-1} \Phi_C\,=\,-\mu_{m-1} \\
k_{m+1} \Phi_R\,=\, \mu_{m+1}\,,\,\,&k_{m+2}\Phi_R\,=\,\mu_{m+2}-\mu_{m+1}\,,\,\, &\ldots,\,\, &k_{n} \Phi_R - \Phi_L\,=\,-\mu_{n}~.
\end{array}
\ee
The sums of the equations in the first, the second and the third lines give
\be \label{condoneJtwocuts}
\begin{split}
\left( \sum_{i=1}^l\,k_i \right) \Phi_L-\Phi_R\,=0~, \quad  \left( \sum_{i=l+1}^m\,k_i \right) \Phi_C =0~, \quad \left( \sum_{i=m+1}^n\,k_i \right) \Phi_R-\Phi_L\,=\,0~.
\end{split}
\ee
Since $\Phi_L$, $\Phi_C$ and $\Phi_R$ are non-vanishing (otherwise, this would violate the assumption that $A_j$ and $\tilde{A}_j$ are non-zero for $j \neq l$), a necessary condition for the existence of a non-trivial solution of the vacuum equation:
\be \label{necesscondtwocuts}
\sum_{i=1}^l\,k_i\,=\sum_{i=m+1}^n\,k_i\,=\,\pm1~, \qquad \sum_{i=l+1}^{m}\,k_i= 0~.
\ee

Let $g_{L}(t, \vec B)$, $g_{C}(t, \vec B)$ and $g_{R}(t, \vec B)$ be baryonic generating functions for the left, central and right sub-quivers, respectively.  Then, the Hilbert series, corresponding to $+$ or $-$ sign in \eref{necesscondtwocuts}, is
\be
\begin{split}
& H(t; z_L, z_C, z_R) \\
&=\, \sum_{m_L \in \mathbb Z}\sum_{m_C\in \mathbb Z}\sum_{m_R\in \mathbb Z}  t^{|m_L-m_C|+|m_C-m_R|} z_L^{m_L} z_C^{m_C} z_R^{m_R} \times \\
& \qquad g_{L}(t, \{ k_1 m_{L}-m_{R}, k_2 m_L,\ldots, k_l m_L \} )\,g_{C}(t, \{ k_{l+1} m_{C},\ldots, k_{m-1} m_C \}) \times \\
& \qquad g_{R}(t, (k_m m_{R}, \ldots, k_{n-1} m_R, k_n m_R-m_L \}) (1-t^2) \delta_{m_R, \pm m_L}~,
\end{split}
\ee
where $z_{L, C, R}$ are fugacities for the topological symmetries. 

\subsubsection*{Cutting at more than two points} 
The above discussion can be easily generalised to the case of cutting the quiver at more than two points.  For the moduli space to be non-trivial, the sum of the CS levels in the two sub-quiver that are connected with $T(U(1))$ must be $\pm 1$, and the sum of the CS levels in the other sub-quiver must be zero. 

\subsubsection*{No cutting at all}  
Assume that $A_i$ and $\tilde{A}_i$ are non-zero for all $i$.  In this case, a necessary condition for the non-trivial moduli space is
\be \label{condnocut}
\sum_{i=1}^n k_i =2~.
\ee
This again can be obtained from the sum of the equations in \eref{quadraticA1}, with $\Phi_i= \Phi = (\varphi, \sigma) \neq 0$ (otherwise we would have $\mu_1=0$ which contradicts our assumption).  The monopole operators $V_{\vec m}$ with fluxes $\vec m = \pm (1,\dots,1)$ are not gauge invariant; however, the following basic dressed monopole operators are gauge invariant
\be
\begin{split}
\bar V_+\,&=\, V_{(1,\dots,1)} \,A_1^{\CK_1}\,A_2^{\CK_2}\,\dots\,A^{\CK_{n-1}}_{n-1}\\
\bar V_-\,&=\, V_{-(1,\ldots,1)} \,\tA_1^{\CK_1}\,\tA_2^{\CK_2}\,\dots\,\tA^{\CK_{n-1}}_{n-1}~,
\end{split}
\ee
for $\CK_i \geq 0$ for all $i=1,\ldots,n-1$, where we define
\be
\kappa_i\,=\,\{k_1-1\,,\,k_2\,,\dots,\,k_{n-1}\,,\,k_n-1\}~, \quad \CK_i\,=\,\sum_{j=1}^i\,\kappa_j~.
\ee
If $\CK_j < 0$ for some $j$, we replace $A_j^{\CK_j}$ by $\tilde{A}_j^{-\CK_j}$ in the first equation and $\tilde{A}_j^{\CK_j}$ by $A_j^{-\CK_j}$ in the second equation.  

Since the $R$-charges of $V_{\pm(1,\ldots,1)}$ are zero, the $R$-charges of $\bar V_{\pm}$ are $\frac{1}{2}\sum_{i=1}^{n-1} |\CK_i|$.  The moduli space is thus generate by the operators $\{\bar V_+\,\, \bar V_-\,,\, \varphi\}$ subject to the quantum relation
\be
\bar V_+\,\bar V_-\,=\,\varphi^{\CK}~, \quad \text{with $\CK = \sum_{i=1}^{n-1} | \CK_i|$}~;
\ee
this is the algebraic definition of:
\be
\mathbb C^2/\mathbb Z_{\CK}~.
\ee 
\paragraph{Example.} Let us consider the following quiver
\be \label{twonodesoneT}
\begin{tikzpicture}[baseline]
\tikzstyle{every node}=[font=\footnotesize]
\node[draw, circle] (node1) at (-1.5,0) {$1_{k_1}$};
\node[draw, circle] (node2) at (1.5,0) {$1_{k_2}$};
\draw[draw=black,solid,thick,-]  (node1) to[bend right=20]   (node2) ;  
\draw[draw=red,solid,thick,-]  (node1) to[bend left=20]  node[midway,above] {{\red $T(U(1))$}}  (node2) ; 
\end{tikzpicture}
\ee
It is not possible to introduce a cut to this quiver.  As a result, from \eref{condnocut}, it is necessary that $k_1 + k_2 =2$ for this theory to have a non-trivial moduli space.  Let us assume this.  Hence $\kappa_i =\{k_1-1, k_2-1\}$, $\CK_i = \{k_1-1, k_1+k_2-2 =0 \}$, and so $\CK = |k_1-1| = |k_2-1|$.  Therefore the moduli space of this theory is $\BC^2/\BZ_{|k_1-1|}$.

\subsection{Adding flavours} \label{sec:addflavours}
Let us now add fundamental flavours to the previous discussion.
\be 
\begin{tikzpicture}[baseline, scale=0.6]
\tikzstyle{every node}=[font=\footnotesize, minimum size=1.2cm]
\node[draw, circle] (node1) at (-2,2) {$1_{k_{i-1}}$};
\node[draw, circle] (node2) at (2,2) {$1_{k_{i}}$};
\node[draw, circle] (node3) at (6,2) {$1_{k_{i+1}}$};
\draw[draw=black,dashed,thick,-]  (-6,2) to  (node1) ; 
\draw[draw=black,solid,thick,-]  (node1) to  (node2) ;
\draw[draw=black,solid,thick,-]  (node2) to  (node3) ; 
\draw[draw=black,dashed,thick,-]  (node3) to  (10, 2) ; 
\node[draw, rectangle] (node4) at (-2, -1) {$f_{i-1}$};
\node[draw, rectangle] (node5) at (2,-1) {$f_{i}$};
\node[draw, rectangle] (node6) at (6,-1) {$f_{i+1}$};
\draw[draw=black,solid,thick,-]  (node1) to  (node4) ;
\draw[draw=black,solid,thick,-]  (node2) to  (node5) ;
\draw[draw=black,solid,thick,-]  (node3) to  (node6) ;
\end{tikzpicture}
\ee
Suppose that there are $n$ gauge groups in total. 
In the $\CN=2$ notation, this quiver can be written as
\be 
\begin{tikzpicture}[baseline, scale=0.8]
\tikzstyle{every node}=[font=\footnotesize, minimum size=1.2cm]
\node[draw, circle] (node1) at (-2,2) {$1_{k_{i-1}}$};
\node[draw, circle] (node2) at (2,2) {$1_{k_{i}}$};
\node[draw, circle] (node3) at (6,2) {$1_{k_{i+1}}$};
\draw[draw=black,dashed,thick,-]  (-6,2) to  (node1) ; 
\draw[draw=black,solid,thick,->]  (node1) to[bend left=20] node[midway,above=-0.2]{$A_{i-1}$} (node2) ;
\draw[draw=black,solid,thick,<-]  (node1) to[bend right=20] node[midway,below=-0.2]{$\tilde{A}_{i-1}$} (node2) ;
\draw[draw=black,solid,thick,->]  (node2) to[bend left=20] node[midway,above=-0.2]{$A_{i}$}   (node3) ; 
\draw[draw=black,solid,thick,<-]  (node2) to[bend right=20] node[midway,below=-0.2]{$A_{i}$}   (node3) ; 
\draw[draw=black,dashed,thick,-]  (node3) to  (10, 2) ; 
\node[draw, rectangle] (node4) at (-2, -1) {$f_{i-1}$};
\node[draw, rectangle] (node5) at (2,-1) {$f_{i}$};
\node[draw, rectangle] (node6) at (6,-1) {$f_{i+1}$};
\draw[draw=black,solid,thick,->]  (node1) to[bend right=20] node[midway,left=-0.2]{$Q_{i-1}$}  (node4) ;
\draw[draw=black,solid,thick,<-]  (node1) to[bend left=20] node[midway,right=-0.2]{$\tQ_{i-1}$}  (node4) ;
\draw[draw=black,solid,thick,->]  (node2) to[bend right=20] node[midway,left=-0.2]{$Q_{i}$}  (node5) ;
\draw[draw=black,solid,thick,<-]  (node2) to[bend left=20] node[midway,right=-0.2]{$\tQ_{i}$}  (node5) ;
\draw[draw=black,solid,thick,->]  (node3) to[bend right=20] node[midway,left=-0.2]{$Q_{i+1}$}  (node6) ;
\draw[draw=black,solid,thick,<-]  (node3) to[bend left=20] node[midway,right=-0.2]{$\tQ_{i+1}$}  (node6) ;
\draw[black] (node1) edge [out=45,in=135,loop,looseness=4] node[midway,above=-0.3cm] {$\varphi_{i-1}$}  (node1);
\draw[black] (node2) edge [out=45,in=135,loop,looseness=4] node[midway,above=-0.3cm] {$\varphi_{i}$}  (node2);
\draw[black] (node3) edge [out=45,in=135,loop,looseness=4] node[midway,above=-0.3cm] {$\varphi_{i+1}$}  (node3);
\end{tikzpicture}
\ee
The vacuum equations read
\be
\label{linearA}
A_{i-1}(\Phi_i-\Phi_{i-1})\,=\,0\,,\quad A_{i}(\Phi_{i+1}-\Phi_{i})\,=\,0,
\ee
also with $A \leftrightarrow \tA$,
\be
\label{linearQ}
Q_{i-1}\,\Phi_{i-1}\,=\,0\,,\quad Q_{i}\,\Phi_{i}\,=\,0\,,\quad Q_{i+1}\,\Phi_{i+1}\,=\,0
\ee
also with $Q \leftrightarrow \tQ$, and
\be
\label{quadraticAQ}
\begin{split}
k_{i-1}\Phi_{i-1}\,&=\, \mu_{i-1}-\mu_{i-2}+\nu_{i-1}\\
k_{i}\Phi_{i}\,&=\, \mu_i -\mu_{i-1}+ \nu_i \\
k_{i+1}\Phi_{i+1}\,&=\, \mu_{i+1}-\mu_i+\nu_{i+1}~.
\end{split}
\ee
where we define
\be
\mu_j = (A_j \tA_j, \, |A_j|^2-|\tA_j|^2)~, \quad \nu_j = (Q_j \tQ_j, \,  |Q_j|^2-|\tQ_j|^2)
\ee
The $R$-charge of the monopole operators $V_{\vec m}$ with flux $\vec m =(m_1, \ldots, m_n)$ is
\be
\label{RchargeFlavour}
R[V_{\vec m} ]\,=\, \dfrac{1}{2}\left( \sum_{i=1}^{n-1}\abs{m_{i+1}-m_{i}} \,+\,\sum_{i=1}^n\,f_i\abs{m_i}\right)
\ee

Equation \eqref{linearQ} admits two non-trivial possibilities:
\be
\Phi_i\,=\,0\quad \text{or} \quad Q_i\,=\tQ_i \, =\,0~.
\ee
If we set $Q_i\,= \tQ_i =\,0$, the analysis is similar to the linear quiver without flavours.  We will instead focus on $\Phi_i\,=\,0$. The remaining constraints in \eref{linearA} and \eqref{linearQ} are thus:
\be
\begin{split}
A_{i-1}\,&\Phi_{i-1}\,=\,0\,,\quad A_{i}\Phi_{i+1}\,=\,0\\
Q_{i-1}\,&\Phi_{i-1}\,=\,0\,,\quad Q_{i+1}\Phi_{i+1}\,=\,0~,
\end{split}
\ee
also with $A \leftrightarrow \tA$, $Q \leftrightarrow \tQ$. Each column of previous set of equations admit two solutions:
\be \label{branches}
\begin{split}
&\Phi_{i-1}\,=\,0\,\quad or\quad \{A_{i-1}\,=\,0\,,\,Q_{i-1}\,=\,0\}\\
&\Phi_{i+1}\,=\,0\,\quad or\quad \{A_{i\,\,\,\,}\,=\,0\,,\,Q_{i+1}\,=\,0\}
\end{split}
\ee
The case $\{A_{i-1}\,=\,0\,,\,Q_{i-1}\,=\,0\}$ obviously induce a cut in the quiver and set to zero the adjacent fundamental matter; the same for $\{A_{i\,\,\,\,}\,=\,0\,,\,Q_{i+1}\,=\,0\}$. Let us focus on $\Phi_{i-1}=\Phi_{i+1} =\,0$.
Now, we have the vacuum equations
\be
\begin{split}
A_{i-2}\,&\Phi_{i-2}\,=\,0\,,\quad A_{i+1}\Phi_{i+2}\,=\,0\\
Q_{i-2}\,&\Phi_{i-2}\,=\,0\,,\quad Q_{i+2}\Phi_{i+2}\,=\,0
\end{split}
\ee
Again, the solutions that do not induce a cut are $\Phi_{i+2}\,=\,\Phi_{i-2}\,=0$ and so on. 

The above procedure divides the initial quiver in ``Higgs" and ``Coulomb" sub-quivers, defined as follows. In the Coulomb one, fundamental matter is set to zero while in Higgs one, all the vector multiplet scalar are set to zero. For instance, we divide the following quiver such that the the first $l$ nodes constitute a Coulomb sub-quiver, the $(l+1)$-th to the $(l+m)$-th nodes constitute a Higgs sub-quiver, and the $(l+m+1)$-th to the $(n)$-th nodes constitute a Coulomb sub-quiver.  
\be  \label{CouHiggsCou}
\begin{tikzpicture}[baseline, scale=0.7]
\tikzstyle{every node}=[font=\footnotesize, minimum size=1.2cm]
\node[draw,circle] (node1) at (-2,2) {$k_1$};
\node[draw, circle] (node2) at (2,2) {$k_l$};
\node[draw, purple,dashed, circle] (node3) at (6,2) {$k_{l+1}$};
\node[draw,purple, dashed,circle] (node4) at (10,2) {$k_{l+m}$};
\node[draw, circle] (node5) at (14,2) {$k_{l+m+1}$};
\node[draw, circle] (node6) at (18,2) {$k_n$};
\node[draw, rectangle] (node7) at (-2,-1) {$f_1$};
\node[draw, rectangle] (node8) at (2,-1) {$f_l$};
\node[draw, rectangle] (node9) at (6,-1) {$f_{l+1}$};
\node[draw, rectangle] (node10) at (10,-1) {$f_{l+m}$};
\node[draw, rectangle] (node11) at (14,-1) {$f_{l+m+1}$};
\node[draw, rectangle] (node12) at (18,-1) {$f_n$};
\draw[draw=black,dashed,thick,-]  (node1) to  (node2) ; 
\draw[draw=red,solid,thick,-]  (node2) to  (node3) ; 
\draw[draw=black,dashed,thick,-]  (node3) to  (node4) ; 
\draw[draw=red,solid,thick,-]  (node4) to  (node5) ; 
\draw[draw=black,dashed,thick,-]  (node5) to  (node6) ; 
\draw[draw=red,solid,thick,-]  (node1) to node[midway,right=-0.1cm] {} (node7) ; 
\draw[draw=red,solid,thick,-]  (node2) to node[midway,right=-0.1cm] {}   (node8) ; 
\draw[draw=black,solid,thick,-]  (node3) to  (node9) ; 
\draw[draw=black,solid,thick,-]  (node4) to  (node10) ; 
\draw[draw=red,solid,thick,-]  (node5) to node[midway,right=-0.1cm] {} (node11) ; 
\draw[draw=red,solid,thick,-]  (node6) to node[midway,right=-0.1cm] {}(node12) ; 
\end{tikzpicture}
\ee
where the purple nodes indicate that $\Phi_i\,=\,0$ (with $i=l+1, \ldots, l+m$), and the red lines indicate that $Q_j = \tQ_j =0$ (with $j=1, \ldots, l, l+m+1, \ldots, n$) and $A_l =\tA_l=A_{l+m} =\tA_{l+m} =0$ (we shall discuss about this later).  For the sake of readability, in the above diagram, we indicate only the CS level in each circular node and omit the rank, which is $1$ for each $U(1)$ gauge group.

Since in the Higgs sub-quiver, $\Phi_i\,=\,0$ for all $i=l+1, \ldots, l+m$; as a consequence, the magnetic flux is set to zero for all gauge nodes in the sub-quiver. Thus, introducing a cut within the Higgs sub-quiver does not produce anything new.   For simplicity, we also assume that there is no further cut in the Coulomb branch sub-quiver.

Moreover, a Higgs sub-quiver cannot end with a node without flavours.  This can be seen as follows. Suppose, on the contrary, that we cut the quiver at the $(l+m)$-th position, namely set $A_{l+m}\,=\,\tA_{l+m}\,=\,0$, with $f_{l+m}\,=\,0$. In this case, \eqref{quadraticAQ} implies:
\be
k_{l+m}\,\Phi_{l+m}\,=\,A_{l+m}\tA_{l+m}-A_{l+m-1}\tA_{l+m-1}\,+\,Q_{l+m}\tQ_{l+m}~.
\ee
Since we cut the quiver at the $(l+m)$-th position, $A_{l+m}\,=\,\tA_{l+m}\,=\,0$. We also have $Q_{l+m}\,=\,\tQ_{l+m}\,=\,0$ since $f_{l+m}\,=\,0$. Also, $\Phi_{l+m}\,=\,0$ since we are looking at the Higgs sub-quiver. Thus the previous condition becomes:
\be
A_{l+m-1}\tA_{l+m-1}\,=\,0
\ee
implying a cut at $A_{l+m-1}$. This procedure must be continued until we have $f_{i}\neq 0$. 

Let us assume that $f_{l+1}$ and $f_{l+m}$ are non-zero. In transiting from the Coulomb sub-quiver to Higgs sub-quiver and vice-versa, we need to introduce a cut at the transition point; this is because from \eref{linearA}, we have, \eg, $0 = A_l (\Phi_{l} - \Phi_{l+1}) = A_l \Phi_l$ which indeed implies $A_l=0$.  Indeed we need to set
\be
A_l = \tA_l =0~, \qquad A_{l+m}=\tA_{l+m} =0~.
\ee

In the Higgs sub-quiver, we have the vacuum equation
\be
\begin{split}
A_{l+1}\tA_{l+1}+Q_{l+1}\tQ_{l+1}\,&=\,0\\
A_{l+2}\tA_{l+2}-A_{l+1}\tA_{l+1}+Q_{l+2}\tQ_{l+2}\,&=\,0\\
&\vdots\\
-A_{l+m}\tA_{l+m}+Q_{l+m}\tQ_{l+m}\,&=\,0~,
\end{split}
\ee
whereas in the Coulomb sub-quiver, we have
\be
\begin{split}
A_{1}\tA_{1} &=\,k_{1} \varphi_L \\
A_{2}\tA_{2} - A_{1} \tA_{1}  &=\,k_{2} \varphi_L \\
&\vdots\\
-A_{l}\tA_{l}\,&=\,k_l \varphi_L~,
\end{split}
\ee
and
\be
\begin{split}
A_{l+m+1}\tA_{l+m+1} &=\,k_{l+m+1} \varphi_R \\
A_{l+m+2}\tA_{l+m+2} - A_{l+m+1} \tA_{l+m+1}  &=\,k_{l+m+2} \varphi_R \\
&\vdots\\
-A_{n-1}\tA_{n-1}\,&=\,k_n \varphi_R
\end{split}
\ee
The sums of these two sets of equations tell us that necessary conditions for the existence of non-trivial moduli spaces of the Coulomb sub-quivers are
\be
\sum_{i=1}^l k_i =0~,  \qquad \sum_{j=l+m+1}^n k_j =0~.
\ee

The gauge charge of the monopole operator $V_{\vec m}$ with flux 
\be \vec m =(\underbrace{\cm_L, \ldots, \cm_L}_{l}, \underbrace{0, \ldots, 0}_{m}, \underbrace{\cm_R, \ldots, \cm_R}_{n-l-m}) \equiv (\cm_L^l, 0^{m}, \cm_R^{n-l-m})~, \ee 
where $0$ is the flux for each gauge group in the Higgs sub-quivers and $\cm$ is the flux for each gauge group in the Coulomb sub-quiver, is
\be
\begin{array}{ll}
q_i[V_{\vec m}]\,=\,-k_i \cm_L &\,\,\, \text{for $i=1,\ldots, l$}~, \\
q_p[V_{\vec m}] \,=\, 0 &\,\,\, \text{for $p=l+1,\ldots, l+m$} \\
q_j[V_{\vec m}]\,=\, -k_j \cm_R &\,\,\, \text{for $j=l+m+1,\ldots, n$}
\end{array}
\ee
The $R$-charge of the monopole operator $V_{\vec m}$ is
\be
\begin{split}
R[V_{\vec m}] &= \frac{1}{2} |\cm_L-0|+\frac{1}{2}  \left( |\cm_L| \sum_{i=1}^l f_i  + |\cm_R| \sum_{j=l+m+1}^n f_j  \right)+ \frac{1}{2} |0-\cm_R| \\
& \equiv \frac{1}{2} |\cm_L| (F_L+1)+ \frac{1}{2} |\cm_R| (F_R+1)~,
\end{split}
\ee
where we define $F_{L,R}$ as the total number of flavours in the left and right Coulomb sub-quivers:
\be
F_L = \sum_{i=1}^l f_i~, \qquad F_R= \sum_{j=l+m+1}^n f_j ~.
\ee

The Hilbert series for the Higgs sub-quiver can be written as
\be \label{HSHiggssubquiv}
\begin{split}
&H_{\text{Higgs}}(t; {\vec x}^{(l+1)}, \ldots, {\vec x}^{(l+m)})\,\\
&=\, (1-t^2)^m \prod_{j=l+1}^{l+m} \oint \dfrac{dq_i}{2\pi i q_j}\PE \left[t\sum_{\alpha=1}^{f_j} \left(q_j (x^{(j)}_\alpha)^{-1}+q_j^{-1}(x^{(j)}_\alpha) \right) \right]\, \\
& \qquad \prod_{i=l+1}^{l+m-1} \PE[t(q_i\,q_{i+1}^{-1}+q^{-1}_i\,q_{i+1})]
\end{split}
\ee
where the first PE is related to fundamental matter and the second one to bi-fundamental matter; the overall $(1-t^2)^m$ is due to the $m$ $F$-term constraints. Observe that the Hilbert series of this sub-quiver does not depend on the CS levels.  It is also worth noting that \eref{HSHiggssubquiv} takes the same form as the Higgs branch Hilbert series of 3d $\CN=4$ $T^{\vec \sigma}_{\vec \rho}(SU(N))$ theory \cite{Gaiotto:2008ak} for some ${\vec \sigma}$ and ${\vec \rho}$ \cite{Cremonesi:2014kwa}; for example, for $m=3$ and $f_{l+1}=f_{l+2}=f_{l+3}=1$, \eref{HSHiggssubquiv} is equal to the Higgs branch Hilbert series of $T^{(3,2,1)}_{(2^2,1^2)}(SU(6))$.

Let us now focus on the Coulomb sub-quiver.  The analysis is very similar to that described in the case without flavours, discussed earlier.  We emphasise that even if all the fundamental matter is set to zero, it still contributes to the dimension of the monopole operators.  For example, if there is no cut in the left and right Coulomb sub-quivers in \eref{CouHiggsCou}, the baryonic generating function of each of these Coulomb sub-quivers are similar to \eref{HSnocutlin}:
\be
\begin{split}
G^{L,R}_{\text{Coulomb}}(t; m) &= \frac{1}{1-t^2} t^{|m| K_{L,R}} 
\end{split}
\ee
where 
\be
K_L = \sum_{i=1}^l |\sum_{j=1}^i k_j |~, \qquad K_R =  \sum_{i=l+m+1}^n |\sum_{j=l+m+1}^i k_j |~.
\ee

The total Hilbert series of \eref{CouHiggsCou} is therefore
\be
\begin{split}
H(t; \vec x) &=H_{\text{Higgs}}(t; \{ \vec x^{(i)} \})  \sum_{\cm_L \in \BZ}\sum_{\cm_R \in \BZ}  t^{(F_L+1) |\cm_L| +(F_R+1) |\cm_R|} z_L^{\cm_L} z_R^{\cm_R} \\
& \qquad \times G^L_{\text{Coulomb}}(t; \cm_L)  G^R_{\text{Coulomb}}(t; \cm_R) \\
&= H_{\text{Higgs}}(t; \{ \vec x^{(\alpha)} \}) \left[ \sum_{\cm_L \in \BZ} \frac{1}{1-t^2} t^{(F_L+K_L+1) |\cm_L|} z_L^{\cm_L} \right] ( L \leftrightarrow R) \\
&= H_{\text{Higgs}}(t; \{ \vec x^{(\alpha)} \}) H[\BC^2/\BZ_{F_L+K_L+1}](t, z_L)  H[\BC^2/\BZ_{F_R+K_R+1}](t, z_R) 
\end{split}
\ee
where
\be
H[\BC^2/\BZ_{F_L+K_L+1}](t, z_L)  = t^2 +( z_L+ z^{-1}_L) t^{F_L+K_L+1} - t^{2(F_L+K_L+1)}~.
\ee
and the same for $(L \leftrightarrow R)$.  The moduli space of quiver \eref{CouHiggsCou} is therefore
\be
(\BC^2/\BZ_{F_L+K_L+1}) \times \CM_{\text{Higgs}}  \times (\BC^2/\BZ_{F_R+K_R+1})~,
\ee
where  $\CM_{\text{Higgs}}$ denotes the moduli space of the Higgs sub-quiver, which is isomorphic to the Higgs branch moduli space of $T^{\vec \sigma}_{\vec \rho}(SU(N))$ for some appropriate $N$, $\vec \sigma$ and $\vec \rho$.

\subsection{Adding flavour with one $J$-fold} \label{sec:addflavourswithoneJ}
Now we want to study the branches of a theory with one $J$-fold and fundamental matter:
\be 
\begin{tikzpicture}[baseline, scale=0.7]
\tikzstyle{every node}=[font=\footnotesize, minimum size=1.2cm]
\node[draw, circle] (node1) at (-2,2) {$1_{k_{1}}$};
\node[draw, circle] (node2) at (2,2) {$1_{k_{2}}$};
\node[draw, circle] (node3) at (6,2) {$1_{k_{n}}$};
\draw[draw=black,solid,thick,-]  (node1) to  (node2) ;
\draw[draw=black,dashed,thick,-]  (node2) to  (node3) ; 
\node[draw, rectangle] (node4) at (-2, -1) {$f_1$};
\node[draw, rectangle] (node5) at (2,-1) {$f_{2}$};
\node[draw, rectangle] (node6) at (6,-1) {$f_{n}$};
\draw[draw=black,solid,thick,-]  (node1) to  (node4) ;
\draw[draw=black,solid,thick,-]  (node2) to  (node5) ;
\draw[draw=black,solid,thick,-]  (node3) to  (node6) ;
\draw[purple] (node1) edge [out=90,in=90,looseness=0.3] node[midway,above=-0.3cm] {$T(U(1))$}  (node3);
\end{tikzpicture}
\ee
If all $\Phi_i$ (with $i=1,\ldots, n$) are set to zero and the presence of the $T(U(1))$ link does not affect the moduli space, the analysis is the same as that discussed in the previous subsection.  On the other hand, if all $Q_i$ and $\tQ_i$ are set to zero, the analysis is similar to that discussed in section \ref{sec:oneJfoldnoflv}; one needs to take into account of the contribution from the fundamental matter to the $R$-charge of the monopole operator.

\paragraph{Example.} Let us consider a simple example with a $U(1)_k$ gauge group, one $T(U(1))$ link and $n$ flavours.
\be \label{onenodenflv}
\begin{tikzpicture}[baseline]
\tikzstyle{every node}=[font=\footnotesize]
\node[draw, circle] (node1) at (0,1) {$1_k$};
\draw[red,thick] (node1) edge [out=45,in=135,loop,looseness=5]  (node1);
\node[draw=none] at (1.3,1.5) {{\red ${T(U(1))}$}};
\node[draw, rectangle] (sqnode) at (0,-0.6) {$n$};
\draw (node1)--(sqnode);
\end{tikzpicture}
\ee
 It is not possible to introduce a cut to this quiver.  $T(U(1))$ is an almost empty theory; it contributes the CS level $-2$ to the $U(1)$ gauge group, so effectively the CS level is $k-2$. 
\be
W= \tilde{Q}_i \varphi Q^i + \frac{1}{2}(k-2) \varphi^2~, \,\,\, i=1, \ldots, n~.
\ee
We have the $F$-term equations:
\be \label{Ftermex1}
\begin{split}
\tilde{Q}_i Q^i +(k-2) \varphi =0~, \qquad \tilde{Q}_i \varphi =0~, \qquad  \varphi Q^i =0~.
\end{split}
\ee
The vacuum equations involving the real scalar field $\sigma$ in the vector multiplet is
\be \label{Ftermex1a}
Q^i\sigma = \sigma \tQ_i =0~.
\ee
The $D$-term equation reads
\be
(Q^\dagger)_i Q^i - \tQ_i (\tQ^\dagger)^i = (k-2) \sigma~.
\ee

If $k=2$, the superpotential and the moduli space are the same as that of 3d $\CN=4$ $U(1)$ gauge theory with $n$ flavours.  The $F$-term with respect to $\phi$ implies that $\tQ_i Q^i=0$.  The Higgs branch is generating by the mesons $M^i_j = Q^i \tQ_j$; this meson matrix has rank at most $1$ and subject to the matrix relation $M^2=0$, which follows from the $F$-term.  Thus, the Higgs branch is isomorphic to the closure of the minimal nilpotent orbit of $SU(n)$.  On the other hand, the Coulomb branch of this theory is $\BC^2/\BZ_n$; this is generated by the monopole operators $V_+$ and $V_-$, carrying the topological charges $\pm1$ and $R$-charges $\frac{1}{2}n$, subject to the relation $V_+ V_- = \varphi^n$.  Note that for $n=1$ and $k = 2$, this theory has no Higgs branch and its Coulomb branch is isomorphic to $\BC^2$.

Let us now suppose that $k \neq 2$.  If $(\varphi, \sigma)$ is non-zero, \eref{Ftermex1} and \eref{Ftermex1a} implies $Q^i$ and $\tQ_i$ are zero, but this is in contradiction with the $D$-term.  Hence $(\varphi, \sigma) =0$ and the Coulomb branch is trivial in this case.  However, there is still the Higgs branch generated by $M^i_j = Q^i \tQ_j$.  As before, this meson matrix has rank at most $1$ and subject to the matrix relation $M^2=0$ (since $\tilde{Q}_i Q^i=0$).  The Higgs branch is therefore isomorphic to the closure of the minimal nilpotent orbit of $SU(n)$.  Note that for $n=1$ and $k \neq 2$, this theory has a trivial moduli space.

\subsubsection*{The case with one cut}
For simplicity, let us first focus on the case of {\it precisely one cut}.  In this case we have two sub-quivers, left and right, connected by the $T(U(1))$ link. We have three possibilities: 
\bi
\item Both the sub-quivers are in the Coulomb sector: this require the usual analysis as in section \ref{sec:oneJfoldnoflv}.

\item Both the sub-quivers are in the Higgs sector: all $\Phi_i$ are set to zero and the T-link does not affect the moduli space.

\item One is a Higgs sub-quiver (say, the left one) and the other is a Coulomb sub-quiver (say, the right one).
\ei

The last case is the interesting one.
\be \label{onecutHC}
\begin{tikzpicture}[baseline, scale=0.7]
\tikzstyle{every node}=[font=\footnotesize, minimum size=1.2cm]
\node[draw, dashed,circle,purple] (node1) at (-2,2) {$1_{k_{1}}$};
\node[draw, dashed,circle,purple] (node2) at (2,2) {$1_{k_{l}}$};
\node[draw,circle] (node3) at (6,2) {$1_{k_{l+1}}$};
\node[draw,circle] (node4) at (10,2) {$1_{k_{n}}$};
\node[draw, rectangle] (sqnode1) at (-2,-1) {$f_{1}$};
\node[draw, rectangle] (sqnode2) at (2,-1) {$f_{l}$};
\node[draw, rectangle] (sqnode3) at (6,-1) {$f_{l+1}$};
\node[draw, rectangle] (sqnode4) at (10,-1) {$f_{n}$};
\draw[draw=black,dashed,thick,-]  (node1) to  (node2) ;
\draw[draw=red,solid,thick,-]  (node2) to  (node3) ;
\draw[draw=black,dashed,thick,-]  (node3) to  (node4) ; 
\draw[draw=black,solid,thick,-]  (node1) to  (sqnode1) ;
\draw[draw=black,solid,thick,-]  (node2) to  (sqnode2) ;
\draw[draw=red,solid,thick,-]  (node3) to  (sqnode3) ;
\draw[draw=red,solid,thick,-]  (node4) to  (sqnode4) ;
\draw[purple] (node1) edge [out=90,in=90,looseness=0.3] node[midway,above=-0.3cm] {$T(U(1))$}  (node4);
\end{tikzpicture}
\ee
where the dashed circles mean that their vector multiplet scalars are zero, and the red lines mean that the hypermultiplets are set to zero:
\be
\begin{split}
& \Phi_1 = \Phi_2 =\ldots = \Phi_l =0~, \\
\end{split}
\ee
The first set of vacuum equations are
\be
\begin{split} \label{fluxlink}
& A_j (\Phi_{j+1}- \Phi_j) = \tA_j (\Phi_{j+1}- \Phi_j) =0~, \quad j=1,\ldots, n-1
\end{split}
\ee
As a consequence, we see that 
\be
\begin{split}
& \Phi_{l+1} = \Phi_{l+2} = \ldots = \Phi_n = \Phi = (\varphi, \sigma) \\
&A_l=\tA_l =0~,
\end{split}
\ee
The latter set of equations say that we need to introduce a cut in transiting from the Higgs sub-quiver to the Coulomb sub-quiver and vice-versa.
The other vacuum equations are
\be
\label{quadraticJ}
\begin{split}
A_{1}\tA_{1}+Q_1\tQ_1\,&=\,{\blue -\varphi}\\
A_{2}\tA_2-A_1\tA_1+Q_2\tQ_2\,&=\,0\\
A_{3}\tA_3-A_2\tA_2+Q_3\tQ_3\,&=\,0\\
&\vdots\\
{\gray A_{l} \tA_{l}}-A_{l-1}\tA_{l-1}+Q_l\tQ_l\,&=\,0\\
\end{split}
\ee
and
\be \label{quadraticJ2}
\begin{split}
A_{l+1}\tA_{l+1} - {\gray A_{l} \tA_{l}} &=\,k_{l+1} \varphi \\
A_{l+2}\tA_{l+2} - A_{l+1} \tA_{l+1}  &=\,k_{l+2} \varphi \\
&\vdots\\
-A_{n-1}\tA_{n-1}\,&=\,k_n \varphi
\end{split}
\ee
where the contribution from the $T(U(1))$ link is denoted in blue.   We denote the vanishing terms in grey in \eref{quadraticJ} and \eref{quadraticJ2}.  The sum of \eref{quadraticJ} gives
\be
\varphi = -\sum_{i=1}^l Q_i \tQ_i~.
\ee
Moreover, a necessary condition for a non-trivial moduli space for the Coulomb sub-quiver can be determined by summing \eref{quadraticJ2} and requiring that $\varphi \neq 0$:
\be
\sum_{i=l+1}^n k_i = 0~.
\ee
The gauge charge of the monopole operator $V_{\vec m}$ with flux 
\be \vec m =(\underbrace{0, \ldots, 0}_{l}, \underbrace{m, \ldots, m}_{n-l}) \equiv (0^l, m^{n-l})~, \ee 
where $0$ is the flux for each gauge group in the Higgs sub-quiver and $m$ is the flux for each gauge group in the Coulomb sub-quiver, is
\be
\begin{split}
&q_1[V_{\vec m}]\,=\,m~, \quad q_j[V_{\vec m}]\,=\,0 \,\,\, \text{for $j=2,\ldots, l$}~, \\
&q_p[V_{\vec m}] \,=\, -k_p m  \,\,\, \text{for $p=l+1,\ldots, n$}~.
\end{split}
\ee
The $R$-charge of the monopole operator $V_{\vec m}$ is
\be
R[V_{\vec m}] = \frac{1}{2} |m-0|+\frac{1}{2} |m|  \sum_{i=l+1}^n f_i \equiv \frac{1}{2} |m| (F_C+1)~,
\ee
where we define $F_C$ as the total number of flavours in the Coulomb sub-quiver:
\be
F_C = \sum_{i=l+1}^n f_i~.
\ee

We can construct the dressed monopole operators that are gauge invariant as follows.
\be
\begin{split}
\bar{V}^{(\alpha)}_+ &= V_{(0^l, 1^{n-l})} (\tQ_\alpha \tA_{\alpha-1} \tA_{\alpha -2} \cdots \tA_1)  \left( A_{l+1}^{K_{l+1}}\,A_{l+2}^{K_{l+2}}\cdots A_{n-1}^{K_{n-1}} \right) \\
\bar{V}^{(\alpha)}_- &= V_{(0^l, 1^{n-l})} (A_1 A_2 \ldots A_{\alpha-1} Q_\alpha)  \left( \tA_{l+1}^{K_{l+1}}\,\tA_{l+2}^{K_{l+2}}\dots \tA_{n-1}^{K_{n-1}} \right)~.
\end{split}
\ee
where $\alpha = 1, \ldots, l$ and 
\be
K_i\,=\, \sum_{p=l+1}^i\,k_p~, \,\,\, \text{for $i =l+1, \ldots, n$}~.
\ee
Note that if $K_j< 0$ for some $j$, we replace $A_j^{K_j}$ in the first equation by $\tilde{A}_j^{-K_j}$, and $\tilde{A}_j^{K_j}$ in the second equation by $A_j^{-K_j}$. 
The $R$-charges of $V^{(\alpha)}_\pm$ are
\be
R[V^{(\alpha)}_\pm] = \frac{1}{2} \left[ (F_C+1) +\alpha +\sum_{p=l+1}^{n-1} |K_p| \right] = \frac{1}{2} \left[ (F_C+1) +\alpha +K \right] ~,
\ee
with 
\be
K \equiv \sum_{p=l+1}^{n-1}\,|K_p|~.
\ee

As in the preceding subsection, if $f_l=0$ (which means $Q_l=\tQ_l=0$), then the Higgs sub-quiver cannot end at the $l$-th position because from \eref{quadraticJ} we have $A_{l-1} \tA_{l-1} =0$, \ie \,\, we need to introduce a cut at the $(l-1)$-th position.  However, if $f_1=0$ (which means $Q_1 =\tQ_1=0$), the Higgs sub-quiver still can end at the 1st position because $A_1 \tilde{A}_1 = -\varphi$.

The Hilbert series of quiver \eref{onecutHC} can be obtained as follows. The baryonic generating function for the Higgs sub-quiver is
\be
\label{higgsg1}
\begin{split}
&G_{\text{Higgs}}(t; {\vec x}^{(1)}, \ldots, {\vec x}^{(l)}; m)\,\\
&=(1-t^2)^{l}\oint \dfrac{dq_1}{2\pi i q_1^{1{\blue+m}}} \PE\left[t \sum_{\alpha=1}^{f_1} \left(q_1 (x^{(1)}_\alpha)^{-1} +q_1^{-1} x^{(1)}_\alpha \right) \right] \times \\
&\prod_{j=2}^{l}\oint \dfrac{dq_j}{2\pi i q_j} \PE\left[t \sum_{\alpha=1}^{f_j} \left(q_j (x^{(j)}_\alpha)^{-1} +q_j^{-1} x^{(j)}_\alpha \right) \right] \,\prod_{i=1}^{l-1} \PE[t(q_i\,q_{i+1}^{-1}+q^{-1}_i\,q_{i+1})]~,
\end{split}
\ee
where we indicated $m$ in blue to emphasise that this is due to the presence of the $T(U(1))$ link.
The baryonic generating for the Coulomb sub-quiver is similar to \eref{HSnocutlin}:
\be
\label{Coulombg1}
\begin{split}
G_{\text{Coulomb}}(t; m) &= \frac{1}{1-t^2} t^{|m| K} ~.
\end{split}
\ee
The total Hilbert series of \eref{onecutHC} is therefore
\be
\begin{split}
&H(t; \{ \vec x^{(i)} \}, z; m) \\
&= \sum_{m \in \BZ} t^{(F_C+1) |m|} G_{\text{Higgs}}(t; m) G_{\text{Coulomb}}(t; m) z^{m} \\
&= \sum_{m \in \BZ} t^{(K+F_C+1) |m|} z^m (1-t^2)^{l-1}\oint \dfrac{dq_1}{2\pi i q_1^{1{+m}}} \PE\left[t \sum_{\alpha=1}^{f_1} \left(q_1 (x^{(1)}_\alpha)^{-1} +q_1^{-1} x^{(1)}_\alpha \right) \right] \times \\
&\prod_{j=2}^{l}\oint \dfrac{dq_j}{2\pi i q_j} \PE\left[t \sum_{\alpha=1}^{f_j} \left(q_j (x^{(j)}_\alpha)^{-1} +q_j^{-1} x^{(j)}_\alpha \right) \right] \,\prod_{i=1}^{l-1}\PE[t(q_i\,q_{i+1}^{-1}+q^{-1}_i\,q_{i+1})]~.\end{split}
\ee

\paragraph{Example.}  Let us consider the following quiver
\be \label{exoneTflv}
\begin{tikzpicture}[baseline, scale=0.7]
\tikzstyle{every node}=[font=\footnotesize, minimum size=1.2cm]
\node[draw, dashed,circle,purple] (node1) at (-2,2) {$1_{k_1}$};
\node[draw, dashed,circle,purple] (node2) at (2,2) {$1_{k_2}$};
\node[draw,circle] (node3) at (6,2) {$1_{k}$};
\node[draw,circle] (node4) at (10,2) {$1_{-k}$};
\node[draw, rectangle] (sqnode2) at (2,-1) {$1$};
\node[draw, rectangle] (sqnode3) at (6,-1) {$f$};
\node[draw, rectangle] (sqnode4) at (10,-1) {$f'$};
\draw[draw=black,solid,thick,-]  (node1) to  (node2) ;
\draw[draw=red,solid,thick,-]  (node2) to  (node3) ;
\draw[draw=black,solid,thick,-]  (node3) to  (node4) ; 
\draw[draw=black,solid,thick,-]  (node2) to  (sqnode2) ;
\draw[draw=red,solid,thick,-]  (node3) to  (sqnode3) ;
\draw[draw=red,solid,thick,-]  (node4) to  (sqnode4) ;
\draw[purple] (node1) edge [out=90,in=90,looseness=0.3] node[midway,above=-0.3cm] {$T(U(1))$}  (node4);
\end{tikzpicture}
\ee
Assume that $k\geq0$. In this case, we have $K=k$ and $F_C =f+f'$.  The Hilbert series is then
\be
\begin{split}
H_{\eref{exoneTflv}}(t; x^{(2)}) &= \sum_{m \in \BZ} t^{(k+F_C+1) |m|} z^m (1-t^2) \oint \dfrac{dq_1}{2\pi i q_1^{1{+m}}}  \oint \dfrac{dq_2}{2\pi i q_2}   \\ 
& \qquad \PE\left[ t \left(q_2 (x^{(2)})^{-1} +q_2^{-1} x^{(2)}\right) \right] \PE[t(q_1\,q_{2}^{-1}+q^{-1}_1\,q_{2})] \\
&= \PE \left[ t^2 + \left(x^{(2)} z^{-1} + (x^{(2)})^{-1} z \right) t^{3+k+F_C} - t^{2(3+k+F_C)} \right]~.
\end{split}
\ee
Hence, the moduli space of this quiver is $\BC^2/\BZ_{3+k+F_C}$.  It is generated by $\varphi$ and $\bar{V}^{(2)}_\pm$,
where
\be
\bar{V}^{(2)}_+ = V_{(0,0,1,1)} \tQ_2 \tA_1 A_3^k ~, \quad \bar{V}^{(2)}_- = V_{(0,0,-1,-1)} A_1 Q_2 \tA_3^k ~,
\ee
subject to the relation
\be
\bar{V}^{(2)}_+ \bar{V}^{(2)}_- = \varphi^{3+k+F_C}~.
\ee

\subsubsection*{The case with more than one cuts}  
In this case, the original quiver is divided into many sub-quivers.  The parts that are not connected to $T(U(1))$ can be analysed as in section \ref{sec:addflavours}, and the parts that are connected to $T(U(1))$ can be analysed as in section \ref{sec:addflavourswithoneJ}.  

\subsection{More examples}
\subsubsection{One $J_k$ fold and one NS5 or D5-brane}
Let us consider the following model:
\be \label{TUNloopk}
\begin{tikzpicture}[baseline]
\tikzstyle{every node}=[font=\footnotesize]
\draw[blue,thick] (0,0) circle (1.5cm) node[midway, right] {$1$ D3};
\draw[thick] (0,-1)--(0,-2) node[right] {NS5};
\tikzset{decoration={snake,amplitude=.4mm,segment length=2mm,
                       post length=0mm,pre length=0mm}}
\draw[decorate,red,thick] (0,1) -- (0,2) node[right] {$J_k$};
\end{tikzpicture}
\qquad\qquad\qquad
\begin{tikzpicture}[baseline]
\tikzstyle{every node}=[font=\footnotesize]
\node[draw, circle] (node1) at (-1.5,0) {$1_k$};
\node[draw, circle] (node2) at (1.5,0) {$1_0$};
\draw[draw=black,solid,thick,-]  (node1) to[bend right=20]   (node2) ;  
\draw[draw=red,solid,thick,-]  (node1) to[bend left=20]  node[midway,above] {{\red $T(U(1))$}}  (node2) ; 
\end{tikzpicture}
\ee
Upon applying S-duality to the above system, we obtain
\be \label{mirrTUNloopk}
\begin{tikzpicture}[baseline]
\tikzstyle{every node}=[font=\footnotesize, node distance=0.45cm]
\tikzset{decoration={snake,amplitude=.4mm,segment length=2mm,
                       post length=0mm,pre length=0mm}}
\draw[blue,thick] (0,0) circle (1.5cm) node[midway, right] {$1$ D3};
\draw[decorate,purple,thick] (0,1) -- (0,2) node[right] {$-J^{-1}_{-k}$};
\node[draw=none] at (0,-2) {D5};
\node[draw=none] at (0,-1.7) {\large{$\bullet$}};
\end{tikzpicture}
\qquad \qquad \qquad
\begin{tikzpicture}[baseline]
\tikzstyle{every node}=[font=\footnotesize]
\node[draw, circle] (node1) at (0,1) {$1_k$};
\draw[purple,thick] (node1) edge [out=45,in=135,loop,looseness=5]  (node1);
\node[draw=none] at (1.3,1.5) {{\purple $\bar{T(U(1))}$}};
\node[draw, rectangle] (sqnode) at (0,-1) {$1$};
\draw (node1)--(sqnode);
\end{tikzpicture}
\ee

Both of these models are analysed in detail around \eref{twonodesoneT} and \eref{onenodenflv}, respectively. The moduli spaces these model are non-trivial if and only if $k=2$. In which case, they are isomorphic to $\BC^2$.

\subsubsection{One $(p,q)$-brane and one NS5-brane} \label{sec:onepqandoneNS5abel}
The techniques that we introduced in the section \ref{sec:abel} are particularly useful to study in a systematic way the moduli space of quiver gauge theories associated to $(p,q)$-brane systems. Let us consider for instance the following brane system  
\be \label{pqandNS}
\begin{tikzpicture}[baseline, scale=0.8]
\tikzstyle{every node}=[font=\footnotesize, node distance=0.45cm]
\draw[blue,thick] (0,0) circle (1.5cm) node[midway, right] {1 D3};
\draw[dashed, thick] (0,1)--(0,2) node[right] {$(p,q)$};
\draw[solid, thick] (0,-1)--(0,-2) node[right] {NS5};
\end{tikzpicture}
\ee
For simplicity, let us take $(p,q)$ to be the following value: $(p,q)\,=\,\bar J_{k_3}\bar J_{k_2}\,\bar J_{k_1}(1,0)$, so that
\be
p = k_1 k_2 k_3 -k_1 -k_3~, \qquad q = k_1 k_2-1~.
\ee
Performing a duality transformation, $\bar J_{k_2}^{\,\,-1}\bar J_{k_3}^{\,\,-1}$, we can study the following $SL(2,\BZ)$ equivalent problem:
\be
\label{pqdual}
\begin{tikzpicture}[baseline, scale=0.8]
\tikzstyle{every node}=[font=\footnotesize, node distance=0.45cm]
\draw[blue,thick] (0,0) circle (1.5cm) node[midway, right] {1 D3};
\draw[dashed, thick] (0,1)--(0,2) node[right] {$(k_1,1)$};
\draw[solid, thick] (0,-1)--(0,-2) node[right] {$(-1, -k_2)$};
\end{tikzpicture}
\qquad\qquad
\begin{tikzpicture}[baseline, scale=0.8]
\tikzstyle{every node}=[font=\footnotesize, node distance=0.45cm]
\draw[blue,thick] (0,0) circle (1.5cm) node[midway, right] {$1$ D3};
\draw[thick] (0,1)--(0,2) node[right] {NS5};
\tikzset{decoration={snake,amplitude=.4mm,segment length=2mm,
                       post length=0mm,pre length=0mm}}
\draw[decorate,red,thick] (-0.75,0.5)--(-1.5,1)node[left] {$\bar{J}_{k_1}^{-1}$};
\draw[decorate,purple,thick] (0.75,0.5)--(1.5,1)node[right] {$\bar{J}_{k_1}$};
\draw[solid, thick] (0,-1)--(0,-2) node[right] {$(-1, -k_2)$};
\def \n {6}
\def \radius {1.2cm}
\def \margin {0} 
\foreach \s in {1,...,10}
{
	\node[draw=none] (\s) at ({360/\n * (\s - 2)+30}:{\radius-10}) {};
}
\end{tikzpicture}
\ee
The associated quiver is
\be \label{modelN3}
\begin{tikzpicture}[baseline, scale=0.7]
\tikzstyle{every node}=[font=\footnotesize, minimum size=1.2cm]
\node[draw, circle] (node1) at (-2,2) {$1_{k_1}$};
\node[draw, circle] (node2) at (2,2) {$1_{-k_1}$};
\node[draw, circle] (node3) at (2,-2) {$1_{-k_2}$};
\node[draw, circle] (node4) at (-2,-2) {$1_{k_2}$};
\draw[draw=purple,solid,thick,-]  (node1) to node[midway, left] {\purple{$\bar{T(U(1))}$}}  (node4) ; 
\draw[draw=black,solid,thick,-]  (node1) to (node2) ; 
\draw[draw=black,solid,thick,-]  (node3) to  (node4) ; 
\draw[draw=red,solid,thick,-]  (node2) to node[midway, right] {\red{${T(U(1))}$}}  (node3) ;
\end{tikzpicture}
\ee
In $\mathcal{N}=2$ language, this can be written as
\be \label{model2a}
\begin{tikzpicture}[baseline, scale=0.7]
\tikzstyle{every node}=[font=\footnotesize, minimum size=1.2cm]
\node[draw, circle] (node1) at (-2,2) {$1_{k_1}$};
\node[draw, circle] (node2) at (2,2) {$1_{-k_1}$};
\node[draw, circle] (node3) at (2,-2) {$1_{-k_2}$};
\node[draw, circle] (node4) at (-2,-2) {$1_{k_2}$};
\draw[draw=purple,solid,thick,-]  (node1) to node[midway, left] {\purple{$\bar{T(U(1))}$}}  (node4) ; 
\draw[draw=black,solid,thick,<->]  (node1) to node[pos=0.2, above=-0.2cm] {$A$} node[pos=0.8, above=-0.2cm] {$\tilde{A}$}  (node2) ; 
\draw[draw=black,solid,thick,<->]  (node3) to node[pos=0.2, below=-0.2cm] {$\tilde{B}$} node[pos=0.8, below=-0.2cm] {$B$}  (node4) ; 
\draw[draw=red,solid,thick,-]  (node2) to node[midway, right] {\red{${T(U(1))}$}}  (node3) ;
\draw[black] (node1) edge [out=45,in=135,loop,looseness=4] node[midway,above=-0.2cm] {$\phi_1$}  (node1);
\draw[black] (node2) edge [out=45,in=135,loop,looseness=4] node[midway,above=-0.2cm] {$\phi_2$}  (node2);
\draw[black] (node3) edge [out=-45,in=-135,loop,looseness=4] node[midway,below=-0.3cm] {$\phi_4$}  (node3);
\draw[black] (node4) edge [out=-45,in=-135,loop,looseness=4] node[midway,below=-0.3cm] {$\phi_3$}  (node4);
\end{tikzpicture}
\ee
The vacuum equations are
\be
\begin{array}{ll}
\label{pqFterms}
A(\varphi_1-\varphi_2)\,=\,0\,=\,\tA(\varphi_1-\varphi_2)\,,&\quad\quad B(\varphi_3-\varphi_4)\,=\,0\,=\,\tB(\varphi_3-\varphi_4)\\
k_1\varphi_1-{\blue{\varphi_3}}\,=\,A\,\tA\,,&\quad\quad k_2\varphi_3-{\blue{\varphi_1}}\,=\,B\,\tB\\
-k_1\varphi_2+{\blue{\varphi_4}}\,=-\,A\,\tA\,,&\quad\quad -k_2\varphi_4+{\blue{\varphi_2}}\,=-\,B\,\tB\,.
\end{array}
\ee
where we emphasised the contributions due to the mixed CS levels in blue. We have two branches as will be analysed as follow.

\subsubsection*{Branch I: $A\tA\neq 0$\, and \,$B\tB\neq0$} In this case the $F$-terms implies:
\be
\varphi_1\,=\,\varphi_2\,=\,\varphi\,,\qquad \varphi_3\,=\,\varphi_4\,=\,\tilde\varphi\,;
\ee
moreover, two constraints are still present, fixing $\varphi\,,\,\tilde\varphi$ in terms of the mesons:
\be
k_1 \varphi-\tilde \varphi\,=\,A\tA\,,\quad k_2\tilde\varphi-\varphi\,=\,B\tB\,.
\ee
An analogous analysis of the D-terms can be performed.  The flux $\vec m$ for the monopole operator $V_{\vec m}$ takes the form  
\be
\vec m = (m,m, \tilde{m}, \tilde{m})~.
\ee
The gauge charges and the $R$-charges of $V_{\vec m}$ are
\be
\begin{split}
&q_1[V_{\vec m}]\,=\,-q_2[V_{\vec m}]\,=\,-(k_1 m-\tilde m)\,,\\
&q_3[V_{\vec m}]\,=\,-q_4[V_{\vec m}]\,=\,-(k_2\tilde m-m)~.
\end{split}
\ee
and
\be
R[V_{\vec m} ] =0~.
\ee

Let us now determine the moduli space and compute the Hilbert series of this theory.  The baryonic generating function is given by 
\be
\begin{split}
G(t; B, \tilde B)&\,=\,\left( \prod_{i=1}^4\oint\frac{dq_i}{2\pi iq_i} \right)  \frac{1}{q_1^B q_2^{-B} q_3^{\tB} q_4^{-\tB}} \PE[t(q_1q_2^{-1}+q_2q_1^{-1})] \PE[t(q_3q_4^{-1}+q_4q_3^{-1})] \\
&\,=\,g_{\text{ABJM}/2}(t;B)\,g_{\text{ABJM}/2}(t;\tB)\,.
\end{split}
\ee
where
\be
g_{\text{ABJM}/2}(t;B) = \frac{t^{|B|}}{1-t^2}~.
\ee
The Hilbert series of \eref{model2a} is thus:
\be
\begin{split}
\label{HSpq}
H_{\eref{model2a}}(t,z)&= \sum_{m\in \mathbb Z} \sum_{\tilde m\in \mathbb Z} z^{m+\tilde m} g_{\text{ABJM}/2}(t; k_1 m-\tilde m )g_{\text{ABJM}/2}(t;k_2 \tilde m-m) \\
&=\,\sum_{m\in \mathbb Z} \sum_{\tilde m\in \mathbb Z} z^{m+\tilde m}\frac{t^{\abs{k_1 m-\tilde m}}}{1-t^2}\,\frac{t^{\abs{k_2 \tilde m-m}}}{1-t^2}~.
\end{split}
\ee
This turns out to be equal to 
\be
\begin{split}
H_{\eref{model2a}}(t,z)&=\,\frac{1}{k_1 k_2 -1 }\sum_{j=1}^{k_1\,k_2-1} \frac{1}{(1-t\,u_j)(1-t\,w_j)}\frac{1}{(1-t/u_j)(1-t/w_j)} \\
&= H[\mathbb{C}^4/\Gamma(k_1,k_1 k_2-1)](t, z)~,
\end{split}
\ee
where
\be
u_j\,=z^{\frac{k_1 +1}{k_1 k_2 -1}}\,e^{j\frac{2\pi i\,k_1}{k_1k_2-1}}\,, \qquad  w_j\,=z^{\frac{k_2 +1}{k_1 k_2 -1}}\,e^{j\frac{2\pi i}{k_1k_2-1}}~.
\ee
This is the Molien formula for the Hilbert series of $\mathbb{C}^4/\Gamma(\mathsf{p},\mathsf{q})$ \cite{Benvenuti:2006qr}, with $\mathsf{p} = k_1$ and $\mathsf{q} =k_1 k_2-1$,  where $\Gamma(\mathsf{p},\mathsf{q})$ is a discrete group acting on the four complex coordinate of $\mathbb{C}^4$ as:
\be
\Gamma(\mathsf{p},\mathsf{q}):\,\,\,(z_1\,,\,z_2\,,\,z_3\,,\,z_4)\,\rightarrow\, (z_1 e^{\frac{2\pi i \mathsf{p}}{\mathsf{q}}}\,,\,z_2e^{\frac{2\pi i}{\mathsf{q}}}\,,\,z_3e^{-\frac{2\pi i \mathsf{p}}{\mathsf{q}}}\,,\,z_4e^{-\frac{2\pi i}{ \mathsf{q}}})~.
\ee
This is in agreement with \cite{Jafferis:2009th, Assel:2017eun}.

\subsubsection*{Branch II: $A\tA\,=\,0$ or $B\tB\,=\,0$}
The second branch appears when we set one of the bi-fundamental hypers to zero, say $A\tA\,=\,0$. In this case, \eref{pqFterms} implies again that:
\be
\varphi_1\,=\,\varphi_2\,=\,\varphi\,,\qquad \varphi_3\,=\,\varphi_4\,=\,\tilde \varphi.
\ee
Moreover, we have\footnote{A special case is $k_1\,=\,k_2\,=\pm 1$. In this case $B\tB\,=\,0$ and we are left with $\varphi$ and the basic monopole operators. The corresponding moduli space is thus simply $\mathbb{C}^2$.}:
\be
k_1\varphi\, \,=\, \tilde \varphi\,,\quad k_2 \tilde \varphi\,-\,\varphi\,=\,B\tB\,.
\ee

Because of $\CN=3$ supersymmetry of the problem, the real scalar in the vector multiplet satisfies the same equation as the complex scalar in the vector multiplet.  As a consequence, the flux $\vec m = (m, m, \tilde{m}, \tilde{m})$ of the monopole operator $V_{\vec m}$ has to satisfy
\be \label{k1mmt}
k_1\,m\,=\, \tilde m
\ee
The gauge charges of $V_{\vec m}$ are
\be
\begin{split}
&q_1[V_{\vec m}]\,=\,-q_2[V_{\vec m}]\,=\,-(k_1 m-\tilde m) = 0\,,\\
&q_3[V_{\vec m}]\,=\,-q_4[V_{\vec m}]\,=\,-(k_2\tilde m-m) = -(k_1k_2 -1)m~.
\end{split}
\ee
The $R$-charge of $V_{\vec m}$ is $R[V_{\vec m}] = 0$. The gauge invariant dressed monopole operators are
\be
\bar{V}_{+}\,=\,V_{(1,1,k_1, k_1)} \,B^{k_1 k_2-1}\,,\quad \bar{V}_{-}\,=\,V_{(-1,-1,-k_1, -k_1)}\,\tB^{k_1 k_2-1}~,
\ee
for $k_1 k_2 -1 > 0$.  If $k_1 k_2 -1 <0$, we replace $B^{k_1 k_2-1}$ by $\tilde{B}^{-(k_1 k_2-1)}$ and $\tB^{k_1 k_2-1}$ by $B^{-(k_1 k_2-1)}$ in the above equations.
They carry $R$-charges $R[\bar{V}_{\pm}]\,=\,\frac{|k_1k_2-1|}{2}$.   Since $(k_1 k_2-1) \varphi = B \tilde{B}$, we see that these dressed monopole operators satisfy the quantum relation
\be
\bar{V}_+\,\bar{V}_-\,=\, \varphi^{|k_1k_2-1|}~.
\ee
Hence the moduli space is $\mathbb{C}^2/\mathbb Z_{|k_1 k_2-1|}$. 

Note that \eref{k1mmt} implies that the magnetic lattice given by $\tilde{m}$ jumps by a multiple of $k_1$, since $m \in \BZ$. If we further require that the magnetic lattice do not jump, we can impose a further condition that $k_1 = \pm 1$.  In this case, the brane system contains a $(\pm 1 , 1)$-brane and a $(-1, - k_2)$-brane. Applying $T^{\mp1}$ to this system, $(\pm 1 , 1)$ becomes $(\pm 1,0)$, and $(-1,k_2)$ becomes $(-1, -k_2 \mp 1)$.  This gives rise to the ABJM theory with CS level $k_2-1$ and $-k_2+1$. Indeed, Branch I (which is $\BC^4/\BZ_{|k_2-1|}$) and Branch II (which is $\BC^2/\BZ_{|k_2-1|}$) are the geometric branch of the ABJM theory and the moduli space of the half-ABJM theory, respectively.

\subsubsection{Multiple $(p,q)$ and NS5-branes}
An interesting generalisation of the example we presented in the previous subsection is the following brane configuration:
\be \label{pqandNS}
\begin{tikzpicture}[baseline, scale=0.8]
\tikzstyle{every node}=[font=\footnotesize, node distance=0.45cm]
\draw[blue,thick] (0,0) circle (1.5cm) node[midway, right] {1 D3};
\draw[dashed, thick] (0,1)--(0,2) node[right] {$l_1$\,\,$(p,q)$};
\draw[solid, thick] (0,-1)--(0,-2) node[right] {$l_2$ NS5};
\end{tikzpicture}
\ee
As before, let us take for simplicity $(p,q)\,=\,\bar J_{k_3}\bar J_{k_2}\,\bar J_{k_1}(1,0)$. Performing a transformation, $\bar J_{k_2}^{\,\,-1}\bar J_{k_3}^{\,\,-1}$, we can study the 
following $SL(2,\BZ)$ equivalent systems:
\be
\label{pqdual}
\begin{tikzpicture}[baseline, scale=0.8]
\tikzstyle{every node}=[font=\footnotesize, node distance=0.45cm]
\draw[blue,thick] (0,0) circle (1.5cm) node[midway, right] {1 D3};
\draw[dashed, thick] (0,1)--(0,2) node[right] {$l_1\,\,(k_1,1)$};
\draw[solid, thick] (0,-1)--(0,-2) node[right] {$l_2\,\,(-1, -k_2)$};
\end{tikzpicture}
\qquad\qquad
\begin{tikzpicture}[baseline, scale=0.8]
\tikzstyle{every node}=[font=\footnotesize, node distance=0.45cm]
\draw[blue,thick] (0,0) circle (1.5cm) node[midway, right] {$1$ D3};
\draw[thick] (0,1)--(0,2) node[right] {$l_1$\,\,NS5};
\tikzset{decoration={snake,amplitude=.4mm,segment length=2mm,
                       post length=0mm,pre length=0mm}}
\draw[decorate,red,thick] (-0.75,0.5)--(-1.5,1)node[left] {$\bar{J}_{k_1}^{-1}$};
\draw[decorate,purple,thick] (0.75,0.5)--(1.5,1)node[right] {$\bar{J}_{k_1}$};
\draw[solid, thick] (0,-1)--(0,-2) node[right] {$l_2\,\,(-1, -k_2)$};
\def \n {6}
\def \radius {1.2cm}
\def \margin {0} 
\foreach \s in {1,...,10}
{
	\node[draw=none] (\s) at ({360/\n * (\s - 2)+30}:{\radius-10}) {};
}
\end{tikzpicture}
\ee
The quiver associated with the brane system on the right is
\be \label{model2La}
\begin{tikzpicture}[baseline, scale=0.7]
\tikzstyle{every node}=[font=\footnotesize, minimum size=1.2cm]
\node[draw, circle] (node1) at (-2,2) {$1_{k_1}$};
\node[draw, circle] (node2) at (10,2) {$1_{-k_1}$};
\node[draw, circle] (node5) at (2,2) {$1_{0}$};
\node[draw, circle] (node6) at (6,2) {$1_{0}$};
\node[draw, circle] (node3) at (10,-2) {$1_{-k_2}$};
\node[draw, circle] (node4) at (-2,-2) {$1_{k_2}$};
\node[draw, circle] (node7) at (6,-2) {$1_{0}$};
\node[draw, circle] (node8) at (2,-2) {$1_{0}$};
\draw[draw=purple,solid,thick,-]  (node1) to node[midway, left] {\purple{$\bar{T(U(1))}$}}  (node4) ; 
\draw[draw=black,solid,thick,-]  (node1) to  (node5) ; 
\draw[draw=black,dashed,thick]  (node5) to   (node6) ; 
\draw[draw=black,solid,thick,-]  (node6) to  (node2) ; 
\draw[draw=black,solid,thick,-]  (node3) to  (node7) ; 
\draw[draw=black,solid,thick,-]  (node4) to (node8) ; 
\draw[draw=black,dashed,thick]  (node7) to (node8) ; 
\draw[draw=red,solid,thick,-]  (node2) to node[midway, right] {\red{${T(U(1))}$}}  (node3) ;
\end{tikzpicture}
\ee
where the numbers of gauge nodes are $l_1+1$ and $l_2+1$ on the upper and the lower sides of the quiver, respectively.
In the $\mathcal{N}=2$ notation, this can be written as
\be \label{model2L}
\begin{tikzpicture}[baseline, scale=0.7]
\tikzstyle{every node}=[font=\footnotesize, minimum size=1.2cm]
\node[draw, circle] (node1) at (-2,2) {$1_{k_1}$};
\node[draw, circle] (node2) at (10,2) {$1_{-k_1}$};
\node[draw, circle] (node5) at (2,2) {$1_{0}$};
\node[draw, circle] (node6) at (6,2) {$1_{0}$};
\node[draw, circle] (node3) at (10,-2) {$1_{-k_2}$};
\node[draw, circle] (node4) at (-2,-2) {$1_{k_2}$};
\node[draw, circle] (node7) at (6,-2) {$1_{0}$};
\node[draw, circle] (node8) at (2,-2) {$1_{0}$};
\draw[draw=purple,solid,thick,-]  (node1) to node[midway, left] {\purple{$\bar{T(U(1))}$}}  (node4) ; 
\draw[draw=black,solid,thick,<->]  (node1) to node[pos=0.2, above=-0.2cm] {$A_1$} node[pos=0.8, above=-0.2cm] {$\tilde{A}_1$}  (node5) ; 
\draw[draw=black,dashed,thick]  (node5) to   (node6) ; 
\draw[draw=black,solid,thick,<->]  (node6) to node[pos=0.2, above=-0.2cm] {$A_{l_1}$} node[pos=0.8, above=-0.2cm] {$\tilde{A}_{l_1}$}  (node2) ; 
\draw[draw=black,solid,thick,<->]  (node3) to node[pos=0.2, above=-0.2cm] {$B_{l_2}$} node[pos=0.8, above=-0.2cm] {$\tilde{B}_{l_2}$}  (node7) ; 
\draw[draw=black,solid,thick,<->]  (node4) to node[pos=0.2, below=-0.2cm] {$B_1$} node[pos=0.8, below=-0.2cm] {$\tilde{B}_1$}  (node8) ; 
\draw[draw=black,dashed,thick]  (node7) to (node8) ; 
\draw[draw=red,solid,thick,-]  (node2) to node[midway, right] {\red{${T(U(1))}$}}  (node3) ;
\draw[black] (node1) edge [out=45,in=135,loop,looseness=4] node[midway,above=-0.2cm] {$\varphi_1$}  (node1);
\draw[black] (node2) edge [out=45,in=135,loop,looseness=4] node[midway,above=-0.2cm] {$\varphi_{l_1+1}$}  (node2);
\draw[black] (node3) edge [out=-45,in=-135,loop,looseness=4] node[midway,below=-0.3cm] {$\tilde{\varphi}_{l_2+1}$}  (node3);
\draw[black] (node4) edge [out=-45,in=-135,loop,looseness=4] node[midway,below=-0.3cm] {$\tilde{\varphi}_{1}$}  (node4);
\draw[black] (node5) edge [out=45,in=135,loop,looseness=4] node[midway,above=-0.2cm] {$\varphi_2$}  (node5);
\draw[black] (node6) edge [out=45,in=135,loop,looseness=4] node[midway,above=-0.2cm] {$\varphi_{l_1}$}  (node6);
\draw[black] (node7) edge [out=-45,in=-135,loop,looseness=4] node[midway,below=-0.3cm] {$\tilde{\varphi}_{l_2}$}  (node7);
\draw[black] (node8) edge [out=-45,in=-135,loop,looseness=4] node[midway,below=-0.3cm] {$\tilde{\varphi}_{2}$}  (node8);
\end{tikzpicture}
\ee
The vacuum equations are
\be 
\begin{array}{ll}
\label{morepq-Fterms}
A(\varphi_i-\varphi_{i+1})\,=\,0\,\,\,i=1,\dots,l_1\,,\quad\quad &B(\tilde\varphi_i-\tilde\varphi_{i+1})\,=\,0\,\,\,i=1,\dots,l_2 \\
k_1\varphi_1-{\blue{\tilde \varphi_1}}\,=\,A_1\,\tA_1\,,\quad\quad &k_2\tilde\varphi_1-{\blue{\varphi_1}}\,=\,B_1\,\tB_1\\
0=A_i\,\tA_i-\,A_{i-1}\,\tA_{i-1}\,\,\,i=2\,,\dots\,,l_1\,,\quad\quad &0=B_i\,\tB_i-\,B_{i-1}\,\tB_{i-1}\,\,\,i=2\,,\dots\,,l_2\,,\\
-k_1\varphi_{l_1+1}+{\blue{\tilde \varphi_{l_2+1}}}\,=-\,A_{l_1}\,\tA_{l_1}\,,\quad\quad &-k_2\varphi_{l_2+1}+{\blue{\tilde \varphi_{l_1+1}}}\,=-\,B_{l_2}\,\tB_{l_2}\,~,
\end{array}
\ee
where we highlighted in blue the contributions from the mixed CS terms due to $T(U(1))$ and $\bar{T(U(1))}$.
We focus on the geometric branch, corresponding to the case $\varphi_i\,=\,\varphi$  for all $i=1\,\dots\,l_1+1$ and $\tilde\varphi_i\,=\,\tilde\varphi$ for all $i=1\,\dots\,l_2+1$. 
Imposing these conditions, we are left with the following constraints of the mesons:
\be
\begin{split}
&k_1\varphi-\tilde\varphi\,=\,A_1\tA_1\,, \quad A_{i+1}\tA_{i+1}-A_{i}\tA_{i}\,=\,0\,,\quad -k_1\varphi+\tilde\varphi\,=\,-A_{l_1}\,\tA_{l_1}\\
&k_2\tilde\varphi-\varphi\,=\,B_1\tB_1\,, \quad B_{i+1}\tB_{i+1}-B_{i}\tB_{i}\,=\,0\,,\quad -k_1\tilde\varphi+\varphi\,=\,-B_{l_1}\,\tB_{l_1}\\
\end{split}
\ee
Let us consider the monopole operator $V_{\vec m}$ with flux
\be
\vec m = ( \underbrace{m, \ldots, m}_{l_1+1}, \underbrace{\tilde{m}, \ldots, \tilde{m}}_{l_2+1}) = (m^{l_1+1} , \tilde{m}^{l_2+1})~.
\ee
The $R$-charge of $V_{\vec m}$ is zero:
\be
R[V_{\vec m}] = 0~.
\ee
and the gauge charges are
\be
\begin{array}{lllll}
q_1[V_{\vec m}]\,=-(k_1\,m\,-\tilde m)\,, &q_2[V_{\vec m}]\,=\,0\,,&\dots\,,&q_{l_1}[V]\,=\,0\,,&q_{l_1+1}[V_{\vec m}]\,=(k_1\,m\,-\,\tilde m)\,,\\
q_{\tilde 1}[V_{\vec m}]\,=-(k_2\,\tilde m\,-\tilde m) &q_{\tilde{2}}[V_{\vec m}]\,=\,0\,,&\dots\,,& q_{\tilde{l_1}}[V_{\vec m}]\,=\,0\,,&\tilde q_{\tilde{l_1+1}}[V_{\vec m}]\,=(k_2\,\tilde m\,-\, m)\,.
\end{array}
\ee
Now we have all the ingredients in order to compute the baryonic generic function:
\be
\begin{split}
g(t; B, \tB)\,&=\,\PE[-t^2]^{l_1-1}\PE[-t^2]^{l_2-1} \oint \frac{d q_1 dq_2\dots dq_{l_1+1}}{(2\pi i)^{l_1+1} q_1^{1+B}q_2\dots q_{l_1}q^{1-B}_{l_1+1}} \\
& \qquad \oint \frac{d\tilde q_1 d\tilde q_2\dots d\tilde q_{l_1+1}}{(2\pi i)^{l_2+1}\tilde q_1^{1+\tB}\tilde q_2\dots \tilde q_{l_1}\tilde q^{1-\tB}_{l_1+1}}  \prod_{i=1}^{l_1}\PE[t(q_i q_{i+1}^{-1}+q_i^{-1} q_{i+1})] \\
& \qquad \prod_{j=1}^{l_2} \PE[t(q_j q_{j+1}^{-1}+q_i^{-1} q_{j+1})] \\
&=g^{\text{ABJM/2}}(t; l_1\,B)\,g^{\text{ABJM/2}}(t; l_2\,\tilde B)\,=\,\frac{t^{l_1 |B|+l_2|\tB|}}{(1-t^2)^2}
\end{split}
\ee
The Hilbert series of the geometric branch of \eref{model2La} is then
\be
\begin{split}
H_{\eref{model2La}}(t,z)\, &=  g(t; k_1 m-\tilde m, k_2 \tilde m-m) \\
&=\,\frac{1}{(1-t^2)^2}\sum_{m\in \mathbb{Z}}\sum_{\tilde m\in \mathbb{Z}} z^{m+\tilde{m}} t^{l_1\abs{k_1 m-\tilde m}+l_2\abs{k_2 \tilde m-m}}~.
\end{split}
\ee
Note that for $l_1=l_2=1$ we recover \eref{HSpq} as expected.

In {\it some} cases, the geometric branch of \eref{model2La} turns out to be isomorphic to $(\mathbb{C}^2/\mathbb{Z}_{l_1}\times \mathbb{C}^2/\mathbb{Z}_{l_2})/\Gamma[k_1\,,\,k_1\, 
k_2-1]$, where the action of $\Gamma[k_1, k_1 k_2-1]$ being
\be \label{actionproduct}
\Gamma[k_1, k_1 k_2-1]: (z_1, z_2; \tilde{z}_1, \tilde{z}_2) \rightarrow (\omega z_1, \omega^{-1} z_2; \,\, \tilde{\omega} \tilde{z}_1, \tilde{\omega}^{-1} \tilde{z}_2)~,
\ee
with $\omega = e^{2 \pi i\frac{k_1}{k_1k_2-1}}$ and $\tilde{\omega} = e^{2 \pi i\frac{1}{k_1 k_2-1}}$; and $(z_1, z_2)$ and $(\tilde{z}_1, \tilde{z}_2)$ are the coordinates of $\mathbb{C}
^2/\mathbb{Z}_{l_1}$ and $\mathbb{C}^2/\mathbb{Z}_{l_2}$ respectively.
For example, when $\{ k_1\,=2\,,k_2\,=\,3\,,\,l_1\,=\,l_2\,=\,2\}$ we have
\be
\begin{split}
H_{\eref{model2La}}(t,z=1)\,& = \frac{1-t^2+4 t^6-t^{10}+t^{12}}{(1-t^2)^3(1-t^{10})} \\
&= H[(\mathbb{C}^2/\mathbb{Z}_{2}\times \mathbb{C}^2/\mathbb{Z}_{2})/\Gamma[2\,,\,5]] (t, z=1)~.
\end{split}
\ee
and when $\{ k_1\,=2\,,k_2\,=\,2\,,\,l_1\,=\,5\,,l_2\,=\,1\}$, we have
\be
\begin{split}
H_{\eref{model2La}}(t,z=1)\, &=\,\frac{(1-t+t^4-t^7+t^8)(1+t^2+t^3+t^6+t^7+t^9)}{(1-t) \left(1-t^2\right) \left(1-t^3\right) \left(1-t^{15}\right)} \\
&= H[(\mathbb{C}^2/\mathbb{Z}_{5}\times \mathbb{C}^2)/\Gamma[2\,,\,3]](t, z=1)~.
\end{split}
\ee
Other cases can be more complicated.  For example, for $\{ k_1\,=2\,,k_2\,=\,3\,,\,l_1\,=\,1\,,l_2\,=\,3\}$, we find that
\be
\begin{split}
H_{\eref{model2La}}(t,z=1)\, &=\frac{1}{(1-t)^2 \left(1-t^5\right) \left(1-t^{15}\right)} \times (1 - t + t^2)(1 - t + t^2 - 2 t^3   \\
& \quad  + 2 t^4 + t^5 + 2 t^6 - 3 t^8 + \text{palindrome up to $t^{16}$}) \\
&= H[(\mathbb{C}^2 \times \mathbb{C}^2/\mathbb{Z}_{3})/\hat{\Gamma}](t, z=1)~,
\end{split}
\ee
where the orbifold $\hat{\Gamma}$ acts as \eref{actionproduct} but with $\omega = e^{2 \pi i\frac{2}{5}}$ and $\tilde{\omega} = e^{2 \pi i\frac{3}{5}} = \omega^{-1}$~.


\subsubsection{One $(p,q)$-brane and one D5-brane}
Let us consider an example of one $(p,q)$-brane and one D5-brane. In particular, let us assume that $(p,q) \,=\,\bar J_{k_1}\,(1,0) = (k_1,1)$:
\be \label{pqandD5}
\begin{tikzpicture}[baseline, scale=0.8]
\tikzstyle{every node}=[font=\footnotesize, node distance=0.45cm]
\draw[blue,thick] (0,0) circle (1.5cm) node[midway, right] {1 D3};
\draw[dashed, thick] (0,1)--(0,2) node[right] {$(k_1,1)$};
\node[draw=none] at (0,-2.2) {D5};
\node[draw=none] at (0,-1.7) {\large{$\bullet$}};
\end{tikzpicture}
\ee
One may apply $SL(2,\BZ)$ action to this configuration and obtain the following configurations:
\be
\label{pqdual1}
\begin{tikzpicture}[baseline, scale=0.8]
\tikzstyle{every node}=[font=\footnotesize, node distance=0.45cm]
\draw[blue,thick] (0,0) circle (1.5cm) node[midway, right] {1 D3};
\draw[solid, thick] (0,1)--(0,2) node[right] {$(1,0)$};
\draw[solid, thick] (0,-1)--(0,-2) node[right] {$(1, k_1)$};
\end{tikzpicture}
\qquad\qquad
\begin{tikzpicture}[baseline, scale=0.8]
\tikzstyle{every node}=[font=\footnotesize, node distance=0.45cm]
\draw[blue,thick] (0,0) circle (1.5cm) node[midway, right] {$1$ D3};
\draw[thick] (0,1)--(0,2) node[right] {NS5};
\tikzset{decoration={snake,amplitude=.4mm,segment length=2mm,
                       post length=0mm,pre length=0mm}}
\draw[decorate,red,thick] (-0.75,0.5)--(-1.5,1)node[left] {$\bar{J}_{k_1}^{-1}$};
\draw[decorate,purple,thick] (0.75,0.5)--(1.5,1)node[right] {$\bar{J}_{k_1}$};
\node[draw=none] at (0,-2.2) {D5};
\node[draw=none] at (0,-1.7) {\large{$\bullet$}};
\def \n {6}
\def \radius {1.2cm}
\def \margin {0} 
\foreach \s in {1,...,10}
{
	\node[draw=none] (\s) at ({360/\n * (\s - 2)+30}:{\radius-10}) {};
}
\end{tikzpicture}
\ee
where the first configuration is obtained by applying a $\bar J_{k_1}^{-1}$ duality transformation to \eref{pqandD5} and using the fact that $\bar J_{k_1}^{-1} (0,1) = (1,k_1)$, and 
for the second configuration we use the fact that  $\bar J_{k_1} (1,0)=(k_1,1)$, so we recover the original set-up \eref{pqandD5}.   

The brane configuration on the left in \eref{pqdual1} is that of the ABJM theory with CS level $(k_1,-k_1)$.  Thus, we expect that the moduli space of the field theories associated 
with these brane configurations has two branches, namely (1) $\mathbb{C}^4/\,\mathbb{Z}_{|k_1|}$, which is the geometric moduli space of the ABJM theory, and (2) $\mathbb{C}
^2/\,\mathbb{Z}_{|k_1|}$, which is the moduli space of the half-ABJM theory, where a pair of bi-fundamental chiral multiplets of the ABJM theory is set to zero. 

Let us derive these moduli spaces for the theory associated with the configuration on the right in \eref{pqdual1}.  The quiver diagram is given by
\be \label{modelpqD5a}
\scalebox{0.8}{
\begin{tikzpicture}[baseline, scale=0.7]
\tikzstyle{every node}=[font=\footnotesize, minimum size=1.2cm]
\node[draw, circle] (node1) at (-2,2) {$1_{-k_1}$};
\node[draw, circle] (node2) at (2,2) {$1_{k_1}$};
\node[draw, circle] (node3) at (0,-1.5) {$1_0$};
\node[draw, rectangle] (node4) at (0,-4.5) {$1$};
\draw[draw=purple,solid,thick,-]  (node1) to node[midway, left] {\purple{$\bar{T(U(1))}$}}  (node3) ; 
\draw[draw=red,solid,thick,-]  (node2) to node[midway, right] {\red{${T(U(1))}$}}  (node3) ;
\draw[draw=black,solid,thick,-]  (node1) to  (node2) ;
\draw[draw=black,solid,thick,-]  (node3) to  (node4) ; 
\end{tikzpicture}}
\ee
In the $\CN=2$ notation, this quiver can be rewritten as
\be \label{modelpqD5}
\scalebox{0.8}{
\begin{tikzpicture}[baseline, scale=0.7]
\tikzstyle{every node}=[font=\footnotesize, minimum size=1.2cm]
\node[draw, circle] (node1) at (-2,2) {$1_{-k_1}$};
\node[draw, circle] (node2) at (2,2) {$1_{k_1}$};
\node[draw, circle] (node3) at (0,-1.5) {$1_0$};
\node[draw, rectangle] (node4) at (0,-4.5) {$1$};
\draw[draw=purple,solid,thick,-]  (node1) to node[midway, left] {\purple{$\bar{T(U(1))}$}}  (node3) ; 
\draw[draw=red,solid,thick,-]  (node2) to node[midway, right] {\red{${T(U(1))}$}}  (node3) ;
\draw[draw=black,solid,thick,<->]  (node1) to node[pos=0.2, below=-0.2cm] {$\tilde{A}$} node[pos=0.8, below=-0.2cm] {$A$}  (node2) ;
\draw[draw=black,solid,thick,<->]  (node3) to node[pos=0.2, right=-0.2cm] {$\tilde{Q}$} node[pos=0.8, right=-0.2cm] {$Q$}  (node4) ; 
\draw[black] (node1) edge [out=45,in=135,loop,looseness=4] node[midway,above=-0.2cm] {$\phi_1$}  (node1);
\draw[black] (node2) edge [out=45,in=135,loop,looseness=4] node[midway,above=-0.2cm] {$\phi_2$}  (node2);
\draw[black] (node3) edge [out=-135,in=135,loop,looseness=4] node[midway,left=-0.3cm] {$\phi_3$}  (node3);
\end{tikzpicture}}
\ee

The vacuum equations are
\be
\begin{array}{ll}
\label{pqD5Fterms}
A(\varphi_1-\varphi_2)\,=\,0\,=\,\tA(\varphi_1-\varphi_2)\,,&\quad\quad Q\,\varphi_3\,=\,0\,=\,\tQ\,\varphi_3\\
-k_1\varphi_1-{\blue{\varphi_3}}\,=\,A\,\tA\,,&\quad\quad Q\tQ\,=\,0\\
k_1\varphi_2+{\blue{\varphi_3}}\,=-\,A\,\tA\,,
\end{array}
\ee
where we indicate the contributions from the mixed CS terms due to $T(U(1))$ and $\bar{T(U(1))}$ in blue.
Let us assume that $A$ and $\tA$ are non-zero.  Therefore $\varphi_1\,=\,\varphi_2\,=\,\varphi$ (and the corresponding magnetic fluxes are set equal: $m_1\,,=\,m_2\,=m$). 
Thus, we have two branches: (1) $Q\,=\,\tilde Q\,=\,0$,  and  (2) $\varphi_3\,=\,0$.

\subsubsection*{Branch I: $Q=\,\tQ\,=\,0$} 
The moduli space is parametrised by $A\tA\,,\,\varphi\,,\,\varphi_3$ and the monopole operators, with the following constraint from the vacuum equations:
\be
-k_1\varphi-\varphi_3\,=\,A\,\tA\,.
\ee
The monopole operator $V_{\vec m}$, with flux $\vec m = (m,m,m_3)$, carries gauge and $R$ charges:
\be
q_1[V_{\vec m}]\,=\,-q_2[V_{\vec m}]\,=\,k_1m\,-\,m_3\,,\quad q_3[V_{\vec m} ]\,=\,0\,\,,\quad R[V_{\vec m}]\,=\, \frac{1}{2}\abs{m_3}
\ee
where we stress that $q_3[V_{\vec m}]\,=\,0$ since $T(U(1))$ and $\bar{T(U(1))}$ contribute $m$ and $-m$ respectively, and the non-trivial contribution to the $R$-charge is 
due to the presence of the flavour.  The baryonic generating function is given by
\be
\begin{split}
g(t;B)\,=&\,\frac{1}{1-t^2}\oint \frac{dq_1}{2\pi\,i\,q_1^{1+B}} \frac{dq_2}{2\pi\,i\,q_2^{1-B}}\frac{dq_3}{2\pi\,i\,q_3}\,\PE[t(q_1^{-1}q_2\,+\,q_1q_2^{-1})]\,=\\
=&\frac{1}{1-t^2}g^{\text{ABJM/2}}(t, B)\,=\,\frac{t^{\abs{B}}}{(1-t^2)^2}
\end{split}
\ee
where the overall $(1-t^2)^{-1}$ is due to the the fact that only one among $\varphi$ and $\varphi_3$ gets fixed. The Hilbert series is thus given by
\be
\begin{split}
H_{\text{I}}(t,z)&\,=\,\sum_{m\,=\,\-\infty}^{+\infty}\sum_{m_3\,=\,\-\infty}^{+\infty}\,z^{m+m_3}\,t^{\abs{m_3}}g(t;-k_1m\,-\,m_3)\,=\,\\
&\,=\,\sum_{m\,=\,\-\infty}^{+\infty}\sum_{m_3\,=\,\-\infty}^{+\infty}\,z^{m+m_3}\,\frac{t^{\abs{m_3}+\abs{k_1 m+m_3}}}{(1-t^2)^2}\,.
\end{split}
\ee
This turns out to be equal to the following Hilbert series of $\mathbb C^4/\mathbb{Z}_{|k_1|}$:
\be
\begin{split}
H_{\text{I}}(t,z)&=\frac{1}{|k_1|}\sum_{j=1}^{|k_1|}\, \frac{1}{(1-t\,w^j)^2(1-t\,/\,w^j)^2}\,,\quad w\,=\,z\,e^{\frac{2\pi\,i}{|k_1|}}\,, \\
&=H[\mathbb C^4/\mathbb{Z}_{|k_1|}](t,z)~.
\end{split}
\ee
This is in agreement with the geometric branch of the ABJM theory.

\subsubsection*{Branch II: $\varphi_3=\,0$} 
In this case, the vacuum equations imply that $Q \tQ =0$. The moduli space is generated by $\varphi = -\frac{1}{k} A \tilde{A}$ and the dressed monopole operators $\bar{V}_+ 
=V_{(1,1,0)} A^{k_1}$ and  $\bar{V}_- = V_{(-1,-1,0)} \tA^{k_1}$ if $k_1 >0$.  If $k_1<0$, we simply change $A^{k_1}$ to $\tA^{-k_1}$ and $\tA^{k_1}$ to $A^{-k_1}$ in these 
equations.  These dressed monopole operators satisfy the quantum relation
\be
\bar{V}_+ \bar{V}_- = \varphi^{|k_1|}~.
\ee
Hence, this branch is isomorphic to  $\mathbb{C}^2/\,\mathbb{Z}_{|k_1|}$, which is the moduli space of the half-ABJM theory.

\subsection{Comments on abelian theories with zero Chern--Simons levels}
Let us now revisit abelian theories with zero CS levels, namely those studied in section \ref{sec:zeroCS} with $N=1$, from the point of view of this section.  

One can start by taking simple examples: comparing \eref{TUNloop} to \eref{onenodenflv}. We set $N=1$ and $n=1$ in the former and set $k=0$ In the latter.   Indeed, as we discussed below \eref{onenodenflv}, such theory has a trivial Coulomb branch, because the scalar in the vector multiplets are set to zero by the vacuum equations.  This is perfectly consistent with the proposal in section \ref{sec:zeroCS}, namely the scalar fields in the vector multiplet of the gauge nodes that are connected by $T(U(N))$ are frozen.  Moreover, from \eref{dimHTUNloop}, we see that when $n=1$ the Higgs branch is also trivial; this is also in accordance with the analysis below \eref{onenodenflv}, where the meson vanishes.   Hence the two approaches, one presented in section \ref{sec:zeroCS} and the other presented in this section, yield the same results.  The same result can be derived easily for the mirror theory \eref{TUNloopmirr}, with $N=1$ and $n=1$, and \eref{twonodesoneT} with $k_1=k_2=0$.

This analysis can be generalised to other models discussed in this section.  When we set all CS levels to zero, the vacuum equations set the scalars in the vector multiplets corresponding to the gauge groups that are connected by $T(U(1))$ to zero.   Other parts of the quiver may still contribute non-trivially to the moduli space.

\section{Non-abelian theories with non-zero Chern--Simons levels} \label{sec:nonabel}
In this section, we focus on non-abelian quiver theories that contain $T(U(N))$ and/or $\bar{T(U(N))}$ theories as edges of the quiver.  In terms of a brane system, these theories
involve multiple D3-branes, along with $J$-folds and possibly with other types of branes.  In contrast to the abelian case, we do not have a general prescription of computing the Hilbert series of the geometric branch of non-abelian theories.  Nevertheless, for theories that arise from $N$ M2-branes probing Calabi-Yau 4-fold singularities, we expect that the geometric branch is the $N$-fold symmetric product of such a Calabi-Yau 4-fold.  In such cases, we can analyse the Hilbert series for each configuration of magnetic fluxes.  Let us demonstrate this in the following example.

\subsection*{One $(k,1)$ and one $(1,k')$ brane}
Let us consider the generalisation of \eref{modelN3} for non-abelian gauge groups.
\be \label{modelN3genN}
\begin{tikzpicture}[baseline, scale=0.7]
\tikzstyle{every node}=[font=\footnotesize, minimum size=1.2cm]
\node[draw, circle] (node1) at (-2,2) {$N_{k_1}$};
\node[draw, circle] (node2) at (2,2) {$N_{-k_1}$};
\node[draw, circle] (node3) at (2,-2) {$N_{-k_2}$};
\node[draw, circle] (node4) at (-2,-2) {$N_{k_2}$};
\draw[draw=purple,solid,thick,-]  (node1) to node[midway, left] {\purple{$\bar{T(U(N))}$}}  (node4) ; 
\draw[draw=black,solid,thick,-]  (node1) to (node2) ; 
\draw[draw=black,solid,thick,-]  (node3) to  (node4) ; 
\draw[draw=red,solid,thick,-]  (node2) to node[midway, right] {\red{${T(U(N))}$}}  (node3) ;
\end{tikzpicture}
\ee
In section \ref{sec:onepqandoneNS5abel}, we see that the geometric branch of the moduli space for the abelian theory ($N=1$) is a Calabi-Yau 4-fold (this is referred to as Branch 
I in that section); the latter is identified to be $\mathbb{C}^4/\Gamma(k_1, k_1 k_2-1)$.  For a general $N$, we expect that the geometric branch of \eref{modelN3genN} is the 
$N$-th fold symmetric product of $\mathbb{C}^4/\Gamma(k_1, k_1 k_2-1)$, namely $\Sym^N \left( \mathbb{C}^4/\Gamma(k_1, k_1 k_2-1) \right)$.  

Let us focus on $N=2$ in the 
following discussion. The Hilbert series of $\Sym^2 \left( \mathbb{C}^4/\Gamma(k_1, k_1 k_2-1) \right)$ is 
given by
\be \label{HmodelN3genN}
H_{\eref{modelN3genN}, \, N=2}(t,z)= \frac{1}{2} \left[ H_{\eref{model2a}}(t,z)^2 + H_{\eref{model2a}}(t^2,z^2) \right]~,
\ee   
where $H_{\eref{model2a}}(t,z)$ is given by \eref{HSpq}.  This computation can be split into five different cases depending on the fluxes and the residual gauge symmetries.
\ben
\item The magnetic fluxes for the two nodes on the upper edge are both $(m,m)$, and the magnetic flux for the two nodes on the lower edge are both $(n,n)$.  In this case, the residual gauge symmetry is $U(2) \times U(2) \times U(2) \times U(2)$.  The Hilbert series in this case can be computed as a second rank symmetric product of the abelian case (which is a product of two half-ABJM theories).  The result is 
\be
\begin{split}
H^{(1)}_{N=2}(t,z) &= \frac{1}{2} \sum_{m, n \in \BZ} \Big[ g^{\text{ABJM/2}}(t; k_1 m{ \blue -n})^2 g^{\text{ABJM/2}}(t; k_2 n{\blue-m})^2 \\
& \qquad + g^{\text{ABJM/2}}(t^2; k_1 m{\blue-n}) g^{\text{ABJM/2}}(t^2; k_2 n{\blue-m})  \Big] z^{2(m+n)}~,
\end{split}
\ee
where the terms indicated in blue are due to the mixed CS terms due to the presence of $T(U(2))$ and $\bar{T(U(2))}$ and
\be
g^{\text{ABJM/2}}(t; B) = \frac{t^{|B|}}{1-t^2}~. 
\ee
Let us report the unrefined Hilbert series, for $k_1=1$ and $k_2=2$, for this case up to order $t^{12}$:
\be \label{unref12}
\begin{split}
H^{(1)}_{N=2, \vec k =(1,2)}(t,z=1) &= 1 + 6 t^2 + 22 t^4 + 62 t^6 + 147 t^8\\
& \quad  + 308 t^{10} + 588 t^{12}+\ldots~.
\end{split}
\ee
In fact, we can also compute \eref{unref12} using the Molien integration \cite{Forcella:2007wk} as follows:
\be \label{Molien}
\begin{split}
&H^{(1)}_{N=2, \vec k=(1,2)}(t,z=1) \\
&= \oint_{|z_1|=1} \frac{\mathrm{d} z_1}{2 \pi i   z_1}  \cdots  \oint_{|z_4|=1}  \frac{\mathrm{d} z_4}{2 \pi i   z_4}   \oint_{|q_1|=1}  \frac{\mathrm{d} q_1}{2 \pi i   q_1} \oint_{|q_2|=1}  \frac{\mathrm{d} q_2}{2 \pi i   q_2} \times \\
& \quad \left( \prod_{j=1}^4 H[\BC^2/\BZ_2](t, z_j) \right)  \PE \Big[  (z_1+z_1^{-1})(z_2+z_2^{-2})(q_1 +q_1^{-1})  t\\
& \quad +  (z_3+z_3^{-1})(z_4+z_4^{-2})(q_2+q_2^{-1}) t \\ 
& \quad  - (z_1^2+1+z_1^{-2})t^2 -  (z_3^2+1+z_3^{-2})t^2  +t^4 -t^8\Big]~.
\end{split}
\ee
We have checked that \eref{Molien} agrees with \eref{unref12} up to order $t^{20}$.
Here $z_1, \ldots, z_4$ are fugacities for the gauge groups $SU(2)_{1,2,3,4}$ that are subgroups of $U(2)_{1,2,3,4}$ gauge groups corresponding to top left, top right, bottom left and bottom right nodes in \eref{modelN3genN} respectively.  The fugacities $q_1$ and $q_2$ corresponds to the two diagonal $U(1)$ gauge groups that are subgroups of $\diag(U(2)_1\times U(2)_2)$ and $\diag(U(2)_3 \times U(2)_4)$ of \eref{modelN3genN} respectively.  $H[\BC^2/\BZ_2](t,z)$ denotes the Hilbert series of the space $\BC^2/\BZ_2$, which is the Higgs and the Coulomb branches of $T(U(2))$ and $\bar{T(U(2))}$, and its expression is given by
\be
H[\BC^2/\BZ_2](t,z) = \PE \left[ (z^2 +1 +z^{-2}) t^2 - t^4\right]~.
\ee
The first and the second terms in the $\PE$ denote the contributions from the bi-fundamental hypermultiplets under $U(2) \times U(2)$. The last line of \eref{Molien} deserves some comments.  For a theory with Lagrangian, these terms would represent the contribution from the $F$-terms.  In this case, however, $T(U(2))$ and $\bar{T(U(2))}$ do not have a manifest Lagrangian description in the quiver.  Nevertheless, such terms can still be interpreted as ``effective $F$-terms'', where at $t^2$ there are relations that transform in the adjoint representations of $\diag(SU(2)_1 \times SU(2)_2)$ and $\diag(SU(2)_3 \times SU(2)_4)$. There is also a relation at order $t^4$ and a syzygy (relation among the relations) at order $t^8$.\footnote{It is instructive to compare this to the following example. Let us consider a 3d $\CN=4$ gauge theory with $U(2) \times U(2)$ gauge group with two bi-fundamental hypermultiplets.  This quiver is an $A_1$ affine Dynkin diagram, so it arises from two M2-branes probing $\BC^2/\BZ_2$ singularity.  We expect the geometric branch of this theory to be $\Sym^2(\BC^2/\BZ_2)$.  The Hilbert series of which can be computed from the Molien integral:
\be
\begin{split}
H(t,x) &= \oint_{|z_1|=1} \frac{d z_1}{2 \pi i z_1} \left( \frac{1-z_1^2}{z_1} \right) \oint_{|z_2|=1} \frac{d z_2}{2 \pi i z_2} \left( \frac{1-z_2^2}{z_2}\right)  \oint_{|q|=1} \frac{d q}{2 \pi i q} \\
& \qquad  \times \PE\left[ (z_1+z^{-1}) (z_2+z^{-2})(q+q^{-1})(x+x^{-1}) - (z_1^2+1+z_1^{-2}+1)t^2 - t^4 \right]~. \nn
 \end{split}
\ee
This is indeed equal to $H[\Sym^2(\BC^2/\BZ_2)](t,x) = \frac{1}{2}\left[ H[\BC^2/\BZ_2](t,x)^2 + H[\BC^2/\BZ_2](t^2,x^2) \right]$.  The first term in the $\PE$ is the contribution from the bi-fundamental hypermultiplets.  Since on the generic point on the moduli space $U(2) \times U(2)$ is not completely broken, but it is broken to the diagonal subgroup $\diag(U(2) \times U(2))$.  The second term indicates the $F$-terms in such a diagonal subgroup.  The last term $-t^4$ is there due to the fact that the $F$-flat moduli space is not a complete intersection because of the unbroken gauge symmetry on the moduli space (see the detailed discussion in \cite{Hanany:2010qu}).
}
\item  The magnetic fluxes for the two nodes on the upper edge are both $(m_1,m_2)$, with $m_1 > m_2$, and the magnetic flux for the two nodes on the lower edge are both $(n,n)$.  In this case, each of the $U(2)$ gauge groups on the upper edge is broken to $U(1)^2$.  Each of the $U(2)$ gauge groups on the lower edge remains unbroken.  In this case, $T(U(2))$ is expected to become $T(U(1))^2$ (and similarly $\bar{T(U(2))}$ becomes $\bar{T(U(1))}^2$).  The Hilbert series in this case is given by
\be
\begin{split}
H^{(2)}_{N=2}(t,z) &= \sum_{m_1>m_2} \sum_{n \in \BZ} g^{\text{ABJM/2}}(t; k_1 m_1-n) g^{\text{ABJM/2}}(t; k_1 m_2-n) \\
& \quad  g^{\text{ABJM/2}}(t; k_2 n-m_1) g^{\text{ABJM/2}}(t; k_2 n-m_1)  z^{m_1+m_2+2n}~.
\end{split}
\ee
As an example, for $k_1=1$ and $k_2=2$, the unrefined Hilbert series up to $t^{12}$ is
\be
H^{(2)}_{N=2, \vec k =(1,2)}(t,z=1) =  4 t^2 + 33 t^4 + 148 t^6 + 483 t^8 + 1288 t^{10} + 2982 t^{12}+\ldots~.
\ee
\item  The magnetic fluxes for the two nodes on the upper edge are both $(m,m)$ and the magnetic flux for the two nodes on the lower edge are both $(n_1,n_2)$, with $n_1>n_2$.  In this case, each of the $U(2)$ gauge groups on the lower edge is broken to $U(1)^2$.   Each of the $U(2)$ gauge groups on the upper edge remains unbroken.  $T(U(2))$ is expected to become $T(U(1))^2$, and similarly $\bar{T(U(2))}$ becomes $\bar{T(U(1))}^2$.  The Hilbert series in this case is given by
\be
\begin{split}
H^{(3)}_{N=2}(t,z) &= \sum_{n_1>n_2} \sum_{m \in \BZ} g^{\text{ABJM/2}}(t; k_1 m-n_1) g^{\text{ABJM/2}}(t; k_1 m-n_2) \\
& \quad  g^{\text{ABJM/2}}(t; k_2 n_1-m) g^{\text{ABJM/2}}(t; k_2 n_2-m)  z^{2m+n_1+n_2}~.
\end{split}
\ee
As an example, for $k_1=1$ and $k_2=2$, the unrefined Hilbert series up to $t^{12}$ is
\be
\begin{split}
H^{(3)}_{N=2, \vec k =(1,2)}(t,z=1) &=  6 t^3 + 34 t^5 + 15 t^6 + 114 t^7 + 76 t^8 + 322 t^9    \\
& \quad  + 234 t^{10}+778 t^{11} + 609 t^{12}+\ldots~.
\end{split}
\ee
\item The magnetic fluxes for the two nodes on the upper edge are both $(m_1,m_2)$, with $m_1 >m_2$. and the magnetic flux for the two nodes on the lower edge are both $(n_1,n_2)$, with $n_1>n_2$.  In this case, each of the $U(2)$ gauge groups in the quiver is broken to $U(1)^2$.  $T(U(2))$ becomes $T(U(1))^2$, and similarly $\bar{T(U(2))}$ becomes $\bar{T(U(1))}^2$.  The Hilbert series in this case is given by
\be
\begin{split}
H^{(4)}_{N=2}(t,z) &= \sum_{n_1>n_2} \sum_{m_1 > m_2} g^{\text{ABJM/2}}(t; k_1 m_1-n_1) g^{\text{ABJM/2}}(t; k_1 m_2-n_2) \\
& \quad  g^{\text{ABJM/2}}(t; k_2 n_1-m_1) g^{\text{ABJM/2}}(t; k_2 n_2-m_2)  z^{m_1+m_2+n_1+n_2}~.
\end{split}
\ee
As an example, for $k_1=1$ and $k_2=2$, the unrefined Hilbert series up to $t^{12}$ is
\be
\begin{split}
H^{(4)}_{N=2, \vec k =(1,2)}(t,z=1) &=  4 t + 10 t^2 + 54 t^3 + 115 t^4 + 350 t^5 + 643 t^6 +  \\
&\quad +1520 t^7 + 2505 t^8 + 5076 t^9 + \\ 
& \quad + 7771 t^{10} + 14142 t^{11} + 20501 t^{12} + \ldots~.
\end{split}
\ee
\item The magnetic fluxes for the two nodes on the upper edge are both $(m_1,m_2)$, with $m_1 <m_2$. and the magnetic flux for the two nodes on the lower edge are both $(n_1,n_2)$, with $n_1>n_2$.  The discussion is very similar to the previous case. The Hilbert series in this case is given by
\be
\begin{split}
H^{(5)}_{N=2}(t,z) &= \sum_{n_1>n_2} \sum_{m_1 < m_2} g^{\text{ABJM/2}}(t; k_1 m_1-n_1) g^{\text{ABJM/2}}(t; k_1 m_2-n_2) \\
& \quad  g^{\text{ABJM/2}}(t; k_2 n_1-m_1) g^{\text{ABJM/2}}(t; k_2 n_2-m_2)  z^{m_1+m_2+n_1+n_2}~.
\end{split}
\ee
As an example, for $k_1=1$ and $k_2=2$, the unrefined Hilbert series up to $t^{12}$ is
\be
\begin{split}
H^{(5)}_{N=2, \vec k =(1,2)}(t,z=1) &=  12 t^5 + 82 t^7 + 24 t^8 + 322 t^9 + 151 t^{10} \\
& \quad + 992 t^{11} + 556 t^{12}+ \ldots~.
\end{split}
\ee
\een
Indeed, the Hilbert series $H_{\eref{modelN3genN}, \, N=2}(t,z)$ given by \eref{HmodelN3genN} is then equal to the sum of the contributions from these five cases:
\be
H_{\eref{modelN3genN}, \, N=2}(t,z)= \sum_{i=1}^5 H^{(i)}_{N=2, \vec k =(1,2)}(t,z) ~.
\ee
For $k_1=1$ and $k_2=2$, we have the unrefined Hilbert series
\be
\begin{split}
H_{\eref{modelN3genN}, \, N=2, \vec k =(1,2)}(t,z=1) &= 1 + 4 t + 20 t^2 + 60 t^3 + 170 t^4 + 396 t^5  \\ 
&\quad + 868 t^6 + 1716 t^7 + 3235 t^8 + 5720 t^9 \\
&\quad + 9752 t^{10} + 15912 t^{11} + 25236 t^{12}+\ldots~.
 \end{split}
\ee
This is indeed an unrefined Hilbert series of $\Sym^2(\BC^4)$.

\section{Conclusion and open questions} \label{sec:conclusion}
In this paper, we study the moduli space of quiver theories arising from the Hanany--Witten brane system, with an insertion of $S$-folds.  In the case of $S$-flips, the quiver contains a $T(U(N))$ links between two $U(N)$ groups both with zero Chern--Simons levels.  We find that such theories have the Higgs and the Coulomb branches. The Higgs branch is given by the hyperK\"ahler quotient described in the beginning of section \ref{sec:zeroCS} and the Coulomb branch can be computed in a similar way to the usual 3d $\CN=4$ gauge theories, with the remark that the vector multiplets of the gauge nodes linked by $T(U(N))$ are frozen and do not contribute to the Coulomb branch.  We check that this proposal is consistent with mirror symmetry.   In the case of $J$-folds, we examine the moduli space of the abelian theories with $T(U(1))$ links and non-zero Chern--Simons levels systematically.  With the inclusion of bifundamental and fundamental hypermultiplets into the quiver, the moduli space can be non-trivial, and in many cases the vacuum equations admit many branches of solutions.  Finally, for the case of non-abelian theories with $T(U(N))$ links and non-zero Chern--Simons levels, we do not have a general prescription to compute the moduli space of such theories.  Nevertheless, we demonstrate the computation of the Hilbert series for an example that belongs to a special class of models arising from multiple M2-branes probing Calabi--Yau 4-fold singularities.

The results in this paper leads to a number of open questions.  First of all, it would be nice to find a general prescription to compute the moduli space of non-abelian theories with $T(U(N))$ links, non-zero Chern--Simons levels and possibly with bifundamental and fundamental hypermultiplets.    Secondly, one could introduce an orientifold place into the brane system and study the corresponding quiver theories.  For example, if we introduce an $\mathrm{O3}^-$ plane on top of the D3 brane segment that passes through the $S$-fold, an expectation is that we should have a quiver that contains a $T(SO(2N))$ link connecting two $SO(2N)$ gauge groups.  Finally, one could ask if one can replace the $T(U(N))$ link between two $U(N)$ gauge groups by the $T^{\vec \sigma}_{\vec \sigma} (U(N))$ link, with an appropriate $\vec \sigma$, between two $G_{\vec \sigma}$ gauge groups (where $G_{\vec \sigma}$ is a subgroup of $U(N)$ that is left unbroken by $\vec \sigma$).  Since $T^{\vec \sigma}_{\vec \sigma} (U(N))$ is invariant under mirror symmetry, we expect this to be a good candidate to replace $T(U(N))$ in the quiver diagram.  We hope to address these problems in future work.

\acknowledgments
We are indebted to Antonio Amariti, Benjamin Assel, Luca Cassia, Matteo Sacchi and Alessandro Tomasiello for valuable discussions.  N.M. gratefully acknowledges support from the Simons Center for Geometry and Physics, Stony Brook University at which some of the research for this paper was performed during the Simons Summer Workshop 2018.  Part of this work was also performed at the Aspen Center for Physics, which is supported by National Science Foundation grant PHY-1607611. N.M. also thanks CERN for the hospitality during the completion of this project.  I.G. acknowledges a partial support from the INFN.

\appendix
\section{Theories with multiple consecutive $J$-folds} \label{app:multiJ}
In this section, we generalise our discussion to theories dual to brane system containing $(m+1)$ consecutive $J$-folds.
\be 
\begin{tikzpicture}[baseline, scale=0.7]
\tikzstyle{every node}=[font=\footnotesize, minimum size=1.2cm]
\node[draw, circle] (node1) at (-2,2) {$1_{k_1}$};
\node[draw, circle] (node2) at (2,2) {$1_{k_2}$};
\node[draw, circle] (node3) at (6,2) {$1_{k_3}$};
\node[draw, circle] (node4) at (10,2) {$1_{k_n}$};
\node[draw, circle] (node5) at (0,4) {$1_{\hat{k}_{1}}$};
\node[draw, circle] (node6) at (4,4) {$1_{\hat{k}_{2}}$};
\node[draw, circle] (node7) at (8,4) {$1_{\hat{k}_{m}}$};
\draw[draw=black,solid,thick,<->]  (node1) to node[pos=0.2, above=-0.2cm] {$\tA_1$} node[pos=0.8, above=-0.2cm] {$A_1$}  (node2) ; 
\draw[draw=black,solid,thick,<->]  (node2) to node[pos=0.2, above=-0.2cm] {$\tA_2$} node[pos=0.8, above=-0.2cm] {$A_2$}  (node3) ; 
\draw[draw=black,dashed,thick,-]  (node3) to  (node4) ;
\draw[black] (node1) edge [out=-45,in=-135,loop,looseness=4] node[midway,below=-0.2cm] {$\phi_1$}  (node1);
\draw[black] (node2) edge [out=-45,in=-135,loop,looseness=4] node[midway,below=-0.2cm] {$\phi_2$}  (node2);
\draw[black] (node3) edge [out=-45,in=-135,loop,looseness=4] node[midway,below=-0.3cm] {$\phi_3$}  (node3);
\draw[black] (node4) edge [out=-45,in=-135,loop,looseness=4] node[midway,below=-0.3cm] {$\phi_n$}  (node4);
\draw[black] (node5) edge [out=45,in=135,loop,looseness=4] node[midway,above=0cm] {$\hat{\phi}_1$}  (node5);
\draw[black] (node6) edge [out=45,in=135,loop,looseness=4] node[midway,above=0cm] {$\hat{\phi}_2$}  (node6);
\draw[black] (node7) edge [out=45,in=135,loop,looseness=4] node[midway,above=0cm] {$\hat{\phi}_m$}  (node7);
\draw[purple] (node1) edge [out=90,in=180,looseness=0.3] node[midway,left =0.1cm] {$T(U(1))$}  (node5);
\draw[purple] (node5) edge [out=0,in=180,looseness=0.3] node[midway,above =0.1cm] {$T(U(1))$}  (node6);
\draw[purple, dashed] (node6) edge [out=0,in=180,looseness=0.3] node[midway,above =0.1cm] {$T(U(1))$}  (node7);
\draw[purple] (node7) edge [out=0,in=90,looseness=0.3] node[midway,right =0.1cm] {$T(U(1))$}  (node4);
\end{tikzpicture}
\ee
The vacuum equations are
\be
\label{linearAn}
A_i(\Phi_{i+1}-\Phi_i)\,=\,0\,,\qquad i\,=\,1,\dots,n-1
\ee
\be
\label{quadraticAn}
\begin{split}
k_1\,\Phi_1-\hat{\Phi}_{1}\,&=\,\mu_1\\
k_i\Phi_i\,&=\,\mu_i-\mu_{i-1}\qquad i\,=\,2\dots n-1\\
k_n\,\Phi_n-\hat{\Phi}_{m}\,&=\,\mu_{n-1}\\
\hat{k}_{1}\hat{\Phi}_{1}\,-\,\Phi_1\,-\,\hat{\Phi}_{2}\,&=\,0\\
\hat{k}_{i} \hat{\Phi}_{i}\,-\,\hat{\Phi}_{i+1}\,-\,\hat{\Phi}_{i-1}\,&=\,0\qquad\qquad\quad i=2\,,\dots,\, m-1\\
\hat{k}_{m}\hat{\Phi}_{m}\,-\,\hat{\Phi}_{m-1}\,-\,\Phi_{n}\,&=\,0\\
\end{split}
\ee
As in the preceding subsection, we analyse the solution of these equations according to the VEVs of bi-fundamental fields that are set to zero (\ie~ the cuts in the quiver).

\subsubsection*{No cut in the quiver}
Let us first focus on the solution in which $A_i$ and $\tilde{A}_i$ are non-zero for all $i=1,\ldots, n-1$.   Equations \eqref{linearAn} are solved as usual imposing $\Phi_1\,=\,\Phi_2\,=\,\dots\,=\Phi_n\,=\,\Phi\,=\,(\varphi,\sigma)$.  The sum of the first three equations in \eqref{quadraticAn} gives
\be \label{condmanyJ}
\left(\sum_{i=1}^n k_i\,\right)\Phi\,-\,\hat{\Phi}_{1}\,-\,\hat{\Phi}_{m}=\,0
\ee
This consistency equation must be added to set of equation formed by the last three in \eqref{quadraticAn}. Calling 
\be
K_n =\sum_{i=1}^n k_i\,
\ee
the above system of equations can be written in a compact way as:
\be
M_{CS} \,
\left(
\begin{matrix}
\Phi\\
\hat{\Phi}_{1}\\
\hat{\Phi}_{2}\\
\vdots\\
\hat{\Phi}_{m}
\end{matrix}
\right)\,=\,0
\ee
where we define the matrix $M_{CS}$ as
\be \label{syseqnsmanyJ}
M_{CS} = \left(\begin{matrix}
K_n &-1&0&0&0&\dots&-1\\
-1&\hat{k}_{1}&-1&0&0&\dots&0\\
0&-1&\hat{k}_{2}&-1&0&\dots&0\\
0&0&-1&\hat{k}_{3}&-1&\ddots&0\\
\vdots&\ddots&\ddots&\ddots&\ddots&\ddots&\vdots\\
-1&0&0&0&\dots&-1&\hat{k}_{m}
\end{matrix}\right)
\ee
Since we assumed that all $A_i$ and $\tilde{A}_i$ are non-zero, we require \eref{syseqnsmanyJ} to have a non-trivial solution; this is the case if and only if
\be \label{detMCSvanishes}
\det\,M_{CS}\,=\,0
\ee
This is a necessary condition for the existence of a non-trivial moduli space.

The magnetic flux has to be of the form
\be
{\bf \mathfrak{m}} = (\underbrace{\cm, \ldots, \cm}_{n \,\, \text{times}}, \hat{\cm}_1, \ldots, \hat{\cm}_m) \equiv (\cm^n, \vec{\hat{m}})~.
\ee
Then, \eref{syseqnsmanyJ} implies that this must satisfy the following condition:
\be
M_{CS} {\bf \mathfrak{m}}^T = 0~.
\ee
In particular,
\be \label{specialconsKn}
K_n \cm \,-\,\hat{\cm}_{1}\,-\,\hat{\cm}_{m}=\,0
\ee
The gauge charges of the monopole operator $V_{\bf \mathfrak{m}}$ are
\be
\begin{split}
q_1[V_{\bf \mathfrak{m}}] &= -(k_1 \cm - \hat{\cm}_1) \\
q_i[V_{\bf \mathfrak{m}}] &= -k_i \cm~, \qquad i=2,\ldots, n-1 \\
q_n[V_{\bf \mathfrak{m}}] &= -(k_n \cm - \hat{\cm}_{m}) \\
q_{\hat{1}}[V_{\bf \mathfrak{m}}] &= -(\hat{k}_{1}\hat{\cm}_{1}\,-\,\cm\,-\,\hat{\cm}_{2}) \\
q_{\hat{i}}[V_{\bf \mathfrak{m}}] &= -(\hat{k}_{i} \hat{\cm}_{i}\,-\,\hat{\cm}_{i+1}\,-\,\hat{\cm}_{i-1})~, \qquad i=2\,,\dots,\, m-1\\
q_{\hat{m}}[V_{\bf \mathfrak{m}}] &= -(\hat{k}_{m}\hat{\cm}_{m}\,-\,\hat{\cm}_{m-1}\,-\,\cm)~.
\end{split}
\ee
Let us now compute gauge invariant dressed monopole operators.  The last three sets of equations, setting to zero, constitute $m$ equations in total; they give a unique solution for $\vec{\hat{m}} = (\hat \cm_1, \ldots, \hat \cm_m)$ in terms of the flux $\cm$.  We denote such a solution by $\vec{\hat{m}}^*(\cm)$.   It should be emphasised that $\cm$, $\hat{\cm}^*_i$ (with $i=1,\ldots,m$), and the CS level $K_n$, must be integers. Such integrality and equations \eref{detMCSvanishes}, \eref{specialconsKn} put a constraint on the possible values of $(\hat{k}_1, \ldots, \hat{k}_m)$, as well as their relation to $K_n$, in order to obtain a non-trivial moduli space.   Note also that $\vec{\hat{m}}^*(1)+ \vec{\hat{m}}^*(-1) =0$. 

For example, in the case of three $J$-folds ($m=2$), we have $\hat{\cm}_1^*(\cm)= \frac{\hat{k}_2+1}{\hat{k}_1\hat{k}_2-1} \cm$ and $\hat{\cm}_2^*(\cm)= \frac{\hat{k}_1+1}{\hat{k}_1\hat{k}_2-1} \cm$.  From \eref{specialconsKn}, we obtain $K_n =  \frac{\hat{k}_1+\hat{k}_2+2}{\hat{k}_1\hat{k}_2-1}$. The integrality of $K_n$, $\hat{m}_1^*(m)$ and $\hat{m}_2^*(m)$ puts constraints on the values of $\hat{k}_1$ and $\hat{k}_2$:  
\be
\frac{\hat{k}_1+\hat{k}_2+2}{\hat{k}_1\hat{k}_2-1} \in \BZ~, \quad \frac{\hat{k}_2+1}{\hat{k}_1\hat{k}_2-1} \in \BZ~, \quad \frac{\hat{k}_1+1}{\hat{k}_1\hat{k}_2-1} \in \BZ~.
\ee
Since $\cm \in \BZ$, we see that the magnetic lattices given by $\hat{\cm}_1^*(\cm)$ and $\hat{\cm}_2^*(\cm)$ ``jump'' by multiples of $\frac{\hat{k}_2+1}{\hat{k}_1\hat{k}_2-1}$ and $\frac{\hat{k}_1+1}{\hat{k}_1\hat{k}_2-1}$ respectively.   If we further require that $\hat{\cm}_1^*(\cm) = \hat{\cm}_2^*(\cm)= \cm$ (\ie~ there is no such jump), we have $\hat{k}_1 = \hat{k}_2 =K_n =2$, assuming that both $\hat{k}_1$ and $\hat{k}_2$ are non-zero.

For convenience, let us define
\be
\kappa_i\,=\,\{k_1- \hat{\cm}^*_1(1) \,,\,k_2\,,\dots,\,k_{n-1}\,,\,k_n- \hat{\cm}^*_m(1)\}~, \quad \CK_i\,=\,\sum_{j=1}^n\,\kappa_j~.
\ee
For $\CK_i >0$ for all $i=1,\dots, n-1$, the basic gauge invariant dressed monopole operators are
\be
\begin{split}
\bar{V}_+ &= V_{(1^n, \vec{\hat{m}}^*(1))} A_1^{\CK_1} A_2^{\CK_2} \cdots A_{n-1}^{\CK_{n-1}} \\
\bar{V}_- &= V_{((-1)^n, -\vec{\hat{m}}^*(1))} \tilde{A}_1^{\CK_1} \tilde{A}_2^{\CK_2} \cdots \tilde{A}_{n-1}^{\CK_{n-1}} ~.
\end{split}
\ee
If $\CK_j < 0$ for some $j$, we replace $A_j^{\CK_j}$ by $\tilde{A}_j^{-\CK_j}$ in the first equation and $\tilde{A}_j^{\CK_j}$ by $A_j^{-\CK_j}$ in the second equation.  Since the $R$-charges of $V_{((\pm 1)^n, \pm \vec{\hat{m}}^*(1))}$ are zero, we have
\be
R[\bar{V}_+] = \frac{1}{2} \sum_{i=1}^{n-1} | \CK_i| \equiv \frac{1}{2} \CK~, \qquad \CK = \sum_{i=1}^{n-1} | \CK_i|~.
\ee
The generators of the moduli space are $\varphi, \bar{V}_\pm$ subject to the quantum relation
\be
\bar{V}_+ \bar{V}_- =\varphi^\CK~.
\ee
The moduli space is indeed $\BC^2/\BZ_\CK$.  We emphasise that the dependence of $\CK$ on $\hat{k}_1, \ldots, \hat{k}_m$ is due to $\hat{\cm}^*_1(1)$.  

\subsubsection*{One cut in the quiver}
Let us analyse the case  $A_l = \tilde{A}_l=0$, \ie~ the quiver is cut at the position $l$.  Equations \eqref{linearAn} implies $\Phi_1\,=\Phi_2\,=\dots=\,\Phi_l\,=\Phi$ and $\Phi_{l+1}\,=\,\Phi_{l+2}\,=\dots=\,\Phi_n\,=\,\tilde\Phi$. The sums of the first $l$ equations and the last $n-l$ ones in the first three sets of equations in \eqref{quadraticAn} imply that
\be \label{sumk}
\begin{split}
(k_1\,+\,k_2\,+\dots+\,k_l)\Phi-\hat{\Phi}_{1}\,&=\,0\\
(k_{l+1}\,+\,k_{l+2}\,+\dots+\,k_{n})\tilde\Phi-\hat{\Phi}_{m}\,&=\,0
\end{split}
\ee
These two condition must be supplemented by the last three sets of equations \eqref{quadraticAn} constraining $\hat{\Phi}_i\,\,i=1\,\dots\,m$
These can be put in a matrix form.  Calling 
\be
\sum_{i=1}^{l}k_i\,=\,K~,  \quad \sum_{i=l+1}^{n} k_i\,=\,\tilde K 
\ee 
we have
\be
M_{CS}
\left(
\begin{matrix}
\Phi\\
\hat{\Phi}_{1}\\
\vdots\\
\hat{\Phi}_{m}\\
\tilde\Phi
\end{matrix}
\right)\,=\,0
\ee
where
\be
M_{CS} = \left(\begin{matrix}
K&-1&0&0&0&\dots&0&0\\
-1&\hat{k}_{1}&-1&0&0&\dots&0&0\\
0&-1&\hat{k}_{2}&-1&0&\dots&0&0\\
0&0&-1&\hat{k}_{3}&-1&\ddots&0&0\\
\vdots&\ddots&\ddots&\ddots&\ddots&\ddots&\ddots&\vdots\\
0&0&0&0&\dots&-1&\hat{k}_{m}&-1\\
0&0&0&0&\dots&0&-1&\tilde K
\end{matrix}\right)\,
\ee
A necessary condition for the existence of the non-trivial moduli space is
\be
\det\,M_{CS}\,=0~.
\ee

The magnetic flux has to be of the form
\be
{\bf \mathfrak{m}} = (\underbrace{\cm, \ldots, \cm}_{l \,\, \text{times}}, \underbrace{\tilde{\cm}, \ldots, \tilde{\cm}}_{n-l \,\, \text{times}} , \hat{\cm}_1, \ldots, \hat{\cm}_m) \equiv (\cm^l, \tilde{\cm}^{n-l}, \vec{\hat{m}})~.
\ee
Then, \eref{syseqnsmanyJ} implies that this must satisfy the following condition:
\be \label{MCSmT}
M_{CS} {\bf \mathfrak{m}}^T = 0~.
\ee
In particular, it follows from \eref{sumk} that
\be \label{specialMCSmT}
\begin{split}
\hat{\cm}_1 &= (k_1\,+\,k_2\,+\dots+\,k_l) \cm = K \cm  \\
\hat{\cm}_m &= (k_{l+1}\,+\,k_{l+2}\,+\dots+\,k_{n}) \tilde{\cm} = \tilde{K} \tilde{\cm}~.
\end{split}
\ee
The gauge charges of the monopole operator $V_{\bf \mathfrak{m}}$ are
\be
\begin{split}
q_1[V_{\bf \mathfrak{m}}] &= -(k_1 \cm - \hat{\cm}_1) \\
q_i[V_{\bf \mathfrak{m}}] &= -k_i \cm~, \qquad i=2,\ldots, l \\
q_j[V_{\bf \mathfrak{m}}] &= -k_j \tilde{\cm}~, \qquad j=l+1,\ldots, n \\
q_n[V_{\bf \mathfrak{m}}] &= -(k_n \tilde{\cm} - \hat{\cm}_{m}) \\
q_{\hat{1}}[V_{\bf \mathfrak{m}}] &= -(\hat{k}_{1}\hat{\cm}_{1}\,-\,\cm\,-\,\hat{\cm}_{2}) \\
q_{\hat{i}}[V_{\bf \mathfrak{m}}] &= -(\hat{k}_{i} \hat{\cm}_{i}\,-\,\hat{\cm}_{i+1}\,-\,\hat{\cm}_{i-1})~, \qquad i=2\,,\dots,\, m-1\\
q_{\hat{m}}[V_{\bf \mathfrak{m}}] &= -(\hat{k}_{m}\hat{\cm}_{m}\,-\,\hat{\cm}_{m-1}\,-\,\tilde{\cm})~.
\end{split}
\ee
Let us now compute gauge invariant dressed monopole operators.  The last three sets of equations, setting to zero, constitute $m$ equations in total; they give a unique solution for $\vec{\hat{m}} = (\hat \cm_1, \ldots, \hat \cm_m)$ in terms of the fluxes $m$ and $\tilde{m}$.  We denote such a solution by $\vec{\hat{m}}^*(\cm, \tilde{\cm})$.  The integrality of such a solution, together with \eref{MCSmT} and in particular \eref{specialMCSmT}, put restrictions on the relation between $K$, $\tilde{K}$ and $\hat{k}_i$ (with $i=1, \ldots, m$).

For example, for the case of three $J$-folds ($m=2$), solving the last three sets of equations gives
\be \label{mhat}
\hat{\cm}^*_1 =\frac{\cm \hat{k}_2+ \tilde{\cm}}{\hat{k}_1 \hat{k}_2 -1}~, \qquad \hat{\cm}^*_2 =\frac{\cm + \tilde{\cm} \hat{k}_1}{\hat{k}_1 \hat{k}_2 -1}
\ee 
Using \eref{specialMCSmT} we have
\be
\cm = -\frac{\tilde{\cm}}{K+\hat{k}_2 - K \hat{k}_1 \hat{k}_2}~, \qquad  \cm =-\tilde{\cm} (\tilde{K}+\hat{k}_1 - \tilde{K} \hat{k}_1 \hat{k}_2)
\ee
Suppose that we look for a solution in which $\cm$ and $\tilde{\cm}$ are non-zero.  The integrality of $K$, $\tilde{K}$, $\hat{k}_{1,2}$ implies that
\be \label{restr1}
K \hat{k}_1 \hat{k}_2- (K+\hat{k}_2) =   \tilde{K} \hat{k}_1 \hat{k}_2 -(\tilde{K}+\hat{k}_1)= \pm 1~.
\ee
The choice $+1$ sets $\cm=\tilde \cm$, whereas the choice $-1$ sets $\cm=- \tilde{\cm}$.  Using these with \eref{mhat}, we also obtain the constriants on $\hat{k}_1$ and $\hat{k}_2$, namely
\be \label{restr2}
\frac{\hat{k}_1 \pm 1}{\hat{k}_1 \hat{k}_2 -1}~, \,\, \frac{\hat{k}_2 \pm 1}{\hat{k}_1 \hat{k}_2 -1} \in \BZ~.
\ee
Since $\cm, \, \tilde{\cm} \in \BZ$, we see that the magnetic lattices given by $\hat{\cm}_1^*$ and $\hat{\cm}_2^*$ ``jump'' by multiples of $\frac{\hat{k}_2 \pm 1}{\hat{k}_1\hat{k}_2-1}$ and $\frac{\hat{k}_1\pm 1}{\hat{k}_1\hat{k}_2-1}$ respectively.   If we further require that $\hat{\cm}_1^* = \hat{\cm}_2^*= \cm$ (\ie~ there is no such jump), we have $\hat{k}_1 = \hat{k}_2 = K = \tilde{K} = \pm2$, assuming that both $\hat{k}_1$ and $\hat{k}_2$ are non-zero.

This can easily be generalised  to an arbitrary number of $J$-folds.  The generalisation of \eref{restr1} is
\be
\mathrm{minor}_{1,1} \, M_{CS} = \mathrm{minor}_{m+1,m+1} \, M_{CS} = \pm 1
\ee
These two choices correspond to $\cm = \pm \tilde{\cm}$.  The integrality of  $\vec{\hat{m}}^*(\cm, \tilde{\cm})$ and  $\vec{\hat{m}}^*(\cm, -\tilde{\cm})$ impose further constraints on $\hat{k}_j$.   The analysis of the moduli space is similar to that presented after \eref{twochoiceskk}.

\subsubsection*{Two or more cuts in the quiver}
The analysis is similar to that of presented around \eref{condoneJtwocuts}.  For the case of two cuts, the quiver is divided into the left, central and right sub-quivers.  The analysis for the central part is presented in section \ref{sec:linquivwoJ}, whereas those for the left and right sub-quivers are as presented above for the one cut case.  One can repeat this procedure for the case with more than two cuts.

\bibliographystyle{ytphys}
\bibliography{ref}

\end{document}